\pdfoutput=1
\documentclass[11pt,letterpaper]{scrartcl}

\makeatletter
\DeclareOldFontCommand{\rm}{\normalfont\rmfamily}{\mathrm}
\DeclareOldFontCommand{\sf}{\normalfont\sffamily}{\mathsf}
\DeclareOldFontCommand{\tt}{\normalfont\ttfamily}{\mathtt}
\DeclareOldFontCommand{\bf}{\normalfont\bfseries}{\mathbf}
\DeclareOldFontCommand{\it}{\normalfont\itshape}{\mathit}
\DeclareOldFontCommand{\sl}{\normalfont\slshape}{\@nomath\sl}
\DeclareOldFontCommand{\sc}{\normalfont\scshape}{\@nomath\sc}
\makeatother

\usepackage[top=2.5cm, bottom=2.5cm, left=3.2cm, right=3cm]{geometry}

\usepackage[T1]{fontenc}
\usepackage[utf8]{inputenc}

\usepackage[hidelinks]{hyperref}
\usepackage[parfill]{parskip}
\usepackage[separate-uncertainty = true,multi-part-units=single]{siunitx}
\usepackage{slashed}
\usepackage{microtype}
\usepackage{csquotes}
\usepackage{graphicx}
\usepackage{amsmath}
\usepackage{amssymb}
\usepackage{braket}
\usepackage{booktabs}
\usepackage{subcaption}

\usepackage[numbers,sort&compress]{natbib}

\usepackage{multirow}

\usepackage{fancyhdr}
\usepackage{cleveref}

\usepackage[auth-sc,affil-it]{authblk}

\usepackage{scalefnt}
\newcommand{\abbrev}{\scalefont{.9}}

\newcommand{\NNLO}{\text{\abbrev NNLO}}
\newcommand{\NLO}{\text{\abbrev NLO}}
\newcommand{\LO}{\text{\abbrev LO}}
\newcommand{\EFT}{\text{\abbrev EFT}}
\newcommand{\SMEFT}{\text{\abbrev SMEFT}}

\newcommand{\SM}{\text{\abbrev SM}}
\newcommand{\BSM}{\text{\abbrev BSM}}
\newcommand{\IR}{\text{\abbrev IR}}

\newcommand{\QCD}{\text{\abbrev QCD}}
\newcommand{\EW}{\text{\abbrev EW}}

\newcommand{\PDF}{\text{\abbrev PDF}}

\newcommand{\LHC}{\text{\abbrev LHC}}
\newcommand{\LHCb}{\text{\abbrev LHCb}}

\newcommand{\ATLAS}{\text{\abbrev ATLAS}}
\newcommand{\HERA}{\text{\abbrev HERA}}

\newcommand{\OMP}{\text{\abbrev OMP}}
\newcommand{\MPI}{\text{\abbrev MPI}}
\newcommand{\MT}{\text{\abbrev MT19937}}
\newcommand{\MCFMNEW}{\text{\abbrev MCFM-9.0}}
\newcommand{\CPU}{\text{\abbrev CPU}}
\newcommand{\SCET}{\text{\abbrev SCET}}

\newcommand{\LHAPDF}{\text{\abbrev LHAPDF}}

\newcommand{\MCFM}{\text{\abbrev MCFM}}
\newcommand{\FEWZ}{\text{\abbrev FEWZ}}
\newcommand{\DYNNLO}{\text{\abbrev DYNNLO}}
\newcommand{\NNLOJET}{\text{\abbrev NNLOJET}}

\newcommand{\taucut}{\ensuremath{\tau_\text{cut}}}

\makeatletter
\renewcommand\maketitle{
	\begin{center}
		{\huge\bfseries\@title\par\vspace{0.3em}}
		{\scshape\@author, \@date}
	\end{center}
}
\makeatother

\fancypagestyle{firstpage}{%
	\lhead{}
	\rhead{FERMILAB-PUB-19-477-T, IIT-CAPP-19-03}
}

\begin{document}

\thispagestyle{firstpage}
\title{\LARGE Precision phenomenology with MCFM}

\author[1]{John Campbell}
\author[1,2]{Tobias Neumann}

\affil{Department of Physics, Illinois Institute of Technology, Chicago, Illinois 60616, USA}
\affil{Fermilab, PO Box 500, Batavia, Illinois 60510, USA}

\maketitle

\vspace{1cm}

\begin{abstract}
        Without proper control of numerical and methodological errors in theoretical predictions at the per mille level
        it is not possible to study the effect of input parameters
        in current hadron-collider measurements at the required precision.
         We present a new version of the parton-level code \MCFM{} that
	achieves this requirement through its highly-parallelized nature, significant performance improvements
        and new features.  An automatic differential cutoff extrapolation is introduced to assess the
        cutoff dependence of all results, thus ensuring their reliability and potentially improving fixed-cutoff
	results by an order of magnitude.        
        The efficient differential study of \PDF{} uncertainties and \PDF{} set differences at \NNLO{},
        for multiple \PDF{} sets simultaneously, is achieved by exploiting correlations.        
        We use these improvements to study uncertainties and \PDF{} sensitivity at  \NNLO{},
        using 371 \PDF{} set members. 
        The work described here permits \NNLO{} studies
        that were previously prohibitively expensive, and
        lays the groundwork necessary for a future implementation
        of \NNLO{} calculations with a jet at Born level in \MCFM{}.
\end{abstract}

\clearpage
\tableofcontents

\section{Introduction}

The production of bosons, either singly or in pairs, provides the bread and butter for \LHC{} analyses
that perform precision tests of the Standard Model (\SM{}).  Consequently, they also serve
as probes of physics beyond the Standard Model (\BSM{}), as well as arenas in which to perform
resilient extractions of fundamental parameters of the \SM{} and non-perturbative inputs
such as parton density functions (\PDF{}s). For this to be the case it is essential that the fixed-order
calculations, upon which these analyses rely, are pushed to as high order in the perturbative expansion
as possible.  After many years of effort, \QCD{} corrections have reached the level of \NNLO{}
and electroweak effects are commonly available at \NLO{}.  Broadly speaking, these push the perturbative
truncation uncertainty of these fixed order predictions to the percent level.  At the same time, uncertainties
related to the determination of \PDF{}s and $\alpha_s$ have been reduced to the same level.
To make further progress it is therefore imperative to have robust tools that can
systematically compute predictions at this level of precision, whilst maintaining sensitivity to
possible differences between theoretical inputs such as the choice of renormalization scale or
\PDF{} set.

Although by now many calculations have been performed at \NNLO{} in \QCD{}, far fewer have resulted
in public codes.\footnote{Fully differential codes for various boson and diboson processes are for example
{\abbrev DYNNLO} \cite{Catani:2009sm}, {\abbrev HNNLO} \cite{Catani:2007vq,Grazzini:2008tf,Grazzini:2013mca}, {\abbrev 
SusHi} \cite{Harlander:2012pb,Harlander:2016hcx}, ggHiggs 
\cite{Bonvini:2014jma}, FehiPro \cite{Anastasiou:2005qj,Anastasiou:2009kn}, proVBFH(H) 
\cite{Cacciari:2015jma,Dreyer:2018rfu}, 
2$\gamma$NNLO \cite{Catani:2011qz}, FEWZ 
\cite{Gavin:2010az,Gavin:2012sy,Li:2012wna} and GENEVA \cite{Alioli:2015toa}.
}  Most of these are restricted to a single process, although some codes do offer additional
specialized features such as the inclusion of electroweak corrections or resummation.
A notable exception is {\abbrev MATRIX} 
\cite{Grazzini:2017mhc}, which in its current release features fixed order \NNLO{} implementations of single boson 
processes
and a subset of diboson processes.
Related, but more complicated, final states that also include the presence of an additional hard
jet have also been computed at \NNLO{} but are only available as private
codes, see for example ref.~\cite{Bellm:2019yyh} and references therein.

\MCFM{} is a publicly available code, with version $1.0$ released in 2001 focusing on
NLO corrections to vector boson pair production processes~\cite{Campbell:1999ah}.  Since then 
the code has been continuously maintained with updated and new processes at \NLO{} and beyond.
In 2015 multi-threading capability using OpenMP
was added \cite{Campbell:2015qma}, enabling multi-core desktop systems to compute
the most complicated \NLO{} processes. In 2016 an initial set of
color singlet \NNLO{} processes was included \cite{Boughezal:2016wmq} together with \MPI{}
capability, allowing full use of cluster systems.
\MCFM{} is now capable of computing $W^\pm$, $Z$, $H$ as well as 
$\gamma\gamma$, $W^\pm H$, $ZH$ and $Z\gamma$ production processes
at \NNLO{}, as well as hundreds of processes at \NLO{}. Some processes include \NLO{} electroweak corrections 
\cite{Campbell:2016dks}, while others account for contributions from \BSM{} sources and anomalous couplings, as well 
as \NLO{} corrections for the \SMEFT{} \cite{Neumann:2019kvk}. All 
leptonic decay channels of $Z$ and $W^\pm$ are included as well as the Higgs boson decays into 
$\gamma\gamma$, $W^+W^-$, $ZZ$, $Z\gamma$, $\tau^+\tau^-$ and $\overline{b}b$.

With its flexibility, ease of use and performance, \MCFM{} has been an indispensable ingredient in hundreds of
experimental studies for the comparison of theory predictions and nature. However, the utility of the code is
not confined to just the hands of experimentalists.
One of the goals of \MCFM{} was always to provide a collection of analytic results to serve as a platform
for further work, where others can extend, modify, or reuse parts
without significant help from the authors. In the past, elements of the \MCFM{} code have been used extensively
for this purpose.  For example, codes such as
{\abbrev DYNNLO} \cite{Catani:2009sm}, {\abbrev DYRes} \cite{Catani:2015vma,Bozzi:2010xn}, 
{\abbrev HNNLO} \cite{Catani:2007vq,Grazzini:2008tf,Grazzini:2013mca}, {\abbrev HRES} 
\cite{deFlorian:2012mx,Grazzini:2013mca} as well as most recently
{\abbrev MCFM-RE} \cite{Arpino:2019fmo} are all based on the \MCFM{} framework. Other public codes benefit from
various parts of \MCFM{} and its efficient implementation of amplitudes.\footnote{See e.g. 
refs.~\cite{Carli:2010rw,Gao:2012ja,Melia:2012zg,Anderson:2013afp} for its use in other codes, and 
\url{https://inspirehep.net/search?ln=en&p=find+fulltext+MCFM} for many more examples. }

This paper describes an update of the parton level code \MCFM{} that includes significant improvements
in its usability, reliability, maintainability and  performance. For example, the integration has been rewritten to 
adaptively reach a specified precision goal. To establish trust in our results at the 
per mille level, we perform a detailed study of the performance of the Vegas~\cite{Lepage:1980dq} integration algorithm 
in our code.  In particular, we focus on the convergence of the integral and the reliability of its error estimates. 

With improvements in the jettiness slicing cutoff ($\taucut$) dependence we improve and update previous benchmark 
results. We furthermore introduce the automatic sampling of additional $\taucut$ values fully differentially, making 
use of the correlations to decrease numerical uncertainties by orders of magnitude. The $\taucut$ dependence and its 
automatic fitting to the known asymptotic behavior can be used to reliably assess systematic $\taucut$ errors.

For the first time, multiple \PDF{} sets can be used at the same time for the evaluation of \PDF{}
uncertainties in the same correlated way. We demonstrate the calculation of \PDF{} set differences and uncertainties at 
\NLO{} and \NNLO{} 
for differential distributions with sub per mille level numerical accuracies.  We then compare \PDF{} uncertainties
obtained using lower order matrix elements for a broad range of processes to understand the level of precision that 
may be expected when estimating \PDF{} uncertainties at \NNLO{} through this procedure.
Making use of correlations, we can furthermore study the
\emph{differences} between any number of \PDF{} sets in our improved setup at the sub per mille level.

In \cref{sec:newfeatures} we introduce the new and improved features of \MCFM{}, and support them with
technical data, details and benchmarks. In \cref{sec:integrationuncertainties} we study the performance of the
Vegas integration routine, comparing the use of a newly introduced low-discrepancy sequence with that of a
pseudo-random number sequence. We focus on issues regarding the estimation of numerical integration uncertainties,
comparing against approaches in the literature and suggesting improvements.
In \cref{sec:benchmark}
we report on the performance gains resulting from using the boosted definition of the jettiness slicing variable and the inclusion 
of power corrections differentially at \NNLO{}. We also present our fully differential automatic $\taucut$ extrapolation 
based on the theoretical 
asymptotic $\taucut$ dependence and discuss its use and limitations. In \cref{sec:physics} we study \PDF{} uncertainties
at \NNLO{} using six \PDF{} sets simultaneously for all \NNLO{} processes in \MCFM{} and compare it against the use
of lower order matrix elements. The correlated multi-\PDF{}-set integration allows for per mille comparisons between 
different \PDF{} sets, which are also covered for the Higgs transverse momentum distribution at large values.
In \cref{sec:WZphysics} we comment on issues in precision studies in $W$ and $Z$-boson physics, where current
experimental data now requires per mille level theory predictions. We compare benchmarks results in the 
literature with our predictions.  Our conclusions are summarized in \cref{sec:conclusions}.

\section{New and improved features in \MCFM{}}
\label{sec:newfeatures}

A number of features have been introduced to simplify the operation
of the \MCFM{} code.  In this section we summarize the most important new and
modified features.
In passing, we note that this version represents
an overhaul of many key components of the code and, for the sake of clarity,
we have removed a number of features introduced in previous versions that had
been largely unused.\footnote{
	These include the ntuple interface, as well as the option to write output to
	LHEF files. These can easily be added back at a later time, according to user
	demand.  In the meantime older versions of \MCFM{}
	can still be downloaded to make use of these features.}
This version aims to be compliant with Fortran~2008 and fully supports GCC
versions newer than $7$ and Intel compilers newer than $19$.  Benchmark
comparisons of compiler versions and optimization flags are described in
\cref{sec:newfeatures-app}. A technical description of the
new features presented in this section and their configuration within \MCFM{} is also given in
\cref{sec:newfeatures-app}. 

\MCFM{} is distributed in a fully self-contained form, where all amplitudes are bundled as optimized explicit 
expressions, making use of the included {\abbrev QCDLoop}~\cite{Ellis:2007qk,Carrazza:2016gav}
distribution for the evaluation of one-loop scalar integrals. Expressions for multi-loop integrals in terms of harmonic 
polylogarithms are evaluated with {\abbrev TDHPL} \cite{Gehrmann:2001jv}.
While a number of \PDF{} sets are included through a native interface, the \LHAPDF{} library \cite{Buckley:2014ana}
can also be linked. This enables using \PDF{} uncertainties and a larger 
number of \PDF{} sets.
Through \LHAPDF{} one also gains the flexibility of easily interchanging grid interpolation routines.
While we do not enforce citations for using \MCFM{} or results obtained with \MCFM{} (\MCFM{} is now released under the 
GPL-3.0 license), we encourage citing at least the papers printed at the beginning of a run. In practice, the 
results in \MCFM{} depend on a deep tree of results (for instance, for the calculation of the matrix elements)
and we hope that users use the appropriate citation depth to acknowledge research appropriately. 

Already with \MCFM{}-8.0 inclusive cross sections can be computed precisely at \NNLO{} on a modern multi-core desktop
computer in a few hours. Achieving sub percent level precision also in tails of distributions requires
more computing resources. Furthermore, adding features like automatic scale variation or automatic
computation of \PDF{} uncertainties increases required computational resources. Apart from buying
more computers, improvements in various parts of the theory predictions and the code, both which are covered
in this study, can be made to
make these computations feasible on smaller sized clusters.

\subsection{User interface}

For this version we have first introduced a new, more flexible input file
format.  The options from the input file can now be over-ridden via command line arguments as well,
which can be useful for batch parameter run scripts.  
Second, we have re-implemented the Vegas algorithm \cite{Lepage1} and the surrounding integration routines, 
including a new alternative to the pseudo-random numbers used in previous versions of code.
By default, we now use the Sobol low discrepancy sequence
\cite{Bratley1988,Fox1986,Antonov1980,Sobol1977,Sobol1976} that is described in detail
in \cref{sec:sobol}.

With the new integration routines all parts of a \NLO{} or \NNLO{} calculation are
now chosen adaptively based on the largest absolute numerical uncertainty. 
A precision goal can be set in the input file as well as a $\chi^2/\text{iteration}$
goal and a precision goal for the warmup run. If the goals for the warmup are not reached, the warmup
repeats with twice the number of calls.  Our new version also allows the
integration to be resumed from any point from a previous run, using snapshots
that save the whole integration state automatically.  This allows the
precision goal to be reduced in a future run without starting the whole integration
from scratch. Further configuration options have been introduced to control the
stages of the integration that can provide benefits over the default settings
in certain situations, such as when calculating PDF uncertainties,
as described in detail in \cref{sec:newfeatures-app}.

\MCFM{} allows to easily specify the most common kinematical cuts in the input file and automatically fills
a pre-defined set of histograms. The cuts from the input file can be modified and augmented in a prepared subroutine 
for user cuts. Additionally, a re-weighting function has been introduced that multiplies
all events in the integration. This feature can be used as a manual importance sampling technique to give
tails of distributions a larger weight so they are integrated with the same relative precision as numerically larger 
contributions. For example in transverse momentum distributions one can reweight with an exponential function of the 
transverse momentum to approximately flatten the distribution and obtain an equal relative precision in all bins. Last, 
histograms of arbitrary kinematical variables can easily be added for full flexibility.

We have implemented a new suite of histogram routines that allows for any number of
histograms with any number of bins, each of which is dynamically allocated. Furthermore, everything is also handled in 
a 
fully multi-threaded approach within the integration. For each \OMP{} thread temporary 
histograms are allocated and filled that are then reduced to a single one after each integration iteration.
These histograms are also written out at every intermediate stage of the integration, and are
updated appropriately when an integration run is resumed with a new precision goal. This allows the user
to inspect the results already during the integration and gives the possibility to interactively stop the 
integration as soon as the results are satisfactory.

\subsection{Correlated calculations with multiple scales, \PDF{} sets and $\taucut$ values.}

When integrating multiple functions at once one can make use of the correlations between the
integrands, and straight away obtain significantly lower integration uncertainties on the differences and ratios
between the different integrands. Generally one expects the difference 
between two perfectly correlated integrands to be computed with the same relative uncertainty
as that of either integral. In practice, we find that the absolute numerical uncertainty
computed for the difference between an integral and the central value turns out to be roughly an order
of magnitude lower than the numerical uncertainty on the central value itself.
This idea had already been employed in previous versions of \MCFM{} for the calculation
of renormalization and scale uncertainties.  In this version we extend the treatment to
the calculation of \NLO{} and \NNLO{} predictions with different values of the $0$-jettiness cutoff $\taucut$,
and to calculations with different \PDF{} sets and members.

\paragraph{\PDF{} uncertainties and \PDF{} set differences.} 

Apart from FEWZ~\cite{Gavin:2010az},
no (public) code is known to the authors that computes \PDF{} uncertainties automatically while 
taking into account  correlations between \PDF{} set members.  Indeed, studies commonly avoid the expensive
calculation of fixed order \NNLO{} results
convoluted with different \PDF{} sets for central values and \PDF{} uncertainties, for example 
\cite{Campbell:2017dqk,Gehrmann-DeRidder:2019avi}.
Instead, relative \PDF{} uncertainties or differences between sets are calculated using fixed order \NLO{} matrix 
elements convoluted with \NNLO{} \PDF{}s.  Other alternatives include frameworks like fastNLO  \cite{Kluge:2006xs}
and applGRID \cite{Carli:2010rw}, which were developed to accelerate \PDF{} fits and to reduce
the burden of computing \PDF{} uncertainties.

In \MCFM{}-8.1 \PDF{} uncertainties could successfully be computed inclusively at \NLO{}. In \MCFMNEW{} we
enable studying calculations of \PDF{} uncertainties differentially at \NNLO{} on smaller sized clusters with just 
a few hundred cores in total.\footnote{The per-mille level differential computations with uncertainties from six \PDF{} 
sets simultaneously in 
this paper were performed on a cluster using at most 16 nodes, where the nodes have 
AMD and Intel CPUs from 2010, for details see \cref{sec:newfeatures-app}.}
This is achieved through a fully parallelized 
\OMP{}+\MPI{} interface to \LHAPDF{} that is based on the new object oriented treatment
of \PDF{}s in \LHAPDF{}~6.  This interface thereby avoids the limitation to have at most one simultaneous
call to the library from the \OMP{} threads. We also added the capability to handle any number of \PDF{} sets, with or 
without \PDF{} uncertainties, limited only by the available system memory. 
Studying precise differences between \PDF{} sets, at the sub per mille level, can then be performed with only little
computational overhead compared to using just one central value.

Note that to enable \PDF{} uncertainties, \MCFM{} has to be compiled with
\LHAPDF{} \cite{Buckley:2014ana} support. Any sets  available
for \LHAPDF{} can be used, and the uncertainties are estimated according to the provided \LHAPDF{}
uncertainty routines for replica and Hessian sets.  In addition, as a further improvement to previous
versions, we also allow for sets with additional members that use different values of $\alpha_s(m_Z)$,
so that combined $\PDF{}+\alpha_s$ uncertainties can be computed at the same time.

\begin{figure}[h]
	\centering
	\includegraphics[]{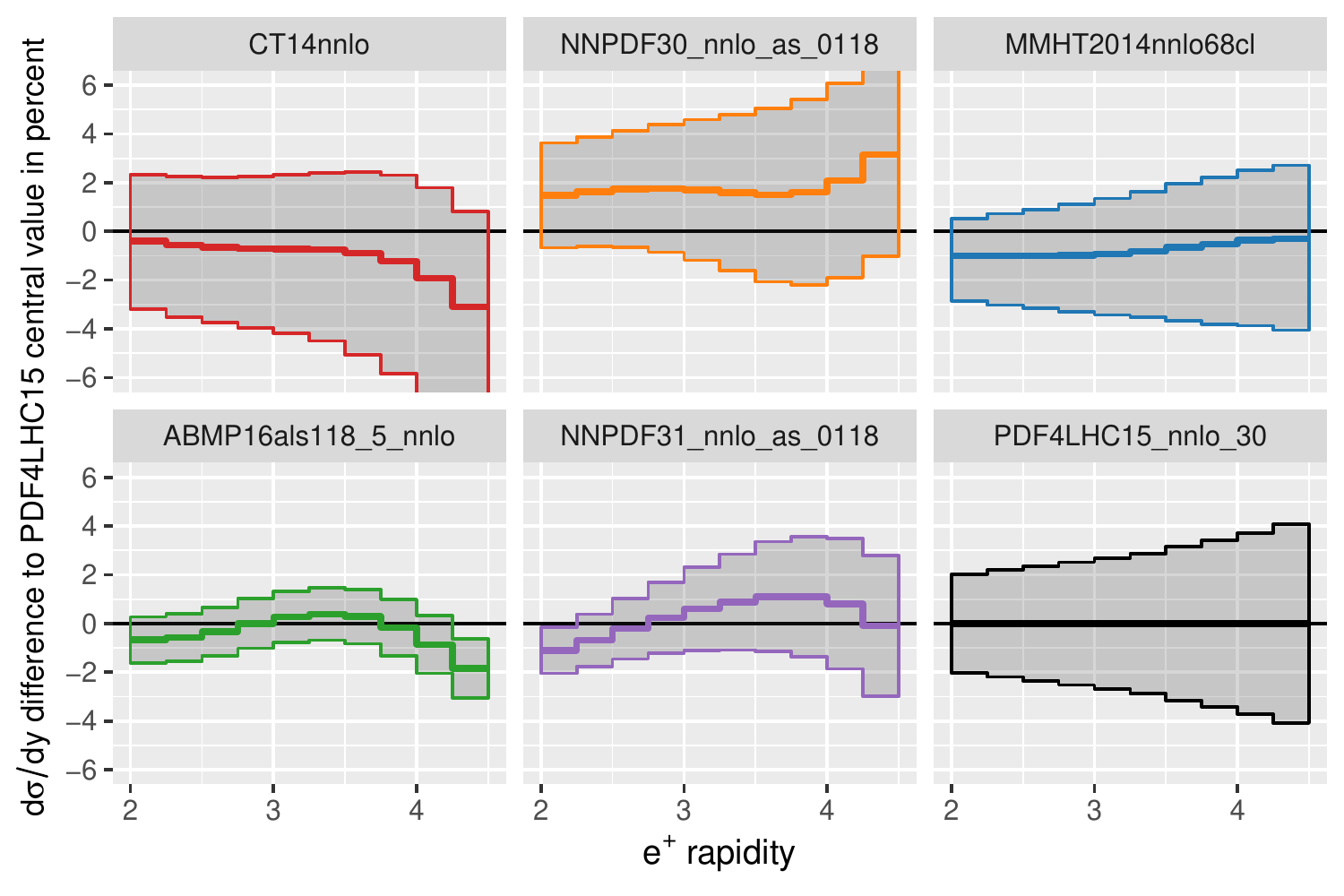}
	\caption{\NNLO{} $e^+$ rapidity distributions for $W^+$ production in the forward region, computed
        with uncertainties from a variety of \PDF{} sets, normalized to the central {\abbrev PDF4LHC} prediction.}
	\label{fig:Wteaser2}
\end{figure}

To demonstrate this procedure at work, in \cref{fig:Wteaser2} we show normalized $e^+$
rapidity distributions in $W^+$ production.  This calculation has been performed at \NNLO{},
using  six \NNLO{} \PDF{} sets simultaneously and with \PDF{} uncertainties for
all of them, so calls to $371$ \PDF{} members for each phase space point. The normalization is with respect to the 
central value of the {\abbrev PDF4LHC}~\cite{Butterworth:2015oua} set \texttt{PDF4LHC15\_nnlo\_30} to show
the differences between this set and the other sets. Cuts and parameters are standard as in \MCFM{}-8.0 and 9.0, 
introduced later in \cref{sec:physics}. Only the positron rapidity $y$ has been constrained to $2.0 < y < 4.5$.
It is clear that this feature allows precision
studies of differences between predictions of various \PDF{} sets and their uncertainty bands. 

We hope that our fast multi-set implementation in \MCFM{}, that can inxpensively study $0.1\%$ differences between
\PDF{} sets and their associated uncertainties, will help facilitate further theoretical work in these directions.
For instance, differences such as those shown in \cref{fig:Wteaser2} and \cref{sec:physics} can be explored
and criticism of the {\abbrev PDF4LHC} paradigm \cite{Accardi:2016ndt} can be supported or refuted directly
at the level of \NNLO{} kinematical predictions.  Furthermore, recent work that 
incorporates theory uncertainties more rigorously in \PDF{} fits \cite{AbdulKhalek:2019bux} can
again be studied directly and efficiently with \MCFMNEW{}.

\paragraph{Multi-$\taucut$ integration and automatic asymptotic fitting.} 

For $0$-jettiness calculations, either at \NLO{} or \NNLO{}, we now allow an array of $\taucut$ values
to be specified in addition to the nominal value of $\taucut$.  The Vegas integration grid is
still adjusted according to the nominal value, and the remaining values are sampled on the fly
with little computational overhead.  This means that values smaller than the nominal value of $\taucut$
are only computed with relatively large uncertainties, although this may still yield useful information.
In contrast, any values larger than the nominal one are computed with approximately the same precision
as the nominal value, and are therefore highly reliable. If no values of additional values of $\taucut$
are specified, the code automatically chooses further values, see \cref{sec:benchmark}.

In all cases
an automatic fit to the known asymptotic behavior is performed for the total cross section as well
as for all histograms differentially. The histograms for the nominal and individual $\taucut$ values are written 
separately from the 
histogram with just the fitted corrections. With this procedure one can quickly and 
easily check the $\taucut$ dependence of the result and estimate the effect of a non-zero value of $\taucut$.
See \cref{sec:benchmark} for details on this procedure.
 
\begin{figure}[h]
	\centering
	\includegraphics[]{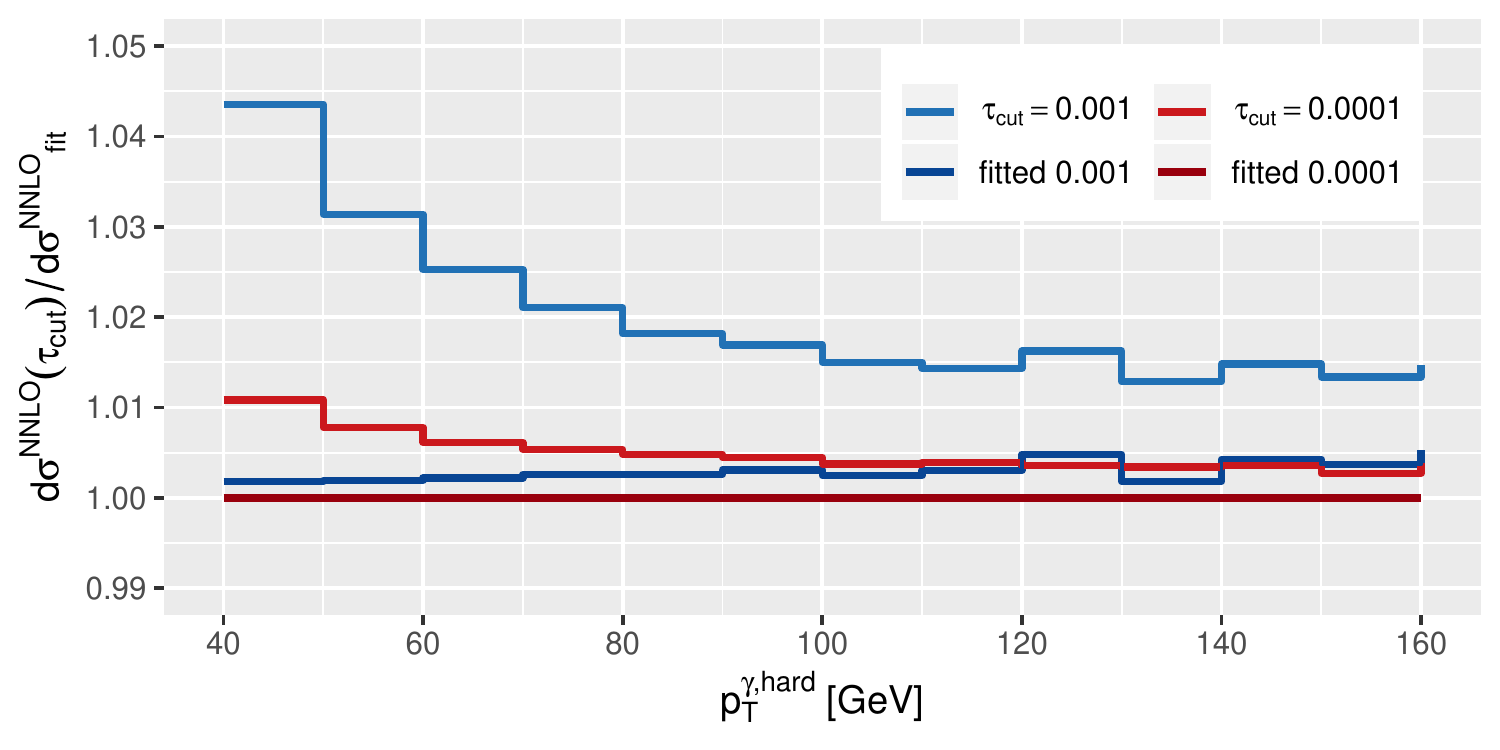}
	\caption{The transverse momentum distribution of the hardest photon in diphoton production, computed at \NNLO{}.
        Results are shown for $\taucut=10^{-3}$\,GeV (blue) and $\taucut=10^{-4}$\,GeV (red), as well as the results obtained
        using automatic fitting (darker blue and red).  All results are normalized to the fit from $\taucut=10^{-4}$\,GeV.}
	\label{fig:fitteaser}
\end{figure}

All $\taucut$ dependence plots in this paper are computed with the multi-$\taucut$ feature and its automatic fitting.
For example \cref{fig:fitteaser} displays the hardest photon $p_T$ spectrum in diphoton production at \NNLO{} using 
the fully automatic differential fit for two runs with nominal
$\taucut$ values of $10^{-3}$\,GeV and $10^{-4}$\,GeV. The latter 
choice is our default and is expected to result in systematic cutoff effects of
less than one percent. In both cases the fit significantly improves the fixed $\taucut$ predictions, and for
$\taucut=10^{-4}$\,GeV it allows one to estimate that indeed the residual dependence is at most $1\%$, while for the choice of
$\taucut=10^{-3}$\,GeV the effects are up to $4.5\%$. The fitted results agree with each other at around the one half
percent level (up to numerical noise), so allow one to choose a nominal $\taucut$ one order of magnitude larger.
For practical applications the purpose of the additionally sampled $\taucut$ values and the fit is that no separate 
runs with different nominal $\taucut$ values have to be run. In order to make use of the fit its quality 
-- separately reported by \MCFM{} -- must be inspected according to \cref{sec:benchmark}.

\section{Integration performance and uncertainty estimation}
\label{sec:integrationuncertainties}

The numerical integration procedure in this version of \MCFM{} has been overhauled and extended,
with the aim of providing a platform with which to perform detailed high-precision studies
of the most complicated \NNLO{} processes.  In this section we describe a number of the new features
and benchmark their performance.

First we study our implementation of the Sobol low discrepancy sequence
as an alternative to  the \MT{} pseudo random number generator.
Although Cuba \cite{Hahn:2004fe}, a popular integration library, 
has used the Sobol sequence as default for a long time already, to our knowledge no systematic
study of its impact in Monte Carlo generators has been performed so far.
We therefore describe a series of benchmarks quantifying the
behavior of the  Sobol low discrepancy sequence in \MCFM{} and compare the performance of 
both sequences.

Second, we investigate the reliability of the result of the integration, including the uncertainty
estimate.  Modern complicated \NLO{} and \NNLO{} calculations are computationally intensive and
typically require of the order of many \CPU{}  months. Since one 
does not want to 
wait months and years to run a \NNLO{} calculation on a single core, parallelization is a straight-forward, easy 
and highly-efficient 
approach for Monte Carlo integrations. An outstanding feature of \MCFM{} is its parallelization in both \OMP{} and 
\MPI{} on multi-core machines and for cluster setups. With this version the 
integration is also fully resumable. Complications arise when the Vegas integration is not parallelized, and one 
tries to combine many independent integrations with low statistics where uncertainties are typically 
underestimated. In order to prevent such problems, various approaches have been taken in the literature
that go beyond a na\"ive combination. We study approaches performed in the literature, suggest improvements, and furthermore compare these with the 
fully parallelized Vegas integration in \MCFM{} that outright avoids such complications.

\subsection{Low discrepancy sequence integration}
\label{sec:sobol}

If an integrand has a so-called bounded 
variation, implying certain 
smoothness properties, see for example the pioneering 
publications for the Koksma--Hlawka inequality \cite{Koksma1942,Hlawka1961}, the Monte 
Carlo integration error has a bound that is proportional to this bounded variation.
The bounded variation is just a property of the integrand and the discrepancy of 
the sequence that defines the points being sampled. Since it is a fixed property, one can try to 
construct sequences that have a lower discrepancy than true or pseudo random numbers
to improve the Monte Carlo integration.
In this sense the asymptotic discrepancy of zero for equidistributed numbers is optimal. The problem for the rectangle 
rule with points $x_i = i/N$ for example, where $N$ is the total number of calls, is that it has a fixed length and any 
increase of the number of calls to get an improved estimate leads to a recomputation that is not statistically 
independent.  However, sequences have been constructed that keep the 
benefits of random numbers (statistically independent estimates) with the additional benefit 
of a lower discrepancy and additional 
uniformity properties in higher dimensions. Their discrepancies are generally bounded by $ \sim 
(\log N)^{d} / N$, where $N$ is the number of integrand evaluations and $d$ the dimensionality.  This is to be compared 
with true random numbers, where the asymptotic uncertainty in a Monte Carlo integration decreases only as 
$1/\sqrt{N}$.

In practice, it is unclear whether a bounded variation for our integrands exists or how large it is. Therefore,
it is also unclear if one can benefit from using a low discrepancy sequence instead of pseudo random numbers.
To test this, we use our implementation of the Sobol sequence that is based on the code \texttt{sobseq} \cite{Vugt2016},
extended to 64~bits with a  maximum sequence length of $2^{63}$ instead of $2^{31} \simeq 2.15
\cdot 10^9$.  The extension to 64~bits is important because the 32-bit limit can quickly be saturated in \NNLO{} 
calculations,
and even in very precise \NLO{} calculations. We use the initialization numbers from  
refs.~\cite{Joe2010,Joe2008,Joe2003}.

To test the performance of the sequence, compared to the usual pseudo random numbers, we run benchmarks for the \NNLO{}
Higgs production double real emission calculation. We focus on the real emission since these 
contributions are usually the most computationally intensive ingredients in higher order calculations. The cuts are 
standard, and described later in \cref{sec:physics}, but do not matter for the discussion here.
The jettiness slicing cutoff $\tau_\text{cut}$ is set to $0.002$\,GeV, corresponding to a systematic precision goal of 
$0.2\%$ for the full cross section with \MCFM{}-8.0 and $\ll 0.1\%$ with \MCFMNEW{}.\footnote{For the default dynamic choice of $\taucut$ 
as in \cref{sec:benchmark} this corresponds to $\taucut/m_\text{Born}=1.6 \times 10^{-5}$.} The dipole subtraction $\alpha$ 
parameter restricting the dipole phase space is set to $1.0$ \cite{Nagy:1998bb,Nagy:2003tz}, which corresponds to the 
original unrestricted formulation. To achieve a $0.2\%$ precision goal also numerically the double real emission 
has to be computed with an uncertainty better than $4$ (fb) or a relative precision of $10^{-4}$.

\begin{figure}
	\includegraphics[width=\columnwidth]{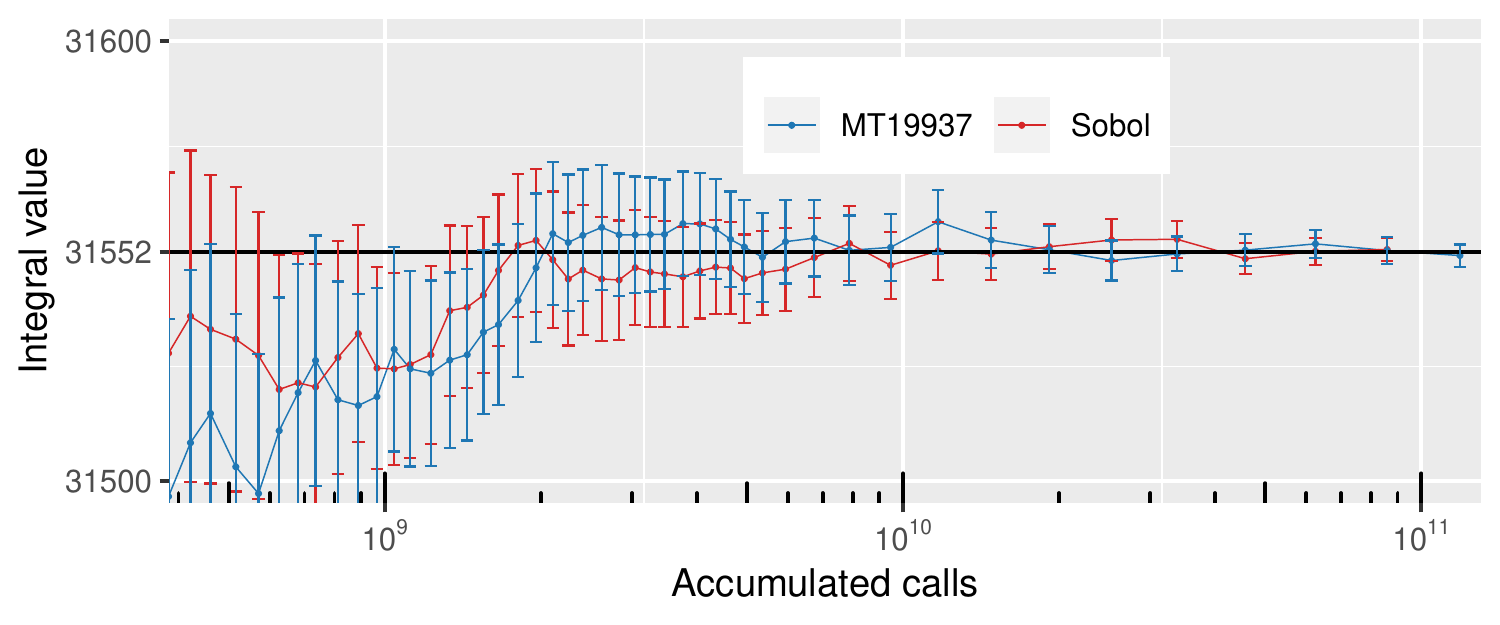}
	\caption{The dependence of the integral for the \NNLO{} Higgs production double real emission contribution
		(for $\alpha=1$ and $\taucut=0.002$\,GeV) on the accumulated number of calls. 
                The red data corresponds to using the Sobol sequence and 
	        blue to a \MT{} sequence. Each point represents a new estimate from a new iteration.}
	\label{fig:nnloRRnumbench}
\end{figure}

\Cref{fig:nnloRRnumbench} shows the results of a comparison between the Sobol and one seeded \MT{} sequence,
as a function of the accumulated number of Vegas calls.  For this study we have multiplied the
number of calls per iteration by $1.4$ after every five iterations up to $500\cdot10^6$ calls per 
iteration, and from that point on we multiply by $1.4$ after every single iteration.\footnote{For the first five warmup 
	iterations the Vegas grid adjustment parameter was set to $1.5$, afterwards to $0.8$, discarding the warmup 
	estimate 
	but keeping the grid. The number of grid subdivisions was kept at the default of $100$.}
By using this factor, after two iterations (or two batches of iterations) the number of calls per iteration has been 
(approximately) doubled. This exponential increase allows successive Vegas iterations to make sufficient
progress in reducing the uncertainty estimate.
From the figure, the approach to the horizontal line -- that represents the final average obtained
using both sequences ($31552$ with an uncertainty of $\pm2.5$) -- is equally good with both sequences.
However, we note that for a smaller number of calls (not displayed in the figure)
the unwritten rule to multiply Monte Carlo integration 
uncertainties by a factor of two is useful to see that the two are equivalent.
We note that for the largest number of accumulated calls shown here, the number of calls per iteration reached $20\cdot10^9$ 
and one might benefit from increasing the number of Vegas grid subdivisions beyond their default value of $100$.

\begin{figure}
	\includegraphics[width=\columnwidth]{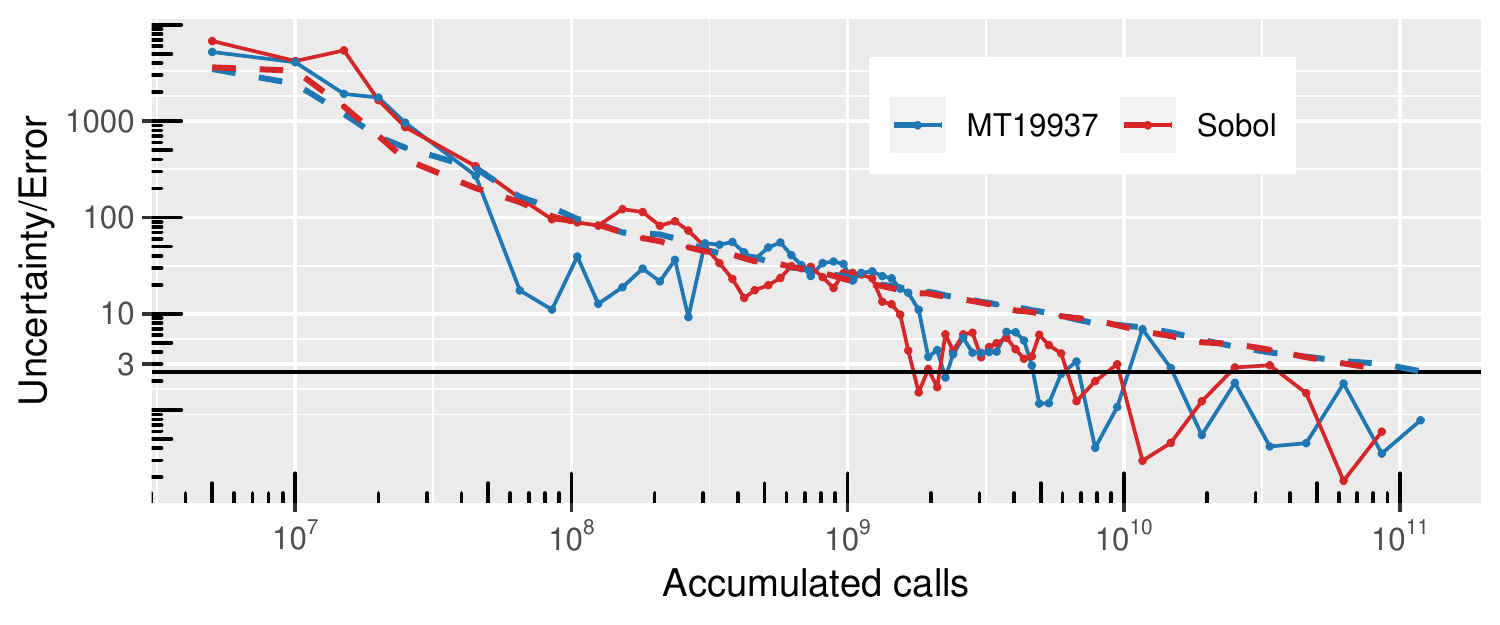}
	\caption{The dependence of the estimated uncertainty of the \NNLO{} Higgs production double real emission contribution
		(for $\alpha=1$ and $\taucut=0.002$\,GeV) on the accumulated number of calls.  The red data corresponds to using the Sobol sequence and 
	blue to the \MT{} sequence. Each point represents a new estimate from a new iteration. The dashed line represents 
	the uncertainty from the Vegas routine, while the points
and solid line represent the ``true'' error, assuming that the average from the last reported numbers of 
sequences is the most precise with an uncertainty of $\pm2.5$.}
	\label{fig:nnloRRnumbench_errors}
\end{figure}

As an alternative measure of performance, in \cref{fig:nnloRRnumbench_errors} we directly show the
dependence of the integration uncertainty on the number of accumulated calls. The dashed line represents
the reported uncertainty estimate from the Vegas routine, while the points and solid lines indicate
the ``true'' error, the distance between the reported integral and the true (final Vegas) value of $31552$.
Again, both sequences perform equally well from this point of view.

In conclusion, we find that the Sobol sequence and the \MT{} pseudo random sequence perform equally well
in practice, at least for the benchmarks performed here for one sequence evaluation each. Additional samples could be 
constructed by using alternative seeds for the \MT{} sequence and by skipping points, or using different initialization numbers,
for the Sobol sequence. Theoretically 
one should benefit from a more even sampling
from the Sobol sequence but we have not been able to identify conclusive evidence for this.
Nevertheless, we expect that this could be the case for some calculations, particularly ones targetting
high-precision kinematical distributions.  With this in mind, and given the fact that it performs at least
as well as the \MT{} pseudo random sequence, we use the 
Sobol sequence as default in \MCFMNEW{}.

\subsection{Parallelization and integration uncertainty estimation}

In many calculations beyond leading order, and in particular at NNLO, it is important
to control numerical uncertainties.  Small differences between contributions that should
exactly cancel can easily lead to incorrect results and misinterpreted conclusions.
To avoid such issues it is essential to have good control of the uncertainties on the
numerical integration.  In \MCFM{} we have chosen to use \OMP{} and \MPI{} to produce
a highly parallelized code, allowing us to reach high levels of precision in a fast and
convenient manner.  In this subsection we compare with possible alternative approaches,
and describe and benchmark the procedure for controlling integration uncertainties.

To assess the benefit of our parallelized code we can contrast it with a possible alternative.
A degree of parallelization can be achieved by simply performing many runs with lower
Monte Carlo statistics, limited by reasonably-achievable run-times. A na\"ively weighted combination is not 
recommended 
since the uncertainties of individual 
runs cannot be trusted for low statistics. We demonstrate this in the following paragraph, then discuss
alternative statistical methods for the case when many low statistics runs have to be combined. We again
choose the \NNLO{} Higgs production double real emission component with $\taucut=0.002$\,GeV and $\alpha=1.0$ as a
representative example for a complicated integration.

\begin{figure}
	\includegraphics[]{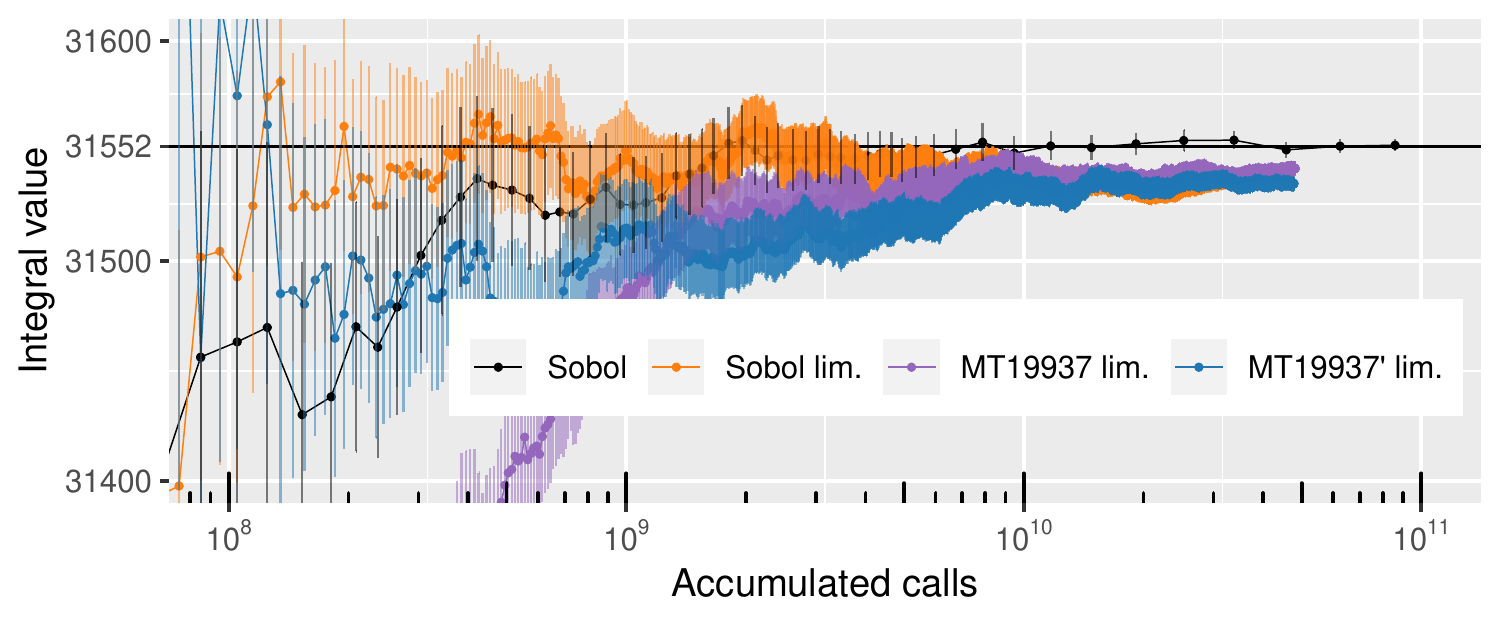}
	\caption{The dependence of the integral for the \NNLO{} Higgs production double real emission contribution
		(for $\alpha=1$ and $\taucut=0.002$\,GeV) on the accumulated number of calls. The black points correspond to
		our default setup, while the others (labelled by ``'lim'') are obtained by limiting the calls per iteration 
		to $10^6$.}
	\label{fig:nnloRRnumbenchlimit}
\end{figure}

\paragraph{Na\"ively weighted combination of low statistics runs.}
Our low statistics runs correspond to 
running the integration for about $4500$ iterations, limiting the calls per iteration to $10$ million calls each for the 
Sobol and \MT{} sequences. This is demonstrated in \cref{fig:nnloRRnumbenchlimit}, where the curves \enquote{lim.} have 
been obtained with the call limit applied. Two runs for the \MT{} sequence with 
different seeds are displayed and one run with the Sobol sequence. For comparison the Sobol run of the previous 
subsection with a steady increase of the number of calls per iteration is also shown.
In this scenario the limited Sobol sequence (orange) vastly outperforms the \MT{} sequences (blue and purple) in the 
region of a low number of 
total calls, still giving reliable error estimates. However, even though individual iteration uncertainties are 
$\lesssim 1\%$, in the final stages 
\emph{both} sequences underestimate the true error.

\begin{figure}
	\includegraphics[width=\columnwidth]{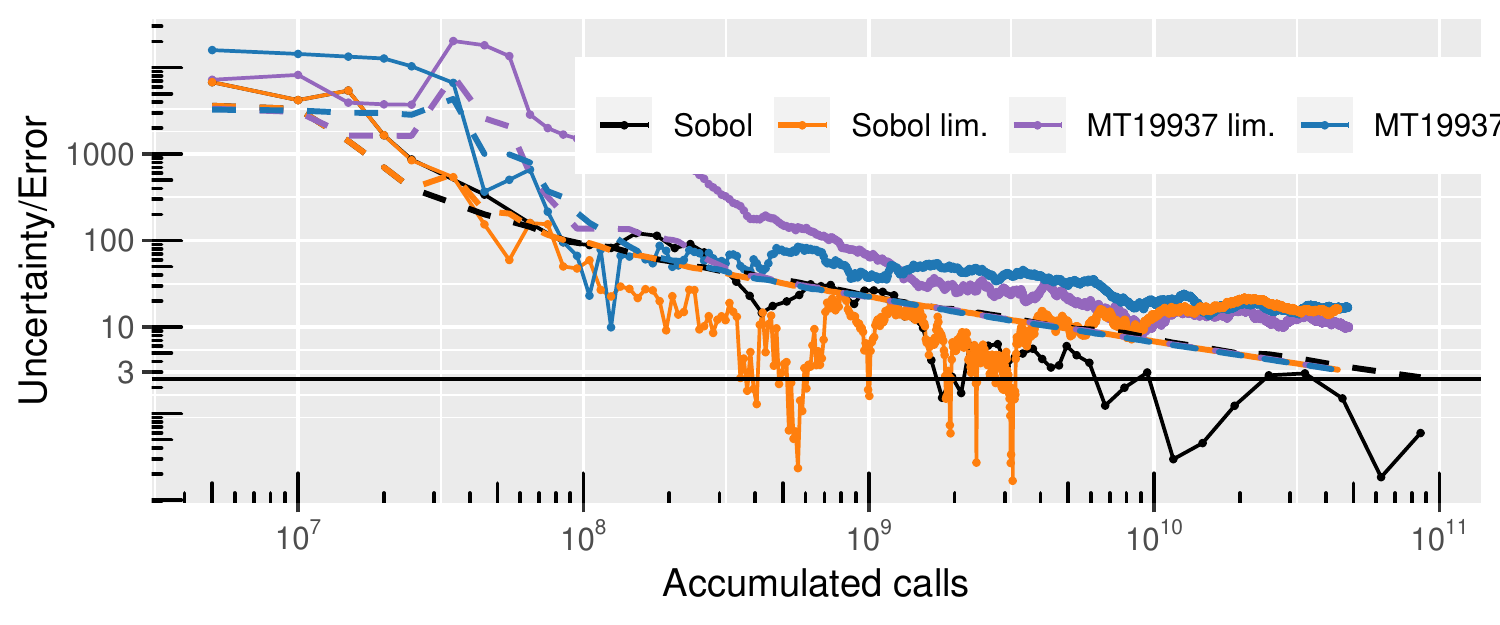}
	\caption{The dependence of the estimated uncertainty of the \NNLO{} Higgs production double real emission contribution
		(for $\alpha=1$ and $\taucut=0.002$\,GeV) on the accumulated number of calls.
                Each point represents a new estimate from a new 
		iteration. The dashed line 	represents 
		the uncertainty from the Vegas routine, while the points
		and solid line represent the ``true'' uncertainty, assuming an estimate of $31552 \pm 2.5$
                from the previous section.}
	\label{fig:nnloRRnumbench_errorslimit}
\end{figure}

When we directly consider the dependence of the integration error on the number of calls, in similar fashion as 
\cref{fig:nnloRRnumbench_errors}, we obtain the results shown in \cref{fig:nnloRRnumbench_errorslimit}. It becomes clear that
when limiting the number of calls the estimated error is severely underestimated for the \MT{} sequences, in comparison with 
the true error. The limited Sobol sequence performs somewhat better in this case. Nevertheless, all limited sequences
lead to significantly underestimated errors and systematically biased results in the final integration stages.
This procedure essentially corresponds to a na\"ively weighted combination of many low statistics runs and its outcome is 
more than clear: the integral estimates lock in at the wrong value with underestimated errors.

Although the absolute effects of numerical precision appear small here, to match the expected systematic cutoff
effect for the Higgs \NNLO{} cross section ($0.2\%$ uncertainty in MCFM-8.0) they must be controlled at the same level. 
This study demonstrates 
that, indeed, limiting the number of calls per iteration is insufficient to achieve the desired precision. For this 
reason we believe that it is important for the integration procedure to not only be highly parallelized, but also 
resumable in order that the number of integration points can be continually increased as required.

\paragraph{Weighted combination of \enquote{pseudoruns}.}
An example of a statistical combination that avoids the weighted combination of runs with underestimated uncertainties 
is provided in ref.~\cite{Ridder:2016rzm}, in the context of a \NNLO{} calculation
of $Z+$jet production using the code \NNLOJET{}.  The prescription for validating and testing the
combination of results is presented in detail there;  here we replicate this methodology in \MCFM{}
and compare it with our usual, fully parallelized, approach.

The study described in ref.~\cite{Ridder:2016rzm} proceeds as follows. After a warmup run to produce a Vegas grid,
this grid is used as the basis for thousands of subsequent low-statistics runs.  From these,
a sample of $k$ individual runs can be combined in an unweighted average to form a ``pseudorun'' with an associated 
variance.
Pseudoruns can then be combined in a statistical manner, using a weighted average, to obtain the
final result.  The dependence of this result on the size of the pseudoruns, i.e. on $k$, is then
used to assess the validity of the computed quantity.  For small $k$ ($k\gtrsim1$) it 
can be very easy to obtain an
incorrect result for the integral since each pseudorun may lack sufficient statistics -- but the final reported
error is small, due to the large number of pseudoruns.  Conversely, for large $k$ the estimate of
the integral is more reliable, but the uncertainty estimate is much larger.  The method is based
on identifying a region of $k$ for which results agree within the computed uncertainties.
Although this appears reasonable, in principle it requires a careful case-by-case study of each observable
to justify the choice of $k$.  

\begin{figure}
	\includegraphics[]{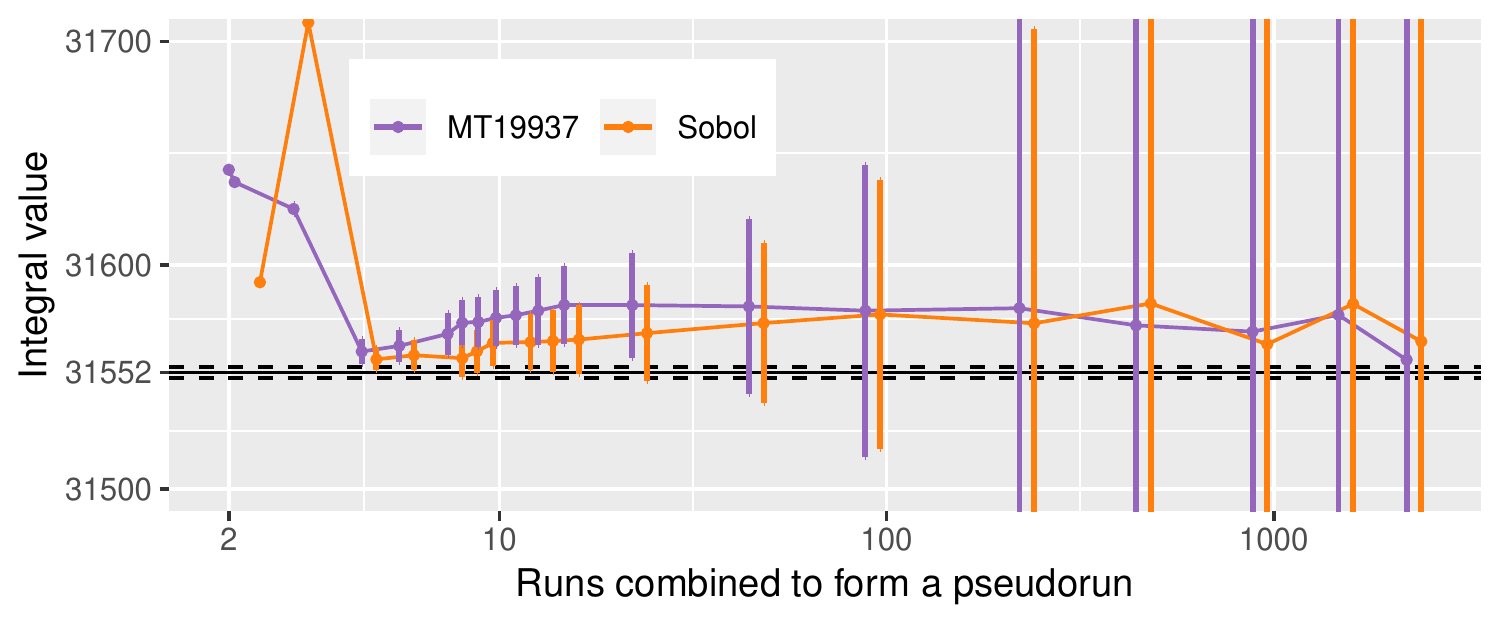}
	\caption{Study similar to ref.~\cite{Ridder:2016rzm}: we recombine our data of about 4500 iterations with 10 million 
	calls 
		each into unweighted pseudoruns, computing the uncertainty using the sample variance. These pseudoruns are then 
		combined into the integral estimate through a weighted sum.  The true result, computed from a separate, longer 
		run, is indicated by the black line with dashed uncertainties.}
	\label{fig:pseudoruns}
\end{figure}

We replicate this method for the calculation of the double real emission contribution to \NNLO{} Higgs production, where
in our study of the previous subsection we generated about 4500 iterations with 10 million calls each for the Sobol sequence 
and one seed of the \MT{} sequence. From these we can then 
combine runs into pseudoruns as described above.  Our results are shown
in \cref{fig:pseudoruns}, as a function of the number of runs combined ($k$).  This displays the features noted
above, at both small and large values of $k$, as well as a relatively stable central plateau.  The true result for
the integral, obtained in the previous subsection, is also displayed.  We see that, while the value of the integral is
in reasonable agreement for a wide range of $k$, it is not clear exactly what value to use. Moreover, it is not clear 
which value of the uncertainty to trust. From \cref{fig:pseudoruns} it is clear that even if a prescription for 
picking the integral value and its uncertainty can be given, any reasonable uncertainty estimate is significantly larger than 
the uncertainty estimate of $\pm 2.5$ from the integration in \MCFM{} with the same number of calls.

\begin{figure}
	\includegraphics[]{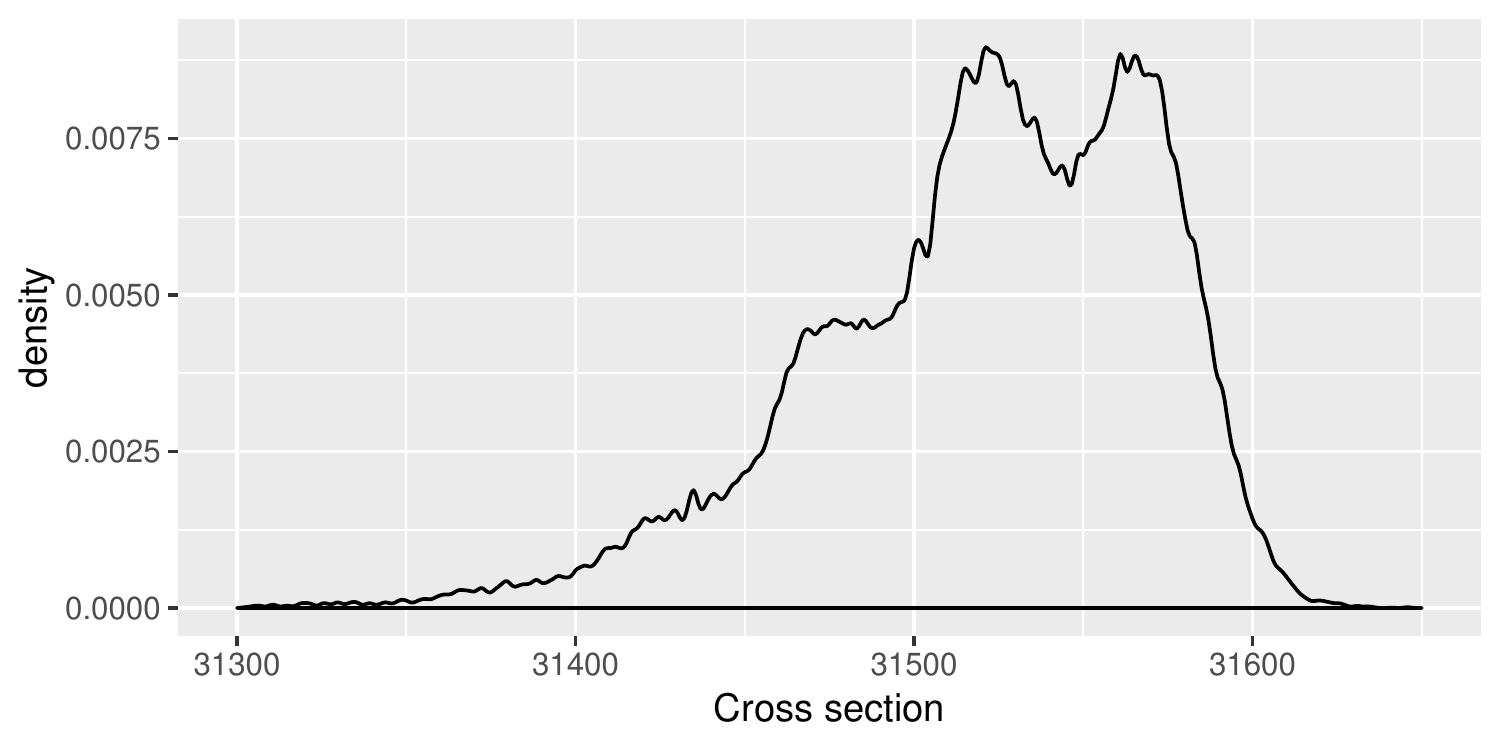}
	\caption{Distribution of the cross section averages from $50,000$ bootstrap replicas.}
	\label{fig:doublegaussian}
	
\end{figure}

\paragraph{Bootstrap.}
To improve upon the method for combining runs into pseudoruns, and to extract a reliable uncertainty, one can 
use the
well-established Bootstrap and Jackknife techniques\footnote{We use the bootstrap functions 
implemented in the GNU R boot package \cite{boot1,boot2,boot3}. }.
Given $N$ estimates of the integral (data points), where $N$ should be sufficiently large
($N\gg10$), within the bootstrap method one randomly resamples $N$ times from the $N$ data points with replacement.
This procedure is repeated $k$ times, so that each one of the $k$ new samples has $N$ resampled data points. The number 
of 
bootstrap samples $k$ should be at least of the order of 
$1000$-$10000$ and we have chosen $k=50000$. From the bootstrap samples one can then obtain unweighted averages of the
data points as estimates for the integral. The generated distribution from the $k$ averages is Gaussian
if $N$ and $k$ are sufficiently large. Bootstrap confidence intervals can be constructed by looking at the percentiles 
of the distribution or using more sophisticated methods \cite{boot1,boot2}. In practice, when $N$ is not asymptotically 
large, one or a few outliers can cause a non-Gaussian distribution. In that case the bootstrap results are not accurate and 
can be sensitive to changes in the data points. 

Such single outliers can be identified with the jackknife-after-bootstrap method \cite{boot3}. Within jackknife 
resampling one leaves out one of the $N$ data points and computes the integral average using the leftover points.
In total $N-1$ resamplings are obtained, generating a distribution from which percentiles can be 
obtained. For the jackknife after bootstrap method the $k$ bootstrapped samples are taken, and for each data point $x$
out of $N$, the samples without $x$ are taken and analyzed with the jackknife technique. One can then directly
analyze the influence each data point $x$ has on the bootstrapped results.   While a detailed description of
the jackknife-after-bootstrap method is beyond the scope of this paper, it is a standard procedure.
We refer to ref.~\cite{boot3} and the software package in ref.~\cite{boot1} for details of the method, and provide
some details for our illustrative example below in \Cref{app:bootstrap}.

\begin{figure}
	\includegraphics[]{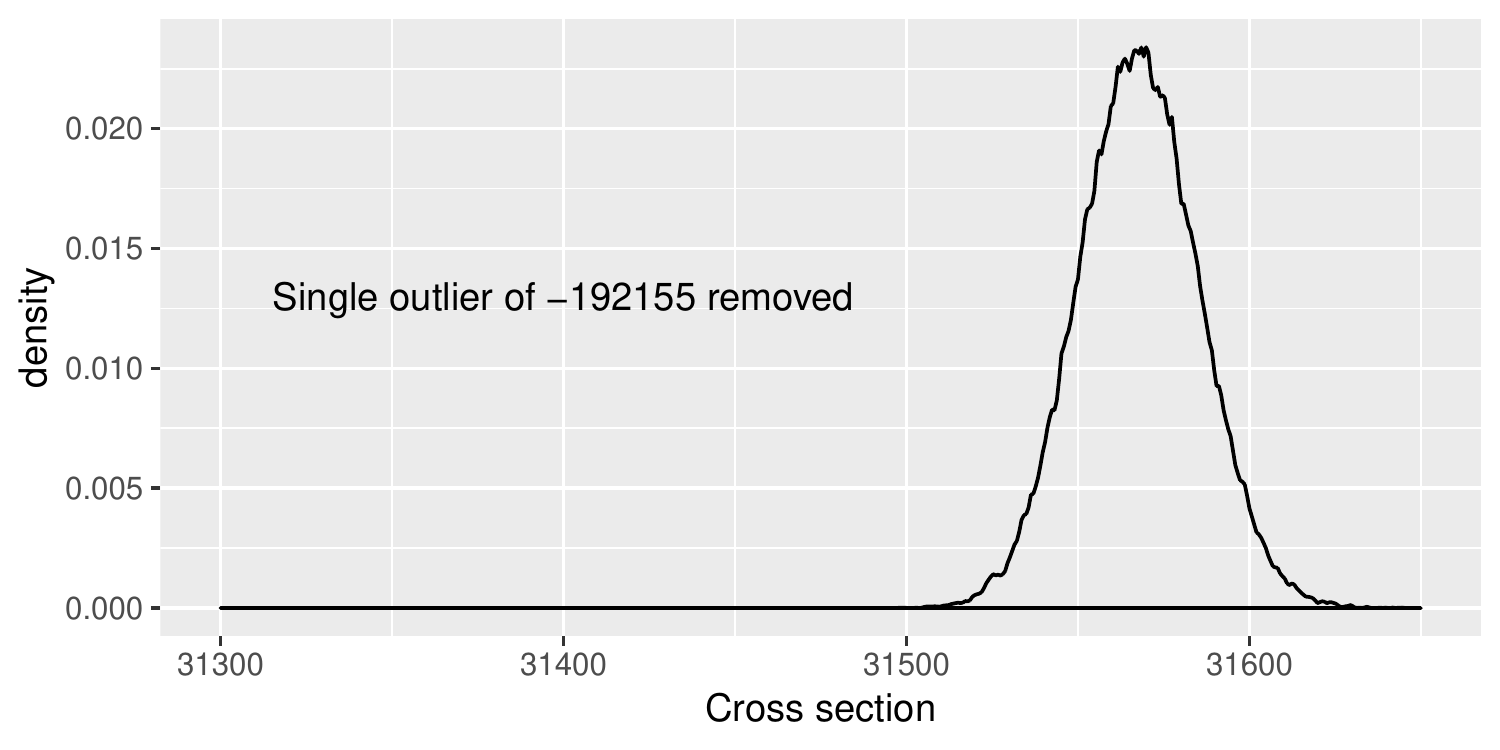}
		\caption{Distribution of the cross section averages from bootstrap replica, after removing
		the worst outlier from the full set in \cref{fig:doublegaussian}.}
	\label{fig:doublegaussian_fixed}
\end{figure}

The distribution of the cross section obtained from a bootstrap with $50,000$ replica samples from the full
$\sim 4500$ \MT{} integral estimates is shown in \cref{fig:doublegaussian}.  We immediately see that this resampled 
distribution
is not Gaussian so that the interpretation of an uncertainty on the average based on percentiles, $31521\pm 49$, is not 
clear.
However, this multi-peaked structure is caused by just a few outliers that can be easily identified 
using the jackknife-after-bootstrap technique described above, and illustrated in more detail
in \cref{fig:bootstrap_mt_first,fig:bootstrap_mt_second} in \cref{app:bootstrap}.
Using such a procedure we can identify a single outlier of $192155 \pm 223717$, whose uncertainty is not
properly taken into account in this method, and remove it.  After doing so we observe the distribution
of cross section averages shown in \cref{fig:doublegaussian_fixed}, that displays a very good Gaussian
shape.  The resulting estimate of the cross section and uncertainty is $31568\pm 17$, where the error
is now expected to be Gaussian and reliable. Removing the worst three outliers gives a further modest
improvement, with an estimate of $31559\pm 13$.
With the $\sim 4500$ integral estimates obtained with the Sobol sequence, we obtain a Gaussian distribution 
outright and an estimate 
of $31570\pm 28$. Removing the worst four outliers in the same way we obtain an estimate of $31572\pm 15$. The 
uncertainty could be reduced further to $10$ by removing 10 outliers, but in the end the sample size should be 
increased.

We conclude that the bootstrap technique is a more robust way to obtain an average and uncertainty when limited to a 
large sample size with a limited number of calls per sample.
For example, it allows for a systematic identification and removal of outliers and it can be used to obtain
Gaussian confidence intervals of various degrees.  Furthermore, it can be automatized straightforwardly and gives
enough indicators to estimate the quality of the results. In fact, we have successfully implemented and tested such an 
approach in \MCFM{} as an alternative to the default Vegas integration uncertainties. 

However, this procedure requires a large number of iterations or independent runs.  Moreover, a completely parallelized
-- and resumable -- approach allows these issues to be avoided, or at least postponed, since a large number of 
integration points can be more easily achieved. By continually increasing the number of points used in each Vegas sweep,
the code can ensure that the grid adjusts sufficiently to the phase space by monitoring the $\chi^2/\text{iteration}$.
This avoids the code ``locking in'' to an incorrect value for the integral, with a \emph{significantly}
underestimated uncertainty. Compared to the other approaches, the few iterations but larger statistics integration 
results in trustworthy uncertainties that are generally significantly smaller.

\clearpage
\section{Updated \NNLO{} benchmarks and $\taucut$-dependence}
\label{sec:benchmark}

The technique employed in \MCFM{} for the calculation of \NNLO{} processes involves a technical jettiness-slicing
parameter $\taucut$ \cite{Boughezal:2015dva,Gaunt:2015pea}, that needs to be chosen small enough to eliminate the 
dependence on it. In practice
one can not choose $\taucut$ arbitrarily small because it induces large cancellations between different
components that need to be integrated separately. The dependence on $\taucut$ also does not need to be
uniform over the whole phase space. So in kinematical distributions the $\taucut$ dependence and
its induced systematic error can be different in each bin. This also holds for cross
sections with different cuts, of course. We ship \MCFM{} with presets for $\taucut$ corresponding to
conservatively-estimated systematic cutoff effects of less than one percent for total inclusive cross sections.

In this section we first demonstrate order of 
magnitude improvements through the use of a boosted definition of $0$-jettiness and the inclusion of 
power corrections. We then introduce automatized methods that allow an assessment of whether the
cutoff effects are indeed as small as suggested, especially in kinematical distributions in the presence of thresholds.

\paragraph{Improved $\taucut$ dependence.}

To improve the performance of the jettiness-slicing method, by default this version of the
code uses a version of $0$-jettiness that incorporates the boost of the Born
system~\cite{Stewart:2010tn,Moult:2016fqy}.  This was introduced already with the \NNLO{} $Z\gamma$ implementation 
\cite{Campbell:2017aul} in \MCFM{}-8.1, but here we benchmark the improvement in all of the \NNLO{}
calculations included in the code.  As a further benefit, we have also implemented the
leading power corrections presented in refs.~\cite{Moult:2016fqy,Ebert:2018lzn} for all appropriate
\NNLO{} processes.  The combination of these two provides a substantial reduction in computational
time required for a given nominal accuracy.  As a result, in this section we update the benchmark
results for color-singlet processes presented in ref.~\cite{Boughezal:2016wmq} and extend
that exercise to the $\gamma\gamma$~\cite{Campbell:2016yrh} and $Z\gamma$~\cite{Campbell:2017aul}
processes.  This allows us to improve the preset $\taucut$ values for our precision goals
in the code accordingly.

We make use of a further feature in this
version of the code that allows $\taucut$ to be defined on an event-by-event basis.  This has
the benefit of elucidating the nature of the power corrections more cleanly, as well as avoiding
potential numerical issues related to very small values of $\taucut$ in events with very
high-energy partons.  This dynamic choice of $\taucut$ is defined by,
\begin{equation}
\taucut = \epsilon \times m \,,
\label{eq:dyntaucut}
\end{equation}
where $m$ is the invariant mass of the Born system, e.g. $m(\ell^- \ell^+)$, $m(\gamma\gamma)$ or $m(Z\gamma)$.

\paragraph{Automatic fitting of the $\taucut$ dependence.}

With this version of \MCFM{} an array of additional $\taucut$ values can be specified that is sampled
in addition to the nominal choice. Using these additional points a fit based on the leading power behavior
of the cross sections' $\taucut$ dependence is automatically performed. When no additional $\taucut$ values
are chosen an automatic choice is made and this will be discussed in \cref{subsec:difftaufit}.

Since we use correlated $\taucut$ values, it is advantageous
to just fit the \emph{correction} with respect to the nominal value with a small uncertainty.
Our weighted fit is then performed using the Minpack package \cite{BurkhardtMinpack,Minpack1,Minpack2}. We weight by 
the (small) uncertainties from the correlated $\taucut$ sampling and thus take into account the different uncertainties 
between larger and smaller values of $\taucut$.

At \NLO{} the leading power $\taucut$ dependence of the cross section starts with $\taucut \log(\taucut)$. We use 
the following form for fitting, including an additional linear term:
\begin{equation}
\sigma(\taucut)^\NLO{} = \sigma_0 + c_1 \cdot \taucut \cdot \log(\taucut/m) + c_2 \cdot \taucut\,,
\label{eq:fitnlo}
\end{equation}
where $m$ is the invariant mass of the Born system as in \cref{eq:dyntaucut}.
At \NNLO{} the dependence starts with $\taucut \log^3(\taucut)$ and we include the subleading term $\taucut 
\log^2(\taucut)$ as well as an optional linear term:
\begin{equation}
\sigma(\taucut)^\NNLO{} = \sigma_0  + c_1 \cdot \taucut \cdot 
\log^3(\taucut/m) + c_2 \cdot \taucut \cdot \log^2(\taucut/m) + c_4 \cdot \taucut\,.
\label{eq:fitnnlo}
\end{equation}
 From the fit
we extract a reduced $\chi^2$ per degree of freedom value that should be small compared to one for a good fit.
The linear subleading term with coefficient $c_4$ is included if the fit without the linear term has a 
$\chi^2/\text{d.o.f.}$ of more than one and the inclusion of the linear term improves the $\chi^2$.
We note that while the inclusion of power corrections in principle removes
the leading coefficient $c_1$ at each order, we still include it in order to remain valid for
processes where such corrections are not known, as well as to be robust against
the effect of cuts.

Our fitting procedure can easily be extended to successively include a tower of 
subleading terms. Such a fit could be relevant for theory applications, for instance to examine residual subleading 
corrections at extremely high precision \cite{Moult:2016fqy}. In practice, to obtain the most reliable results at the 
highest level of precision it is better to choose a smaller value of  $\taucut$, closer to the asymptotic behavior,
rather than using very high numerical precision far from the asymptotic region.

\subsection{Inclusive $\taucut$ benchmarks}

\begin{figure}
\includegraphics[width=0.5\textwidth]{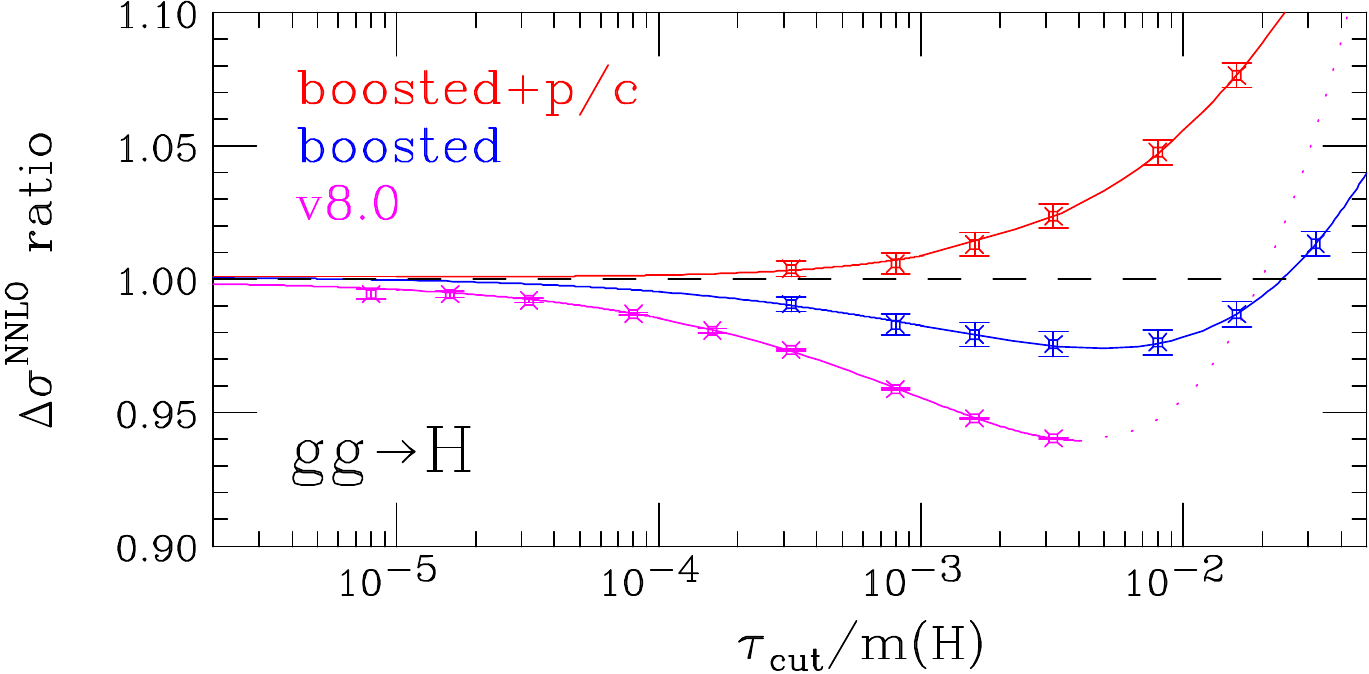}
 \includegraphics[width=0.5\textwidth]{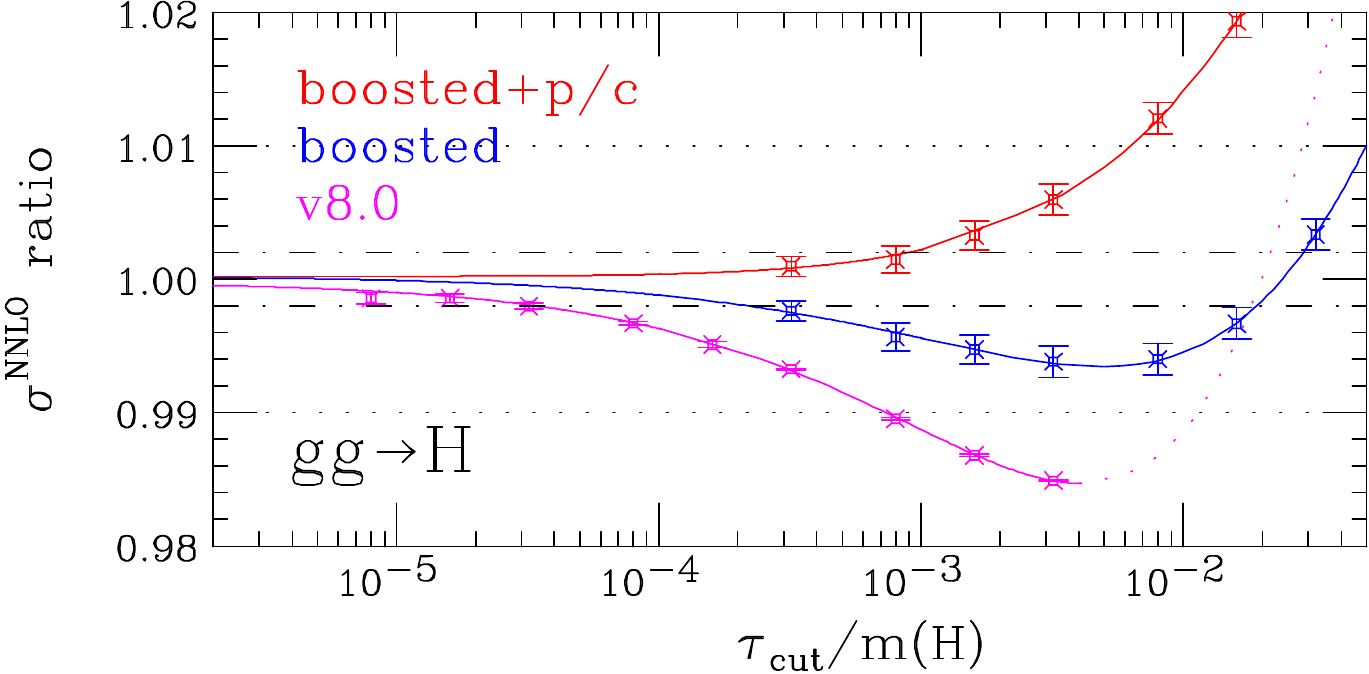} \\
\includegraphics[width=0.5\textwidth]{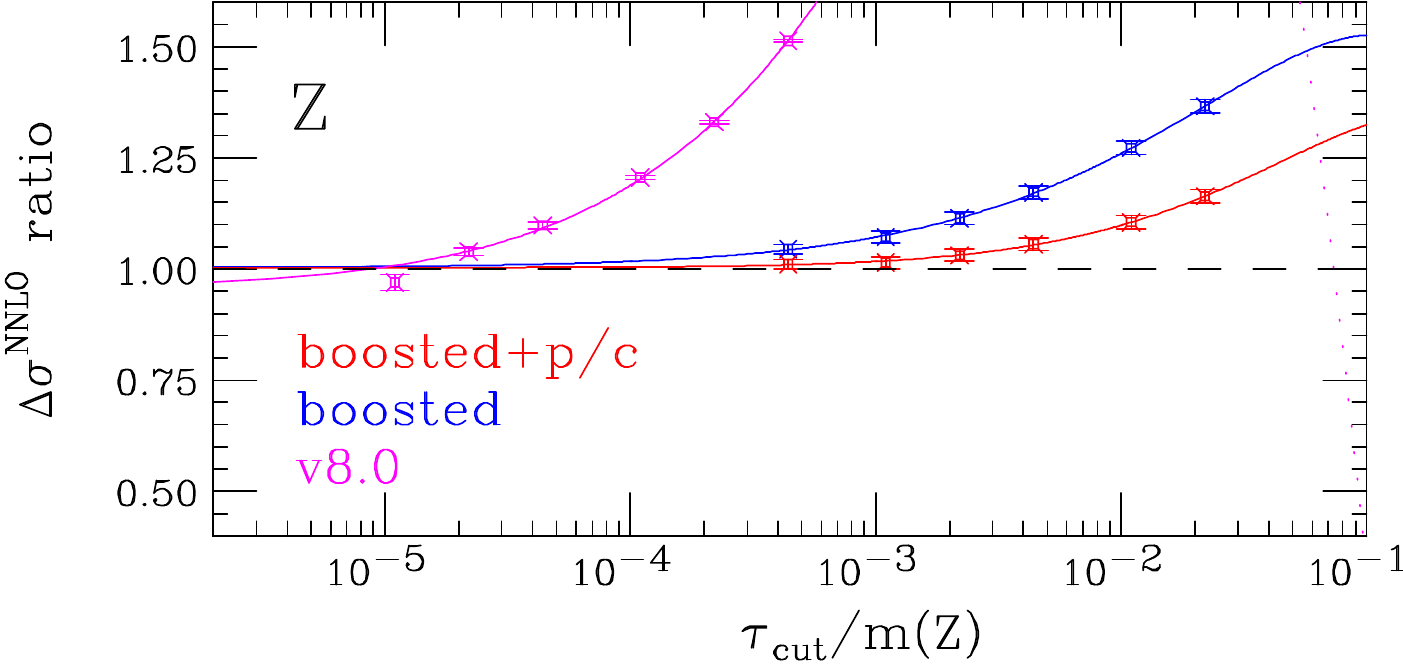}
 \includegraphics[width=0.5\textwidth]{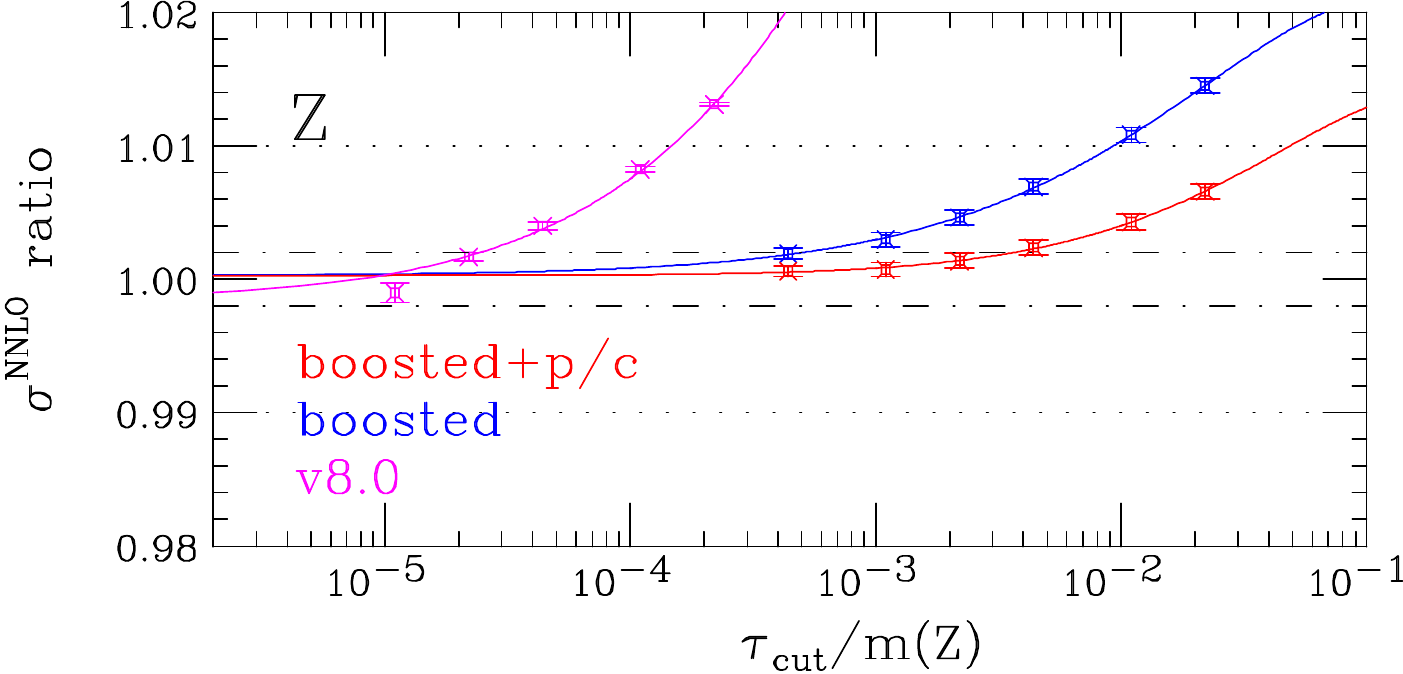} \\
\includegraphics[width=0.5\textwidth]{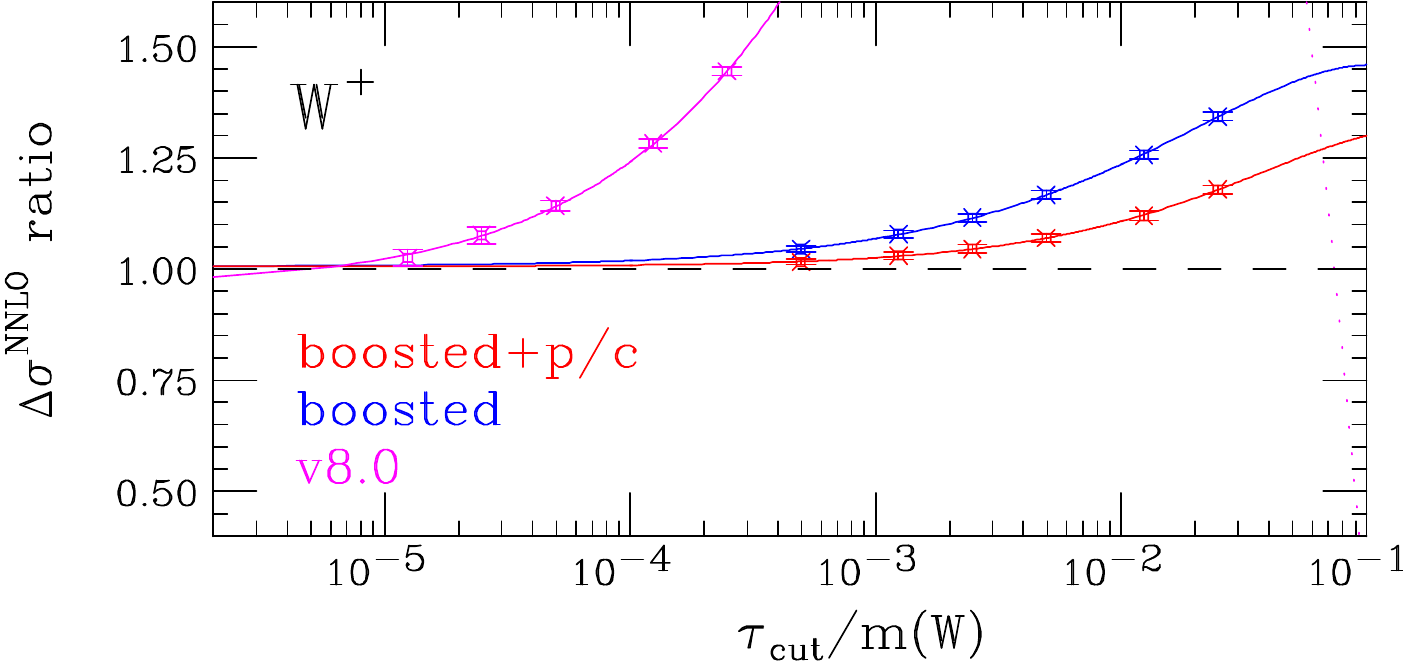}
 \includegraphics[width=0.5\textwidth]{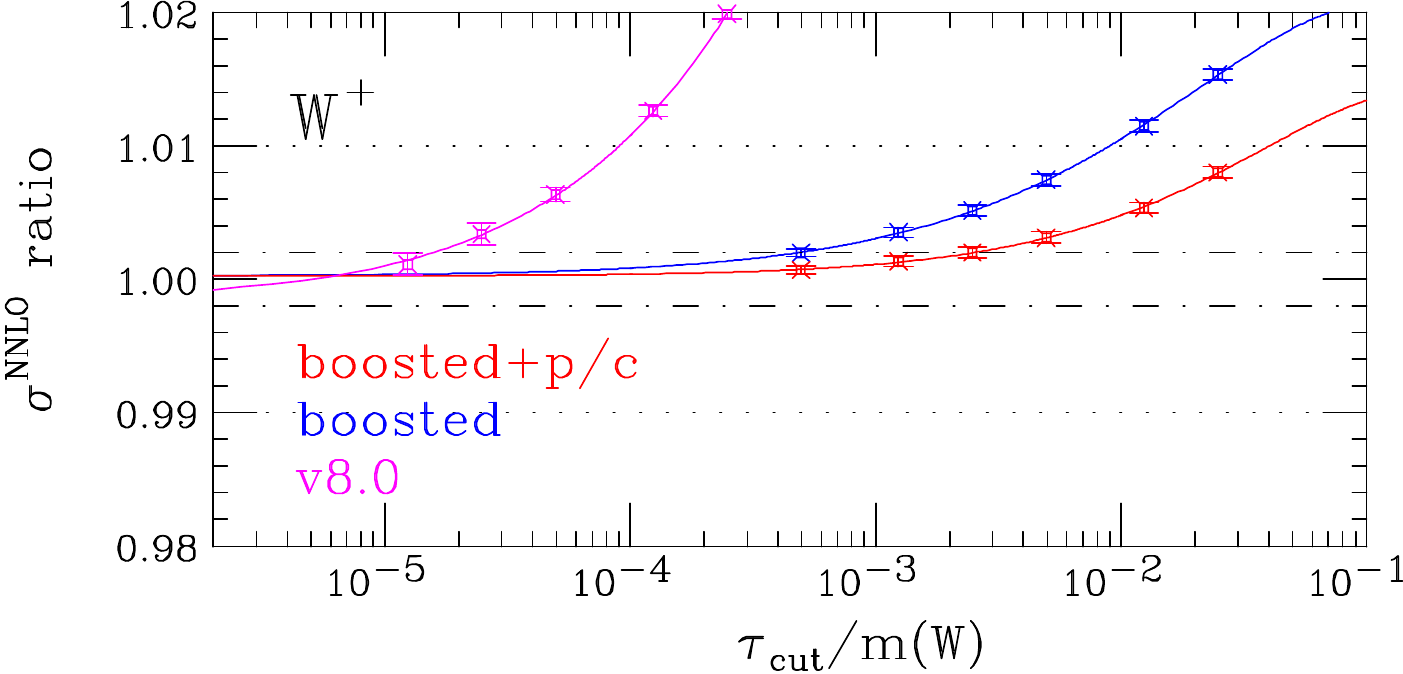} \\
\includegraphics[width=0.5\textwidth]{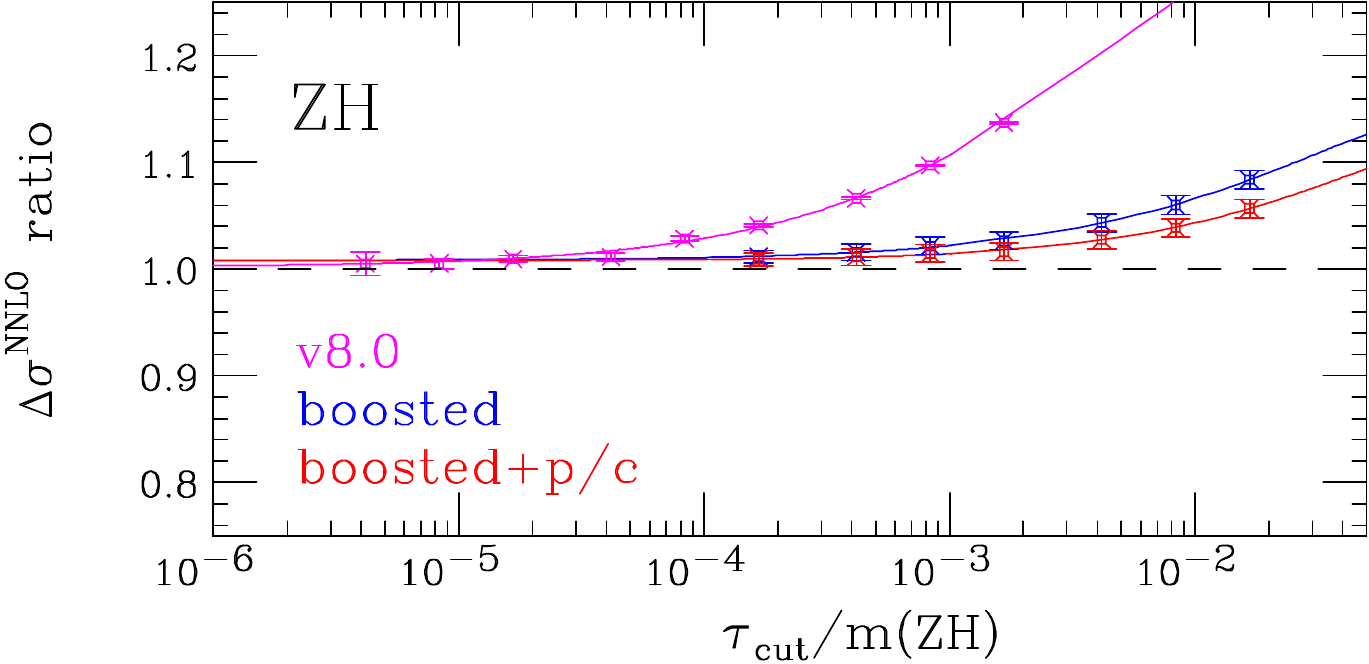}
\includegraphics[width=0.5\textwidth]{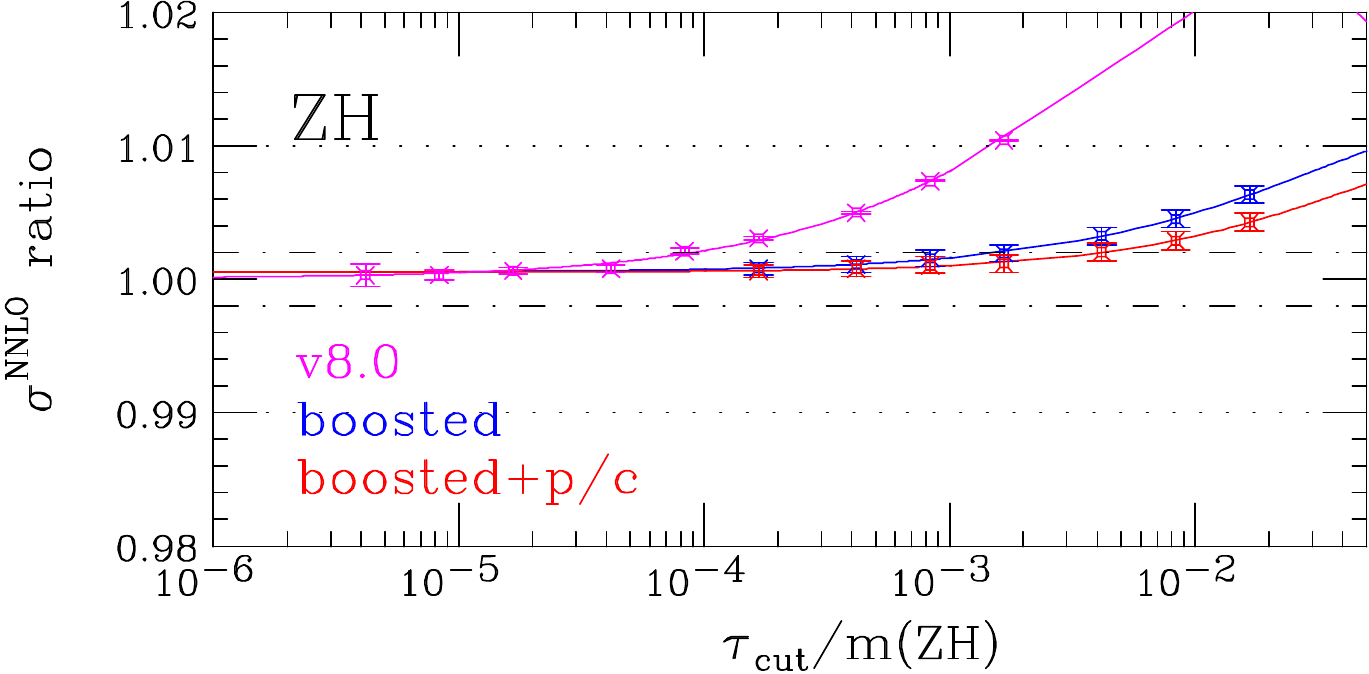} \\
\includegraphics[width=0.5\textwidth]{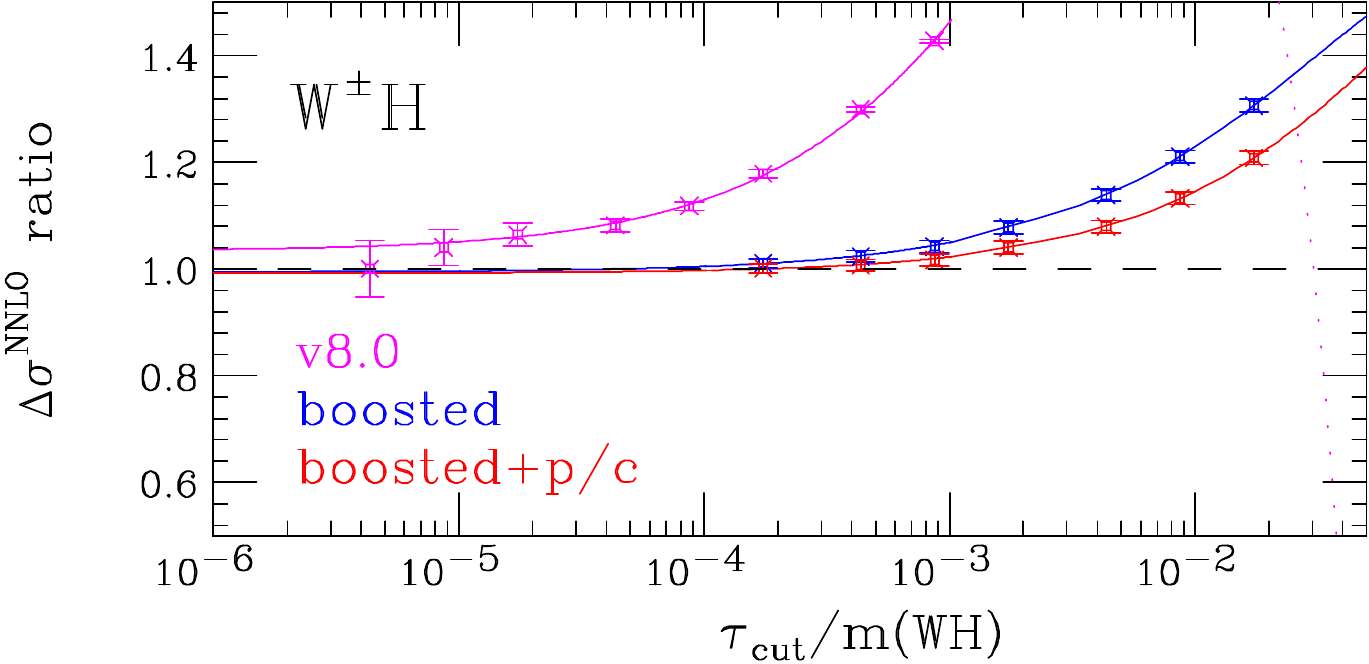}
\includegraphics[width=0.5\textwidth]{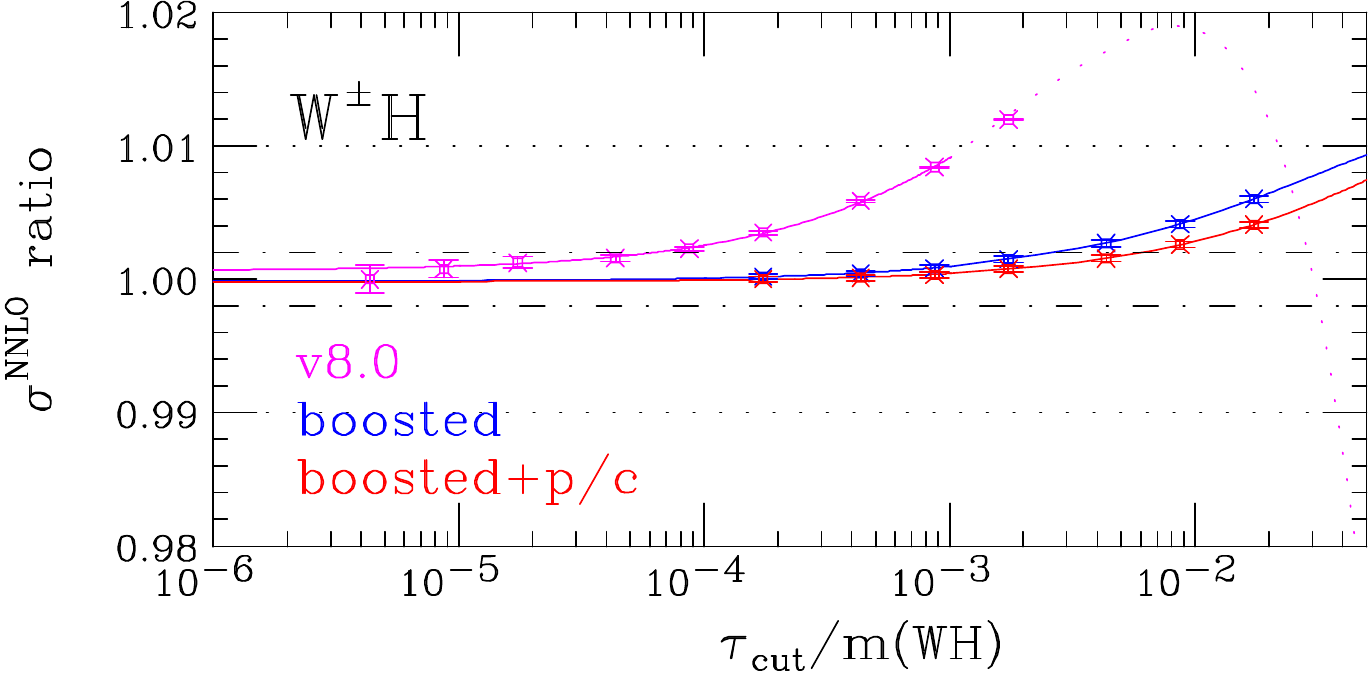}
\caption{\label{fig:taudepnnlo-all} The ratio of the NNLO correction (left) and total
NNLO (right), calculated using jettiness slicing, to the known result.
The ratios are plotted as a function of the jettiness resolution parameter, normalized to
the mass of the Born system.   The results using the previous (unboosted)
definition of $0$-jettiness, as implemented in MCFM v8.0, are shown in magenta.
Results using the boosted definition of $0$-jettiness are shown in blue,
while results after the further addition of power corrections are shown in red.
For reference, horizontal lines are shown in the total NNLO plots that represent
accuracies of $1\%$ and $2$ per mille.}
\end{figure}

We first show results for the processes for which the effect of leading power corrections is known
-- namely $gg \to H$, $W$, $Z$, $ZH$ and $WH$ production -- in Fig.~\ref{fig:taudepnnlo-all}.
For this inclusive study all the settings and parameters are chosen according to
ref.~\cite{Boughezal:2016wmq}. The ratio of the \MCFM{} calculations to the known \NNLO{} results,
for the correction itself and for the total rate,
are shown as a function of the $0$-jettiness variable $\taucut$.  In each case the dependence
on $\taucut$ is shown using the unboosted definition of $0$-jettiness as in \MCFM{}-8.0
(corresponding exactly to the results in Ref.~\cite{Boughezal:2016wmq}), with the 
boosted definition implemented in this version, and after the further addition of the leading
power corrections.

In all cases the improvement from v8.0 to this version is
dramatic. Equivalent residual cutoff effects are obtained for $\taucut$ values that are around two orders of 
magnitude larger. The improvement from using the boosted definition is most dramatic for
$W$ and $Z$ production, for which a significant portion of the inclusive
cross-section results from events at large rapidities where the unboosted
definition performs poorly. 

\begin{figure}
\includegraphics[width=0.5\textwidth]{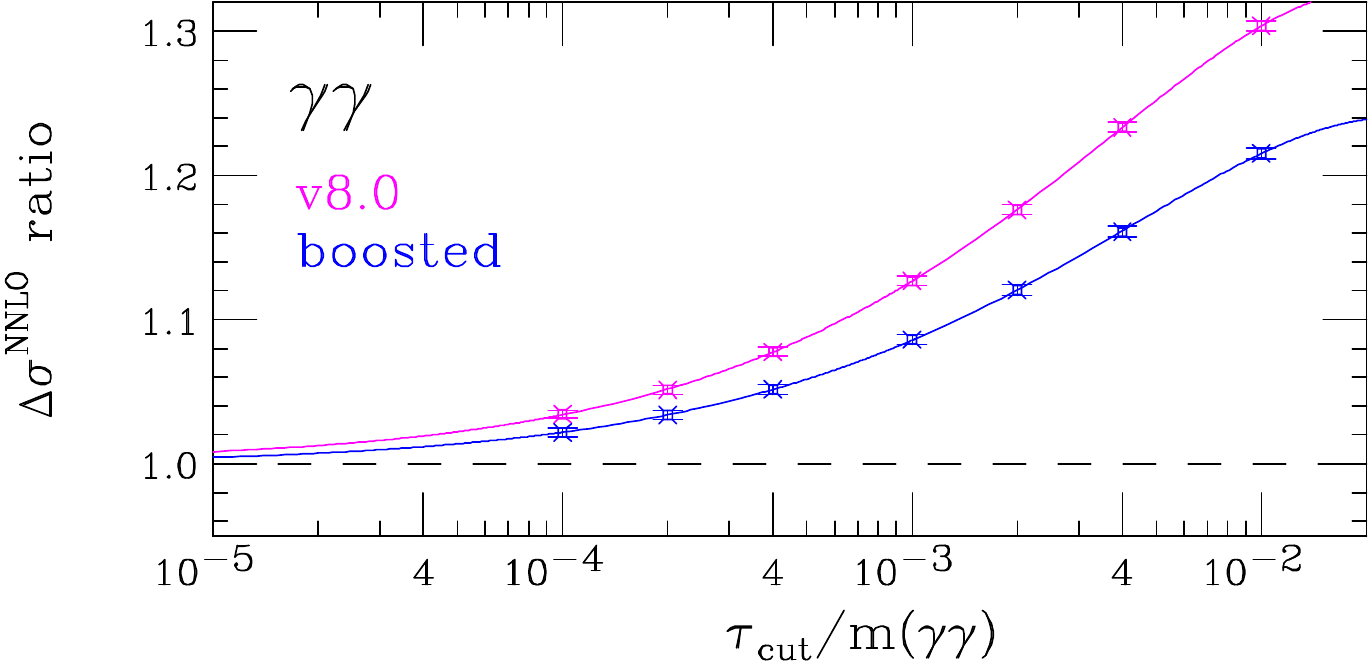} 
\includegraphics[width=0.5\textwidth]{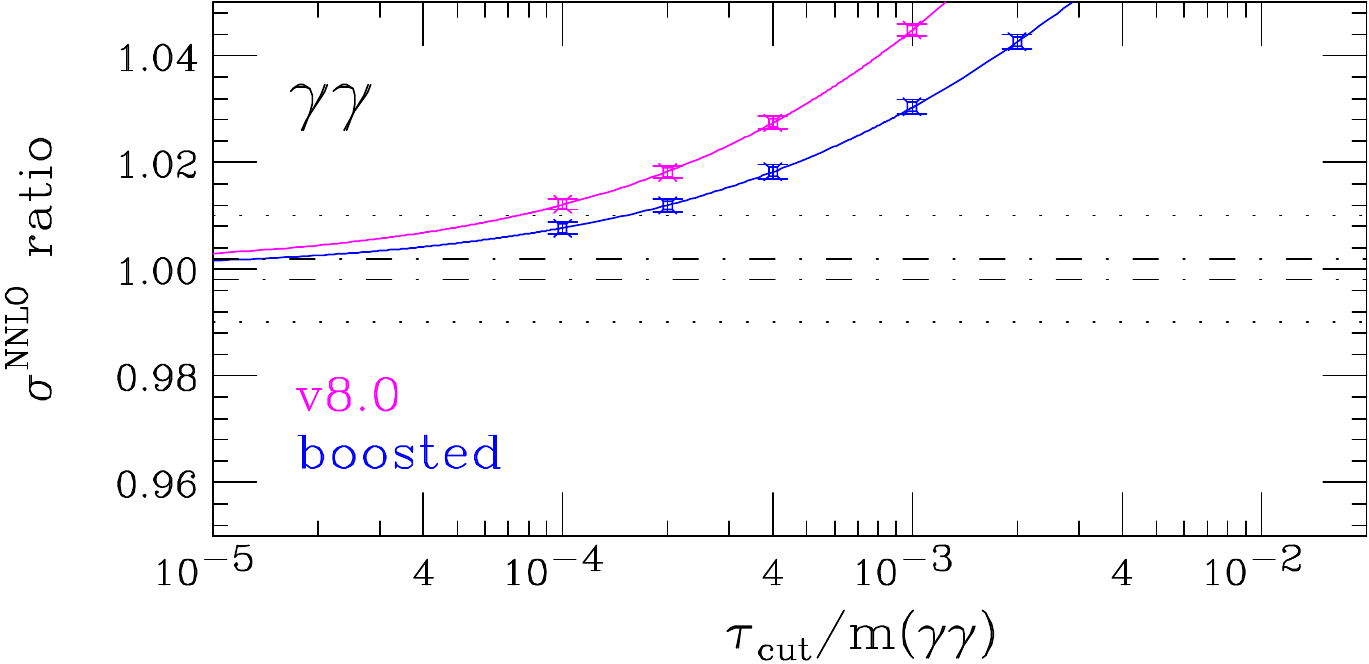} \\
\includegraphics[width=0.5\textwidth]{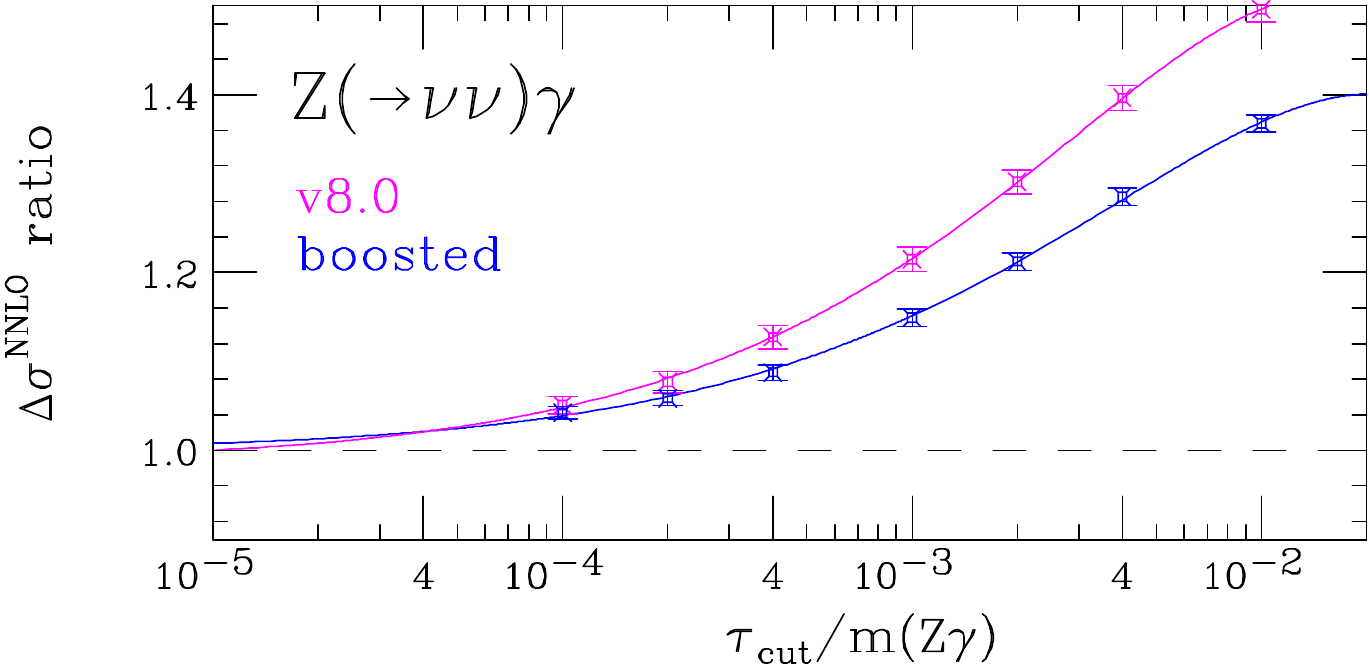} 
\includegraphics[width=0.5\textwidth]{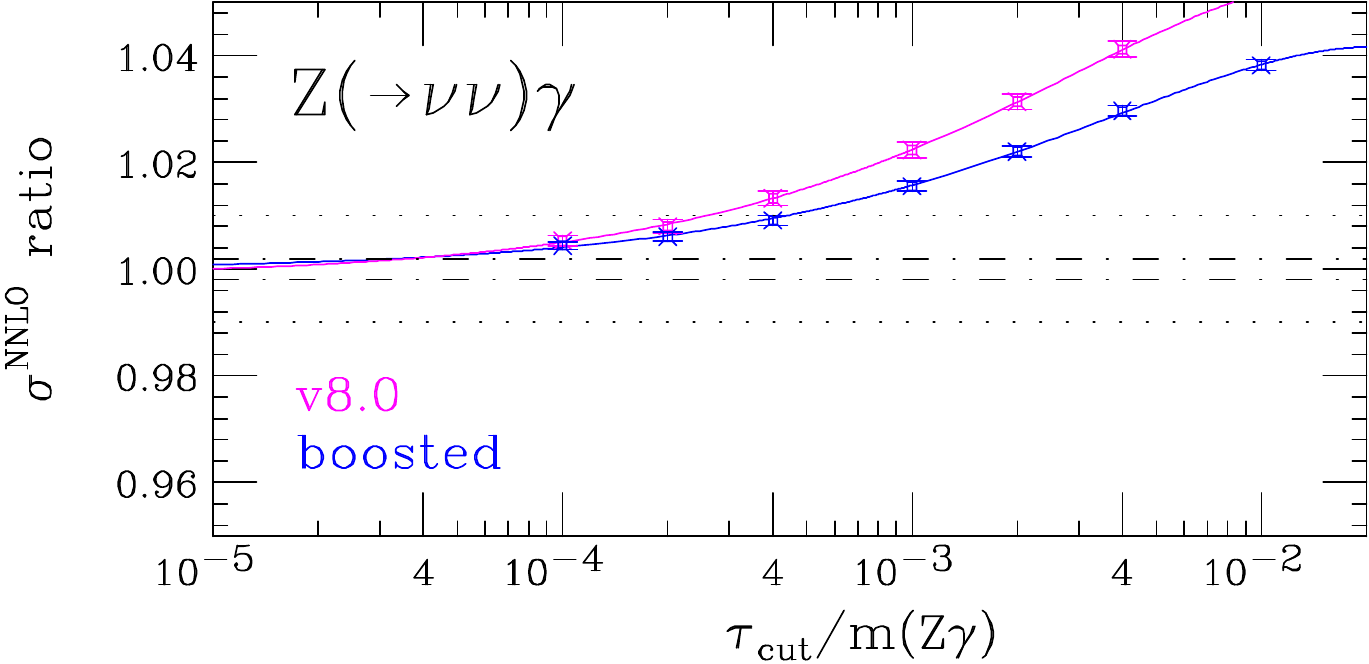}
\caption{\label{fig:taudepnnlo-all2} The ratio of the NNLO correction (left)
and total NNLO (right), calculated
using jettiness slicing, to the asymptotic result.
The ratios are plotted as a function of the $0$-jettiness
resolution parameter scaled to the invariant mass of the Born system,
c.f. Eq.~(\ref{eq:dyntaucut}).   The results using the previous (unboosted)
definition of $0$-jettiness, as implemented in MCFM v8.0, are shown in magenta
while results using the boosted definition of $0$-jettiness are shown in blue.
For reference, horizontal lines are shown in the total NNLO plots representing
accuracies of $1\%$ and $2$ per mille.}
\end{figure}

We now turn to the remaining processes, $\gamma\gamma$ and $Z\gamma$ production, for
which the leading power corrections are unknown due to their $t$-channel
nature~\cite{Campbell:2017aul}.  However, we can still examine the improvement due to the
use of the boosted definition of $0$-jettiness. 
For this study we use the setup of the validation section of ref.~\cite{Campbell:2016yrh}
for the $\gamma\gamma$ process and the validation cuts for the $Z \to \nu\bar\nu$ decay
specified in ref.~\cite{Campbell:2017aul} for the $Z\gamma$ process.  The results of this
study are shown in \cref{fig:taudepnnlo-all2}.  Here the improvement from using the
boosted definition is relatively mild due to the fact that the Born system tends to
be quite centrally produced.

We now summarize our findings by specifying the default
values of the dynamic $\taucut$ values that are used in the code. Previous version of \MCFM{} included
presets for $1\%$ and $0.2\%$ cutoff effects for the total inclusive cross sections.
In this version we devise a different scheme that improves the quality and allows for an
automatic inclusive and differential $\taucut$ fitting with automatically chosen $\taucut$ values
to assess the $\taucut$ uncertainty.
Our process-dependent preset $\taucut$ values are overall chosen to first satisfy better than $1\%$ cutoff effects in 
the total inclusive cross section. While a one percent error on the total cross section can be considered small for 
many processes with larger theory uncertainties, this is not a sufficient criterion that allows for an automatic 
reliable estimation of systematic $\taucut$ errors.
In \MCFM{} the \NLO{} component is independent of the value of $\taucut$ and only
the \NNLO{} corrections are computed with jettiness subtractions that include a $\taucut$ dependence.
Since the size of the \NNLO{} corrections varies
widely between different processes, this means
that the \NNLO{} coefficient itself must be probed in the asymptotic regime described by \cref{eq:fitnnlo}, independent
of the size of the contribution to the total cross section.

For example in Drell-Yan production the \NNLO{} coefficient is less than one percent of the total cross section, while 
for Higgs production it is about $25\%$. 
Our goal is to automatically include a sufficient range of larger $\taucut$ values in the fit to 
improve results and allow for a systematic assessment of $\taucut$ dependence and uncertainties.
By default, when enabling a fit, in addition to the nominal value $\taucut$ we use an array
of additional values $\kappa \times \taucut$ with $\kappa \in \{ 2, 4, 8, 20, 40 \}$.
Therefore, we further 
require that $\taucut$ values as large as $40$ times the nominal choice, used for the fitting of the \NNLO{} 
coefficient, are still in the asymptotic regime and thus described by \cref{eq:fitnnlo}. 

Our settings are given in \Cref{tab:default-taucut-new}. With these criteria we obtain
$\taucut$ presets that always give precision better than $1\%$, yet are orders of magnitude larger than the presets in 
\MCFM{}-8.0 due to the boosted definition and power corrections. To achieve an even smaller systematic
error one can choose values of $\taucut$ according to \cref{fig:taudepnnlo-all,fig:taudepnnlo-all2}
above.

\begin{table}[]
	\centering
	\caption{Default $\taucut$ values for each process, together with the expected level of precision when computing
		the total cross section with that value.  As indicated, all $\taucut$ values are specified relative to the 
		invariant 
		mass of the Born system as in Eq.~(\ref{eq:dyntaucut}).
		\label{tab:default-taucut-new}\vspace{0.5em}}
	\begin{tabular}{@{}lcc@{}}
		\toprule
                Process & $\taucut / m_\text{Born}$ & $\sigma^{NNLO}$ precision \\
		\midrule
		$H$            & $4 \times 10^{-3}$ & $0.7\%$ \\
		$Z$            & $6 \times 10^{-3}$ & $0.4\%$ \\
		$W$            & $6 \times 10^{-3}$ & $0.4\%$ \\
		$ZH$           & $3 \times 10^{-3}$ & $0.2\%$ \\
		$WH$           & $3 \times 10^{-3}$ & $0.2\%$ \\
		$\gamma\gamma$ & $10^{-4}$          & $0.8\%$ \\
		$Z\gamma$      & $3\times 10^{-4}$  & $0.8\%$ \\
	\end{tabular}

\end{table}

\subsection{Differential $\taucut$ extrapolation and assessment of $\taucut$ uncertainty}
\label{subsec:difftaufit}
While the results in the previous section were highly inclusive, and we believe that in most circumstances
the presets of less than one percent $\taucut$-dependence also apply to the most typically-used cuts and 
distributions, modern applications and benchmark comparisons require true per mille precision differentially. Then a 
very careful 
assessment of the numerical uncertainties and $\taucut$ dependence has to be performed. We addressed the former case in 
\cref{sec:integrationuncertainties}. Here we address the latter.

In the previous section the fitted inclusive results were obtained using a wide range of $\taucut$ values to study and 
display the asymptotic behavior. In practice, a good fit can improve results by an order of magnitude compared to 
using just the smallest value of $\taucut$. Here we benchmark such improvements fully differentially and discuss the 
reliability of the fit. With such a fit one can 
then also estimate how large the true residual 
$\taucut$ dependence is and whether it matches the expected uncertainty estimate of the presets in 
\Cref{tab:default-taucut-new}. A successful fit depends on the process, the smallest value of $\taucut$ chosen,
the kinematical cuts applied, and the precision of the input data. However, even a failed fit still 
delivers valuable information that can be used
to further assess the quality of the prediction. Here we discuss both cases, describe the output of the \MCFM{} $\taucut$ 
fitting procedure and provide guidance on how to obtain such improvements and assess the systematic $\taucut$ error.

To our knowledge this is the first (public) code that allows the user to automatically and  systematically evaluate the 
reliability of the slicing calculation and assess true per mille level precision also for kinematical distributions. 
To perform such analyses in previous versions of \MCFM{} required many more orders of magnitude of computational
resources. Nevertheless, some manual inspection is required to be certain about the systematic 
$\taucut$ uncertainties. This is because
any procedure with hard-coded reliability thresholds will eventually fail for some combination of
processes and kinematic cuts. So instead of hard-coding fixed warnings we instead outline a systematic 
procedure that can always be applied to assess the reliability of the results.

Both for the inclusive cross section and fully differentially for the output histograms of the fitted corrections we 
include gauges for a good fit. The first gauge is the maximum relative numerical uncertainty on the additionally 
sampled $\taucut$ corrections, describing the quality, or rather uncertainty of the fit's input data. Second, we output 
the reduced $\chi^2$ 
of the standard-deviation-weighted fitting, as a direct measure of the fit quality. These two numbers are in addition 
to the fitted correction itself and its 
corresponding uncertainty. A more detailed technical description of the \MCFM{} output files is provided in 
\cref{sec:newfeatures-app}. All of these data points should be used to establish a chain 
of trust as described in the following three steps.

\begin{enumerate}
	\item  The 
	associated uncertainties of the additionally sampled $\taucut$ corrections, as direct input to the fit, have to be 
	reliable 
	for the subsequent steps and the fit to be meaningful.
    If the uncertainties of the additional $\taucut$ corrections are severely underestimated, one might be led to the 
    wrong conclusion through a fit that appears to work reliably but which is flawed from the outset. 
    
    To estimate this, we include the maximum relative numerical uncertainty on the sampled $\taucut$ corrections as 
    information with the fit. While a 
	$\chi^2/\text{iteration}$ could be a useful quantity to consider for that purpose, we do not store the necessary 
	information to compute it fully differentially. Instead, we generally simply find that relative uncertainties of 
	the integrals are underestimated when they are still large. So while a $10\%$ uncertainty on a result indicates
	only questionable precision, once percent-level	precision is reached we deem results generally reliable. In 
	practice one can also check whether the smoothness of differential distribution agrees with the reported 
	uncertainties.
	
	\item Assuming from the first step that the additional $\taucut$ corrections and their uncertainties are reliably 
	estimated, the input data to the fitting procedure can be trusted. For the uncertainty-weighted fit we report the 
	$\chi^2$ per degree of freedom (reduced $\chi^2$), which should 
	generally be close to one, or smaller, for the fit to be reliable. A large $\chi^2$ value means that
	the fitting formula does not work well. This can either be caused by data that is so precise that additional
	subleading terms in $\taucut$ need to be included in the fitting form, or that additional terms should be included
	because the $\taucut$-dependence is so large that it has not yet reached the asymptotic regime of 
	\cref{eq:fitnlo,eq:fitnnlo}. 
	In practice, for 
	phenomenology the first case is not relevant when \enquote{only} per-mille level precision is required, but
	additional terms can easily be added if required.\footnote{Also, instead 
	of 	improving the fit with more precise data, one might rather choose a set of smaller $\taucut$ to begin with. 
		Nevertheless, one can of course easily adjust the fitting function in the code of \MCFM{}-9.0 for theory 
		applications.}
	
	This leaves the second
	case for usual applications, where the large $\chi^2$ value of the fit then means that the $\taucut$ dependence is 
	large and
	the default $\taucut$ setting for the total inclusive cross section can likely not be trusted in the considered 
	kinematic 
	region. The values of $\taucut$ being probed are not asymptotic and are not described by \cref{eq:fitnnlo}. This 
	does not necessarily 
	mean that the full \NNLO{} result has a large $\taucut$ dependence, as only the \NNLO{} coefficient is computed 
	using the jettiness slicing with $\taucut$ 	dependence. In the case of $W$ or $Z$ production the \NNLO{} coefficients 
	are generally tiny, so larger values of $\taucut$ can be chosen. Nevertheless, one should always ensure that
        one remains in the 
	asymptotic regime to obtain a reliable estimate of $\taucut$ corrections.	
	
	\item When finally the reduced $\chi^2$ is small or close to $1$, after possibly decreasing the nominal and 
	smallest value 
	of $\taucut$, the fitted correction and its uncertainty can be trusted. When the uncertainty on the fitted 
	correction is small compared to the fitted correction, the fit can often improve results with a $1\%$ cutoff error 
	to the per-mille level. We recommend applying the fitted correction when this chain of trust has been established. 
	When the uncertainty on the fitted correction is large, and after the previous steps it can nevertheless be 
	trusted, it still gives an estimate of the maximum $\taucut$ dependence that can be expected.
\end{enumerate}

In the next subsection we give examples where
results have been obtained with sufficient numerical precision and goodness of fit to allow for order of magnitude 
improvements and assessment of per mille level precision fully differentially. Performing this example first at \NLO{} 
allows us to compare against cutoff-independent results. We then proceed to two examples where the automatic fitting 
fails. In the first example the input data is unreliable (large maximum numerical uncertainty on the additionally 
sampled $\taucut$ values). In the second example we consider a kinematical distribution at \NNLO{} that contains both a 
threshold region with large $\taucut$ dependence and a region that has no logarithmic $\taucut$ dependence and is fully 
finite for $\taucut\to0$. These examples are constructed to show that with the above procedure and all the 
information given together with the fitted corrections, the cutoff dependence can be studied without further runs even 
when the fit fails.

\subsection{Examples of automatic differential $\taucut$ fitting.}
First we demonstrate our implementation on \NLO{} results of diphoton production, where we can compare with
the cutoff-independent dipole subtraction~\cite{Catani:1996vz} implementation. At \NNLO{} we can compare fitted results using larger values of 
$\taucut$ with small values of $\taucut$ that can be considered accurate at the per-mille level. Cuts and parameters 
are as described later in \cref{sec:physics}, but do not matter for the discussion here.

\begin{figure}
	\includegraphics[]{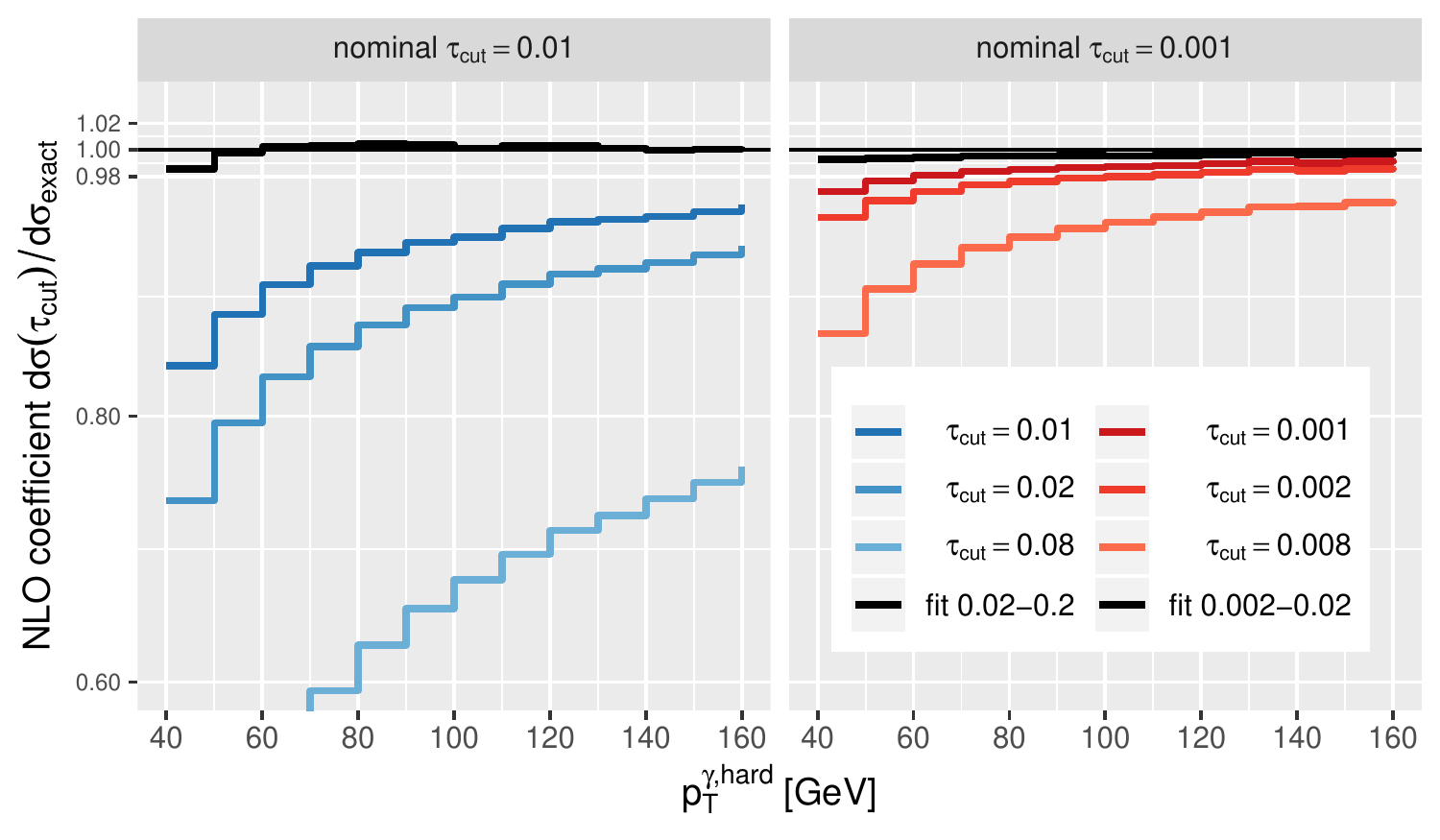}
	\caption{The \NLO{} coefficient of the transverse momentum distribution of the hardest photon in diphoton production.
		The left panel shows the $\taucut$ dependence for $\taucut/m_{\gamma\gamma}$ ranging from $0.01$ to $0.08$, where the fitted result
		has been obtained by using four points $\taucut/m_{\gamma\gamma}=0.02,0.04,0.08,0.2$, not using the nominal value of $0.01$.
		The right panel shows the corresponding results for cutoff parameters a factor of ten smaller.
	}
	\label{fig:ptga_hard_nlo}
\end{figure}

In \cref{fig:ptga_hard_nlo} we show the \NLO{} \emph{coefficient} for the transverse momentum distribution
of the hardest photon normalized to the result of the exact computation. In the left panel we show
results for $\taucut/m_{\gamma\gamma}$ values ranging between $0.01$ and $0.08$, and the fitted result using four points
$\taucut/m_{\gamma\gamma}=0.02,0.04,0.08,0.2$ (not including the nominal value $0.01$). The right panel shows results with
cutoffs smaller by a factor of ten. For the nominal $\taucut/m_{\gamma\gamma}=0.01$ panel the fitting improves the residual
systematic uncertainties by over $15\%$, and its error when compared to the exact result is at the level
of at most $1.5\%$. The improvements when choosing a lower $\taucut$ as in the right panel are smaller, but the fitted
result again is within $1\%$ of the cutoff-independent result, while $\taucut/m_{\gamma\gamma}=0.001$ deviates as much as $4\%$. 

From this we conclude that the fit can improve even large values of $\taucut$ with systematic errors of $10\%$ and 
more. Furthermore, and quite importantly, the fit makes the $\taucut$ dependence uniform over the whole range. Without 
such a fitting
procedure, to guarantee a systematic $\taucut$ error differentially, one has to carefully choose $\taucut$
according to the worst region, which is at small $p_T$. For the example above this would mean running at $\taucut/m_{\gamma\gamma}=10^{-4}$
instead of $0.01$ to achieve the same uniform level of precision for the \NLO{} coefficient.

\begin{figure}
	\includegraphics[]{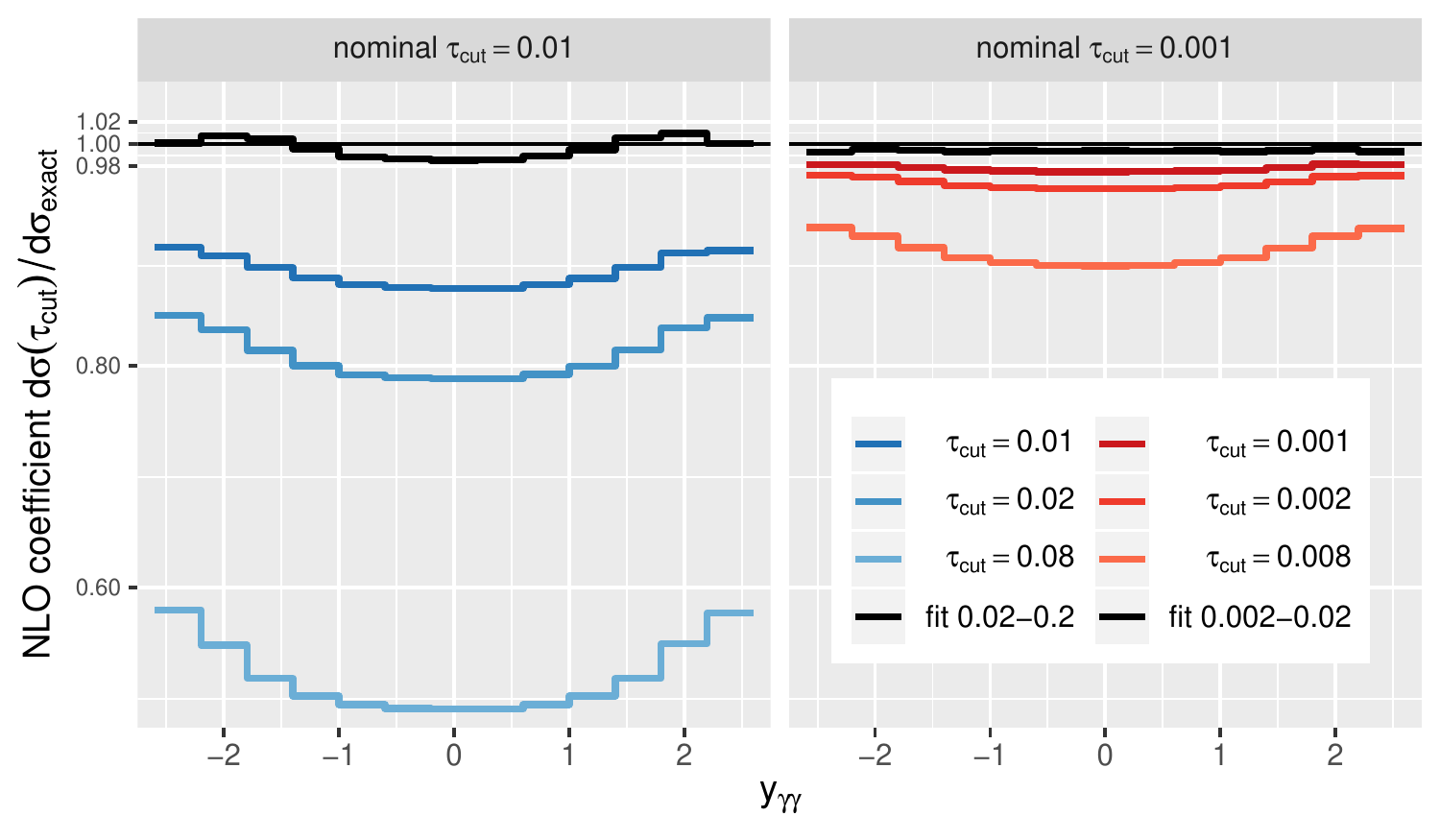}
	\caption{\NLO{} coefficient for the diphoton rapidity distribution.
		The left panel shows the dependence for $\taucut/m_{\gamma\gamma}$ values ranging from $0.01$ to $0.08$, where the fitted result
		has been obtained by using four points $\taucut/m_{\gamma\gamma}=0.02,0.04,0.08,0.2$, not using the nominal value of $0.01$.
		The right panel shows the corresponding results for cutoff parameters a factor of ten smaller.
	}
	\label{fig:ygaga_nlo}
\end{figure}

Similarly, we show in \cref{fig:ygaga_nlo} the \NLO{} coefficient for the diphoton rapidity distribution. While
the $\taucut$ dependence is clearly more uniform in the beginning, the conclusions drawn are the same.

For the total \NLO{} result, but also the coefficient alone, we see that one can use the fitted result as the most precise result 
(following the chain of trust to ensure its reliability) and compare it with the result using the smallest and nominal 
$\taucut$ value. If the difference is larger than the desired or expected systematic error
one should consider the use of a smaller nominal $\taucut$ value.

\paragraph{NNLO results.}

\begin{figure}
	\includegraphics[]{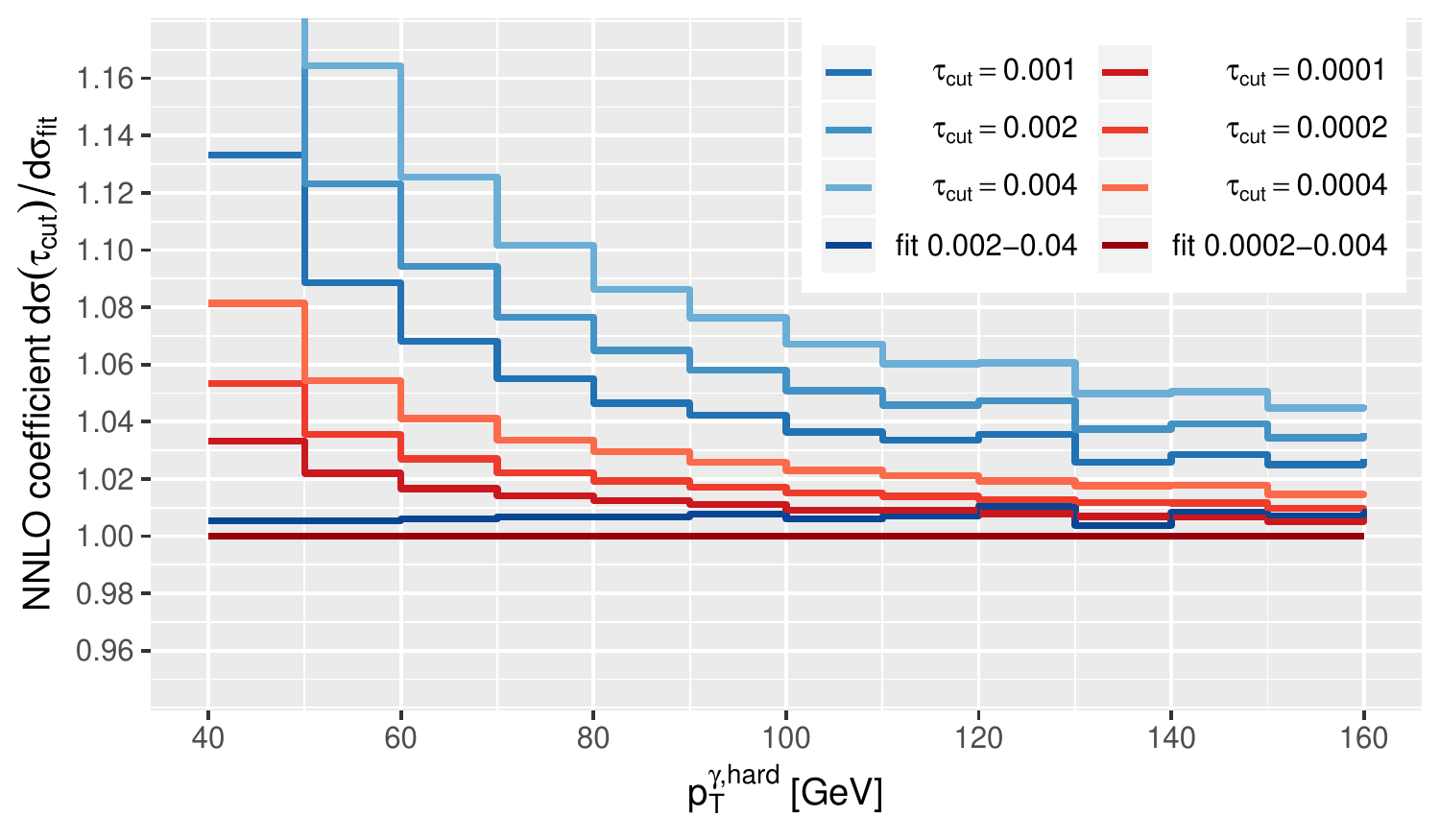}
	\caption{The \NNLO{} coefficient of the transverse momentum distribution of the hardest photon in diphoton
        production.  All curves are normalized to the fitted result with the smaller set of five $\taucut/m_{\gamma\gamma}$ values.}
	\label{fig:ptga_hard_nnlo}
\end{figure}

Corresponding to the \NLO{} coefficient plot in \cref{fig:ptga_hard_nlo}, we show in \cref{fig:ptga_hard_nnlo} the 
results for the \NNLO{} \emph{coefficient}. Since this time we do not have a cutoff-independent result available
we directly compare results obtained with two sets of $\taucut$. The first set ranges from the nominal value of
$\taucut/m_{\gamma\gamma}=0.001$ through $\taucut/m_{\gamma\gamma}=0.04$, where the five values $\taucut/m_{\gamma\gamma}=0.002,0.004,0.008,0.02,0.04$ have been
used for the asymptotic fit according to \cref{eq:fitnnlo}. A second set of $\taucut$ values smaller by a
factor of ten is also shown.

\begin{figure}
	\includegraphics[]{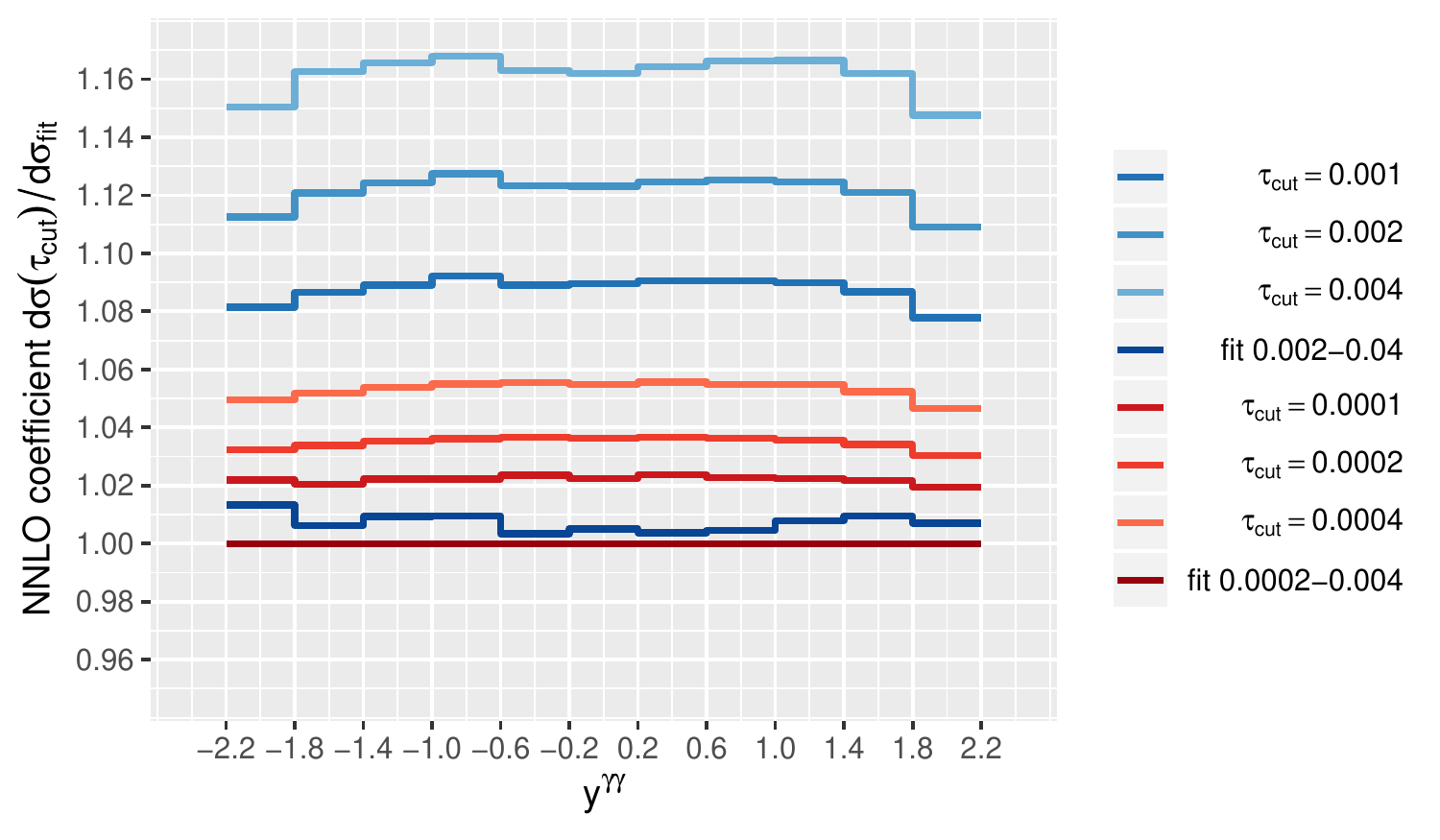}
	\caption{The \NNLO{} coefficient of the diphoton rapidity distribution.
        All curves are normalized to the fitted result with the smaller set of five $\taucut/m_{\gamma\gamma}$ values.}
	\label{fig:ygaga_nnlo}
\end{figure}

We first see that, as for the \NLO{} coefficient, there is a non-uniform $\taucut$ dependence. Second,
both fits agree within $1\%$, such that we can trust that the fit results are accurate within $1\%$. In both cases they
significantly improve the prediction and make the $\taucut$ dependence uniform. The improvement from the fit
is dramatic especially in the first bin, where for $\taucut/m_{\gamma\gamma}=0.001$ the systematic uncertainty without a fit is over
$10\%$. Similar conclusions are drawn from the diphoton rapidity distribution \NNLO{} coefficient in 
\cref{fig:ygaga_nnlo}.

\begin{figure}
	\includegraphics[]{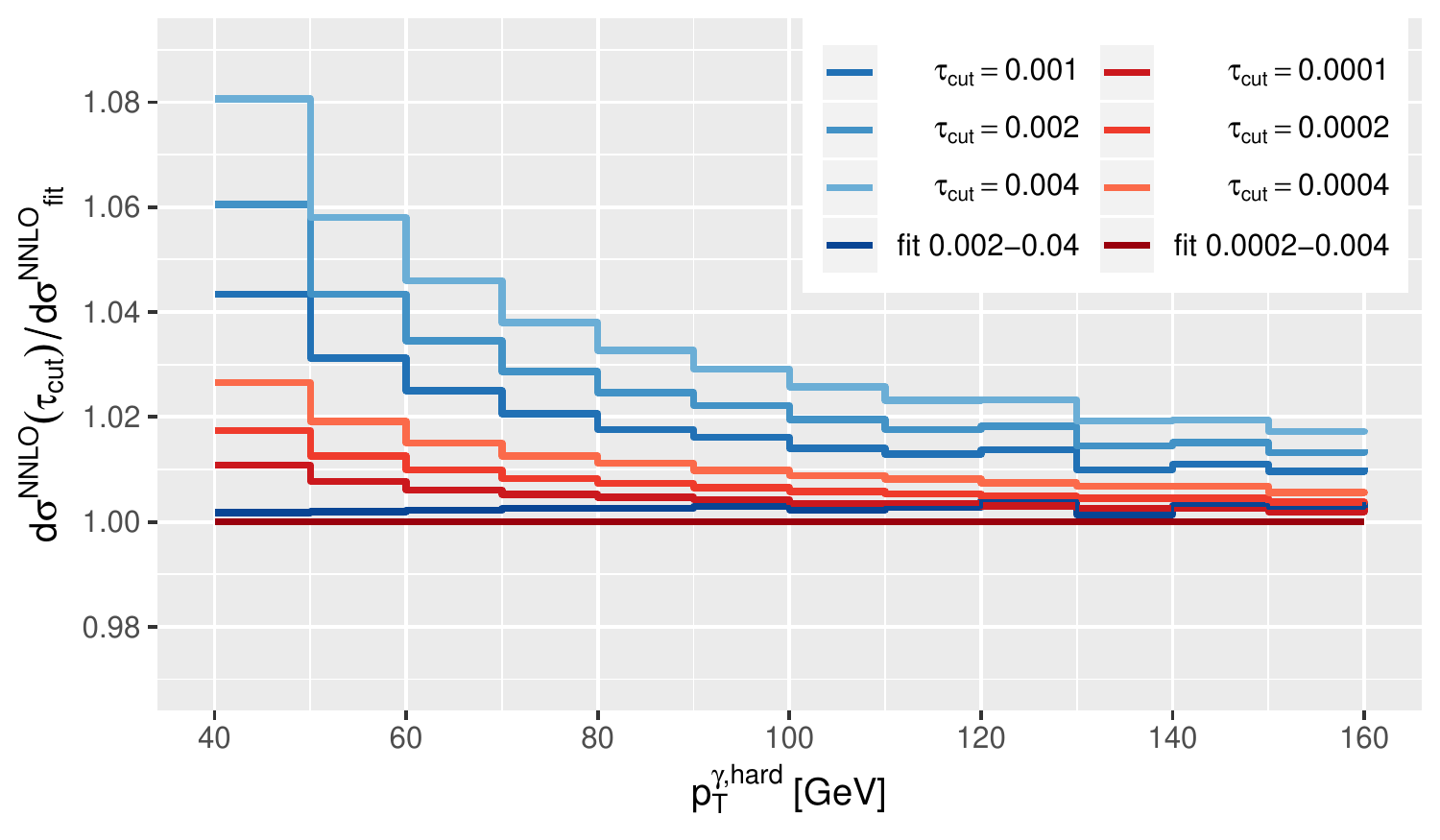}
	\caption{The full \NNLO{} prediction for the transverse momentum distribution of the hardest photon in
        diphoton production.
        All curves are normalized to the fitted result with the smaller set of five $\taucut/m_{\gamma\gamma}$ values.}
	\label{fig:ptga_hard_nnlo_full}
\end{figure}

Since in diphoton production the \NNLO{} corrections are large, we show in \cref{fig:ptga_hard_nnlo_full} the full
\NNLO{} result with corresponding $\taucut$ dependence and fitted results. Our default setting, with a residual
cutoff dependence of $0.8\%$, is $\taucut/m_{\gamma\gamma}=10^{-4}$. The figure shows that, up to numerical artifacts, the same
level of precision can be reached using the results of a fit with $\taucut/m_{\gamma\gamma}\simeq 10^{-3}$.
Nevertheless, we prefer to use the $\taucut$ values in \Cref{tab:default-taucut-new} as our default settings.
The additional asymptotic fitting can then be used to check that the cutoff dependence is indeed at the
percent level.  In the case of a good fit it can also result in a substantial 
improvement over that, with the added benefit of a uniform systematic cutoff error.

\begin{figure}
	\centering
	\includegraphics[]{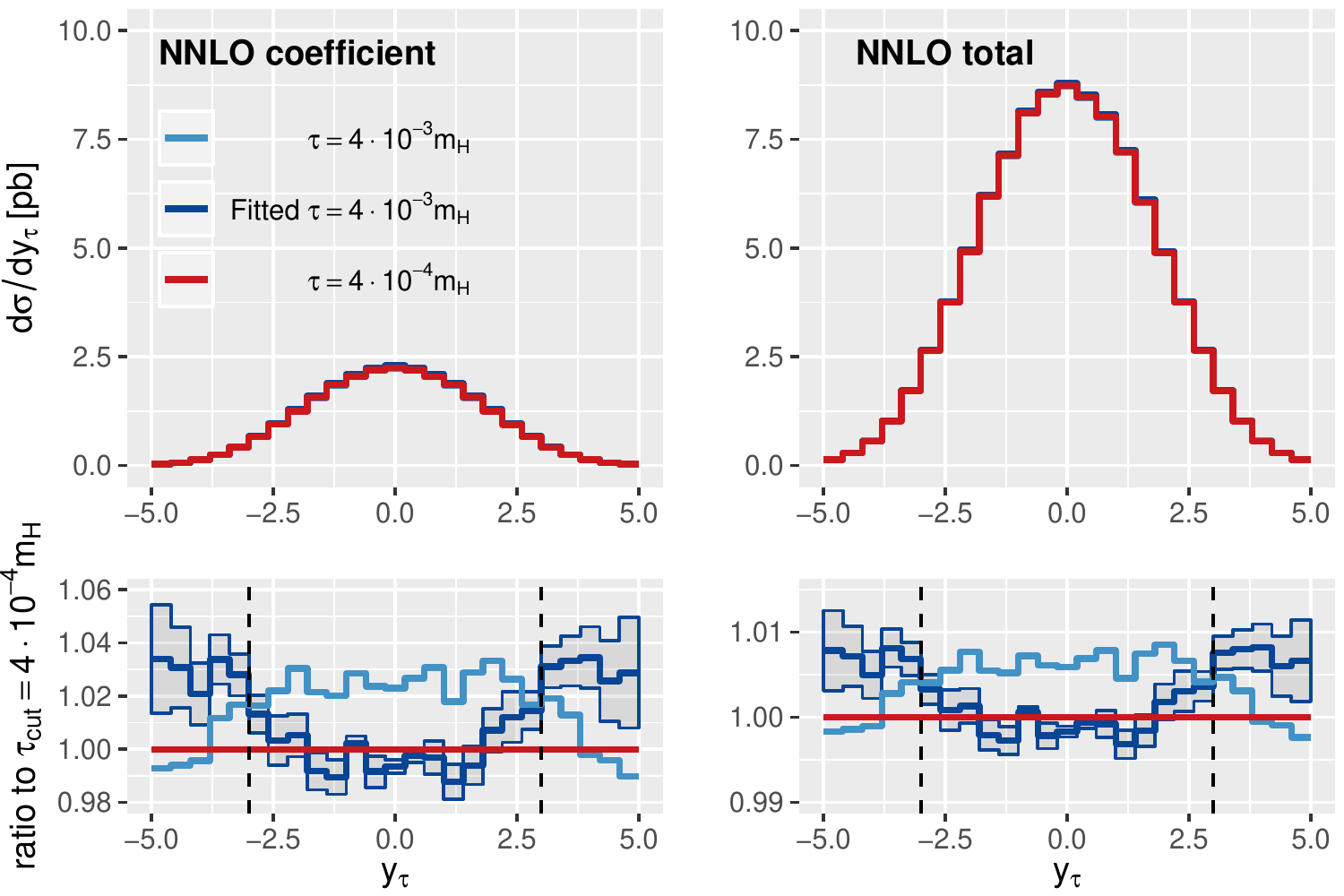}
	\caption{$\tau^-$ rapidity distributions in $pp \to H \to \tau^- \tau^+$,
		corresponding to the \NNLO{} coefficient (left) and the \NNLO{} total (right).
		The histograms for the fit result are included with the fit uncertainties as a shaded band, not including the 
		numerical uncertainties (represented by the noise in the data).
		The dashed line indicates the point at which the fit is no longer trustworthy (see text).}
	\label{fig:ytauH}
\end{figure}

We now consider another representative example, the calculation of the $\tau^-$ rapidity spectrum 
in the process $pp \to H \to \tau^- \tau^+$.  We use the same setup as for our benchmark 
calculation of the inclusive $gg \to H$ cross section.  In \cref{fig:ytauH} we compare the prediction
for this distribution, both in the \NNLO{} coefficient and in the \NNLO{} total, for two choices
of $\taucut$ -- our default value ($\taucut/m_H=4 \times 10^{-3}$) and one corresponding to
an expected $0.2\%$ accuracy in the inclusive cross section ($\taucut/m_H=4 \times 10^{-4}$).
We first see that the residual $\taucut$-dependence of these two choices of $\taucut$ agrees within the expectation 
from the presets in \cref{tab:default-taucut-new},
although this is of course not guaranteed since the table is based only on the inclusive cross
section. 

The figure also indicates the result of the fit that is automatically provided by
\MCFM{} for the case of the default value of $\taucut$. The shaded band corresponds to the fit uncertainty and does not 
include other numerical uncertainties. In the bulk of the rapidity range the fit
is in excellent agreement with the $0.2\%$ result, indicating a clear improvement from the
non-fitted calculation.
In the tails of the distribution, for $|y|>3$, the fit is no longer reliable
since the input data to the fit is calculated with insufficient precision only at the order of $5$--$10\%$, corresponding
to a failure of step 1 of our prescription above. We have denoted this with a dashed line where the numerical 
uncertainty on the input data becomes more than $5\%$. One clearly observes that the reported fit uncertainty is too 
small to explain the discrepancy between the two fixed $\taucut$ values. However, when taking into account the numerical 
uncertainties of all contributions one still observes agreement. 
In this instance, instead of increasing the 
numerical precision in the tails for a reliable fit, either through an additional run with cuts or a reweighting 
function, it may be less expensive to simply run the code with a smaller value 
of $\taucut$ to draw better conclusions about the residual $\taucut$ dependence.

\begin{figure}
	\centering
	\includegraphics[]{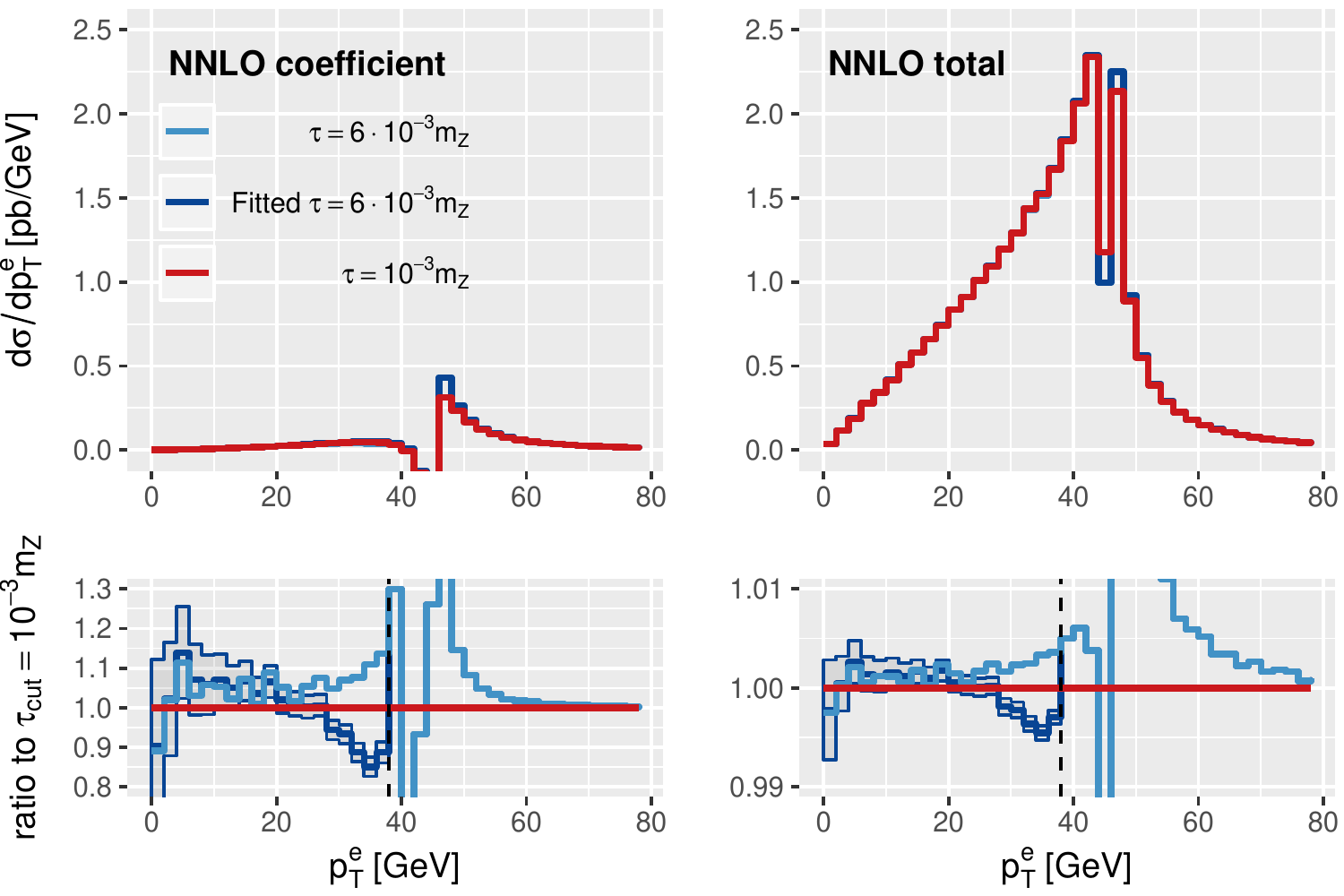}
	\caption{$e^-$ transverse momentum distributions in $pp \to Z \to e^- e^+$,
		corresponding to the \NNLO{} coefficient (left) and the \NNLO{} total (right).	The histograms for the fit 
		result are included with the fit uncertainties as a shaded band, not including the 
		numerical uncertainties (represented by the noise in the data).
		The dashed line indicates the point at which the fit is no longer trustworthy (see text).}
	\label{fig:pteZ}
\end{figure}

Finally, in \cref{fig:pteZ} we show the transverse momentum distribution of the electron
produced in $pp \to Z \to e^- e^+$, again using our benchmark setup.  As above, we compare
results using the default value $\taucut/m_Z=6 \times 10^{-3}$ with results targeting
$0.2\%$ precision, $\taucut/m_Z=10^{-3}$.  At low transverse momenta, $p_T(e^-) \lesssim 35$\,GeV
the situation is much the same as observed above in \cref{fig:ytauH} -- the two results are in
agreement within expectations. Although the quality of the fit is fine, that is we successfully checked
that the input data for the fit has reliable uncertainties and that the $\chi^2/\text{d.o.f.}$ from the fit is small,
the fitted corrections still have large numerical uncertainties. To improve upon this and improve the predictions with 
the fit would require more numerical precision. But we have still gained valuable information: since the fit can be 
trusted, the fitted corrections including their uncertainty (the shaded band) give an estimate of the
residual $\taucut$ dependence. 

In the region 
around $m_Z/2$ the fixed order predictions are not reliable since this distribution is exactly zero
for $p_T(e^-) > m_Z/2$ in the leading order calculation. This is because the benchmark setup of 
ref.~\cite{Boughezal:2016wmq}
uses the zero-width approximation. For the same reason, the 
only $\taucut$ dependence beyond $m_Z/2$ comes from subleading power corrections to jettiness factorization, so
that there is no logarithmic dependence on $\taucut$ in this region.  Since the residual $\taucut$ dependence
is not captured by our fit form, the code reports poor fits in the region
$p_T(e^-) \gtrsim \SI{38}{\GeV}$.  As a result we have only displayed the fit up to this point
(the dashed line in \cref{fig:pteZ}).   Following our own advice we have repeated this study with an
even smaller cutoff, $\taucut/m_Z=10^{-4}$, and verified that the result shown in \cref{fig:pteZ} is accurate
at the per mille level for $p_T(e^-) \gtrsim \SI{50}{\GeV}$.

\paragraph{Summary.} While in all the studies in this section we have compared results using two choices of 
nominal $\taucut$, in most cases
conclusions for the reliability and magnitude of residual finite  $\taucut$ effects could be drawn from the fit and 
the automatically sampled $\taucut$ dependence alone. No further runs with additional nominal $\taucut$ values are 
necessary for most practical applications. This is in contrast to previous versions of \MCFM{} and other codes that 
provide no means to assess reliability of results fully differentially without additional extra runs.

\MCFM{} outputs histograms for the additionally sampled $\taucut$ values separately in addition to the nominal 
$\taucut$ histogram. Furthermore, the fitted corrections are also output separately. In the best case the $\taucut$ 
dependence is 
asymptotic and the numerical precision is good, so that the fitted corrections improve the nominal $\taucut$ result and 
at the same time estimate the error made by using the nominal $\taucut$ value. If 
the numerical 
precision is insufficient the fit might still be good, but the fitted corrections have large uncertainties, adding 
noise. This is still valuable information, since it estimates the maximum $\taucut$ dependence.

The fitting procedure can fail for the following reasons. Either the input data cannot be trusted (generally 
uncertainties are then still large, around $5$-$10\%$), or the input data uncertainties are small, but the fit itself is poor with a 
large $\chi^2/\text{d.o.f.}$. In the 
first case
one can increase the numerical precision or manually inspect the $\taucut$ dependence instead of relying on the fit. In 
the latter case, the calculation is not in the asymptotic regime where the logarithmic form of the fit function
applies. This can signify an 
unusually large $\taucut$ dependence, which requires a manual inspection of the $\taucut$ dependence and/or
a further run for this 
kinematical region with a lower nominal $\taucut$ value. It can also signify that the region of phase space is only 
populated starting at \NLO{}. For this region there is no logarithmic 
$\taucut$ dependence and so the fit fails. In this case a manual inspection should demonstrate
a small $\taucut$ dependence.

\section{Precision \PDF{} uncertainties and \PDF{} set differences}
\label{sec:physics}

Within the realm of uncertainties for cross section predictions, \PDF{} uncertainties are often large compared to
residual theoretical uncertainties.  For instance, uncertainties from missing higher order corrections are 
generally no longer leading and, even for inclusive Higgs production, the \PDF{}+$\alpha_s$ uncertainty estimates range from three 
\cite{Anastasiou:2016cez} to thirteen percent \cite{Accardi:2016ndt}. They are in the same range as the combined 
theory uncertainties of about seven percent \cite{Anastasiou:2016cez}. Since the \LHC{} is able to collect data on far 
more than just total cross sections, the real goal is to compute \PDF{}
uncertainties differentially for kinematic distributions at \NNLO{} precision.

However, \PDF{} uncertainties are among the most computationally expensive to study. This is due to the need to sample
many different \PDF{} sets, or \PDF{} set members. With \PDF{} uncertainties at the few percent level,
one wishes to have per mille level numerical uncertainties that are sufficiently small in comparison.
 The most complicated
\NNLO{} processes (in \MCFM{}) require hours up to days on a $10$-$20$ node cluster to reach the desired $1$-$2$ per mille 
level of numerical precision also differentially. The most na\"ive implementation of a \PDF{} uncertainty calculation 
then requires an individual evaluation for every \PDF{} set member,
which for our demonstrations below with \PDF{} uncertainties for six chosen \PDF{} sets, multiplies required resources 
by $371$, the number of individual \PDF{} set members. The small cluster is now no longer sufficient, and either 
one has to run the calculation for months on the small system or days on thousands of nodes.

It is clear that the na\"ive evaluation of \PDF{} uncertainties represents a substantial additional burden.
To avoid this situation, in \MCFM{} we take into account the correlations between matrix elements
evaluated with different \PDF{}s by integrating over all \PDF{}s at once, storing only difference information. This not 
only holds for one \PDF{} set,
but for any number of \PDF{} sets, only limited by the available system memory, see \cref{subsec:performance}. In 
practice one then gains several orders of magnitude in performance. 
Per mille level \NNLO{} \PDF{} uncertainty predictions and per mille level differences between the different \PDF{} 
sets for hundreds of \PDF{}s members in total can be computed in the same time necessary to calculate the nominal 
central prediction to the per mille level. 

\begin{figure}[h]
	\centering
	\includegraphics[]{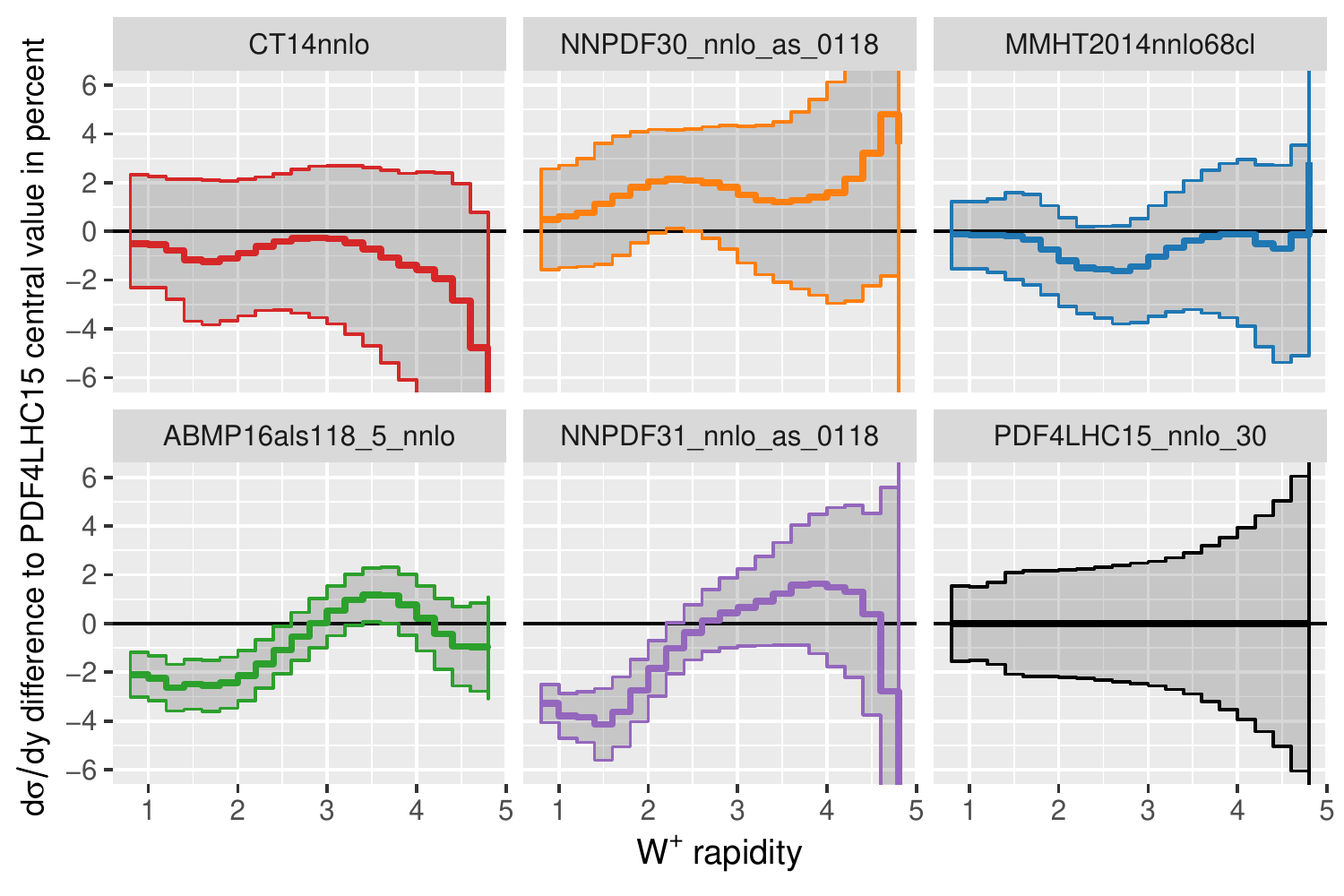}
	\caption{\NNLO{} $W^+$ rapidity distributions with \PDF{} uncertainty bands after applying 
		a lepton rapidity cut, $2.0 < y_l < 4.5$.  All curves
                are normalized to the central {\abbrev PDF4LHC} prediction.}
	\label{fig:wrapidity_nnlo_teaser}
\end{figure}

The capability of our efficient implementation of multi-set \PDF{} uncertainties is demonstrated
in \cref{fig:wrapidity_nnlo_teaser}. Shown are the \NNLO{} $W^+$ rapidity distributions for a total of six different 
\PDF{} sets, each with their own \PDF{} uncertainties, normalized to the {\abbrev PDF4LHC} central value.\footnote{See 
also \cref{fig:Wteaser2} for the positron rapidity distributions obtained in the same run.}
Note that we have included the full $W^+$ boson decay in the integration, but we choose to show the $W^+$ 
rapidity distribution since it is more directly related to the momentum fraction variables that parametrize the \PDF{}s.
These \NNLO{} results were obtained by running less than a day 
on our cluster using 16 nodes (2010 generation) with 12 cores each. A total of $371$ \PDF{}s were evaluated for each 
phase space point.
 Not only are the \PDF{} uncertainties predicted within numerical
sub per mille level precision, but one can read off sub per mille level differences between different \PDF{} sets fully
differentially. This is an invaluable benefit for detailed studies on the impact of different data and methods on 
predictions at \NNLO{}. 
For instance we note that while use of the {\abbrev PDF4LHC} envelope set gives a first estimate of the uncertainties,
\cref{fig:wrapidity_nnlo_teaser} also clearly shows its shortcomings:  the averaging obscures the rich structure
of the individual \PDF{} determinations shown in the upper row of the figure.  Precision studies such as this one
can help to illuminate the differences between the central values, and the corresponding uncertainty bands, of all the sets.

\paragraph{\NNLO{} \PDF{} uncertainties using lower order matrix elements.}
Depending on the number of \PDF{} sets, and considering bounds on \CPU{} or memory availability, the gain from correlations
might not be enough to perform a calculation on a local desktop computer.  In this case it is natural to ask
whether it is sufficient to compute relative \PDF{} uncertainties at a lower perturbative order.
This would cut the computing time by another few orders of magnitude, especially if it were
possible to use \LO{} matrix elements together with \NNLO{} \PDF{}s to estimate the \PDF{} uncertainties for the full
\NNLO{} result at the per mille level.
The difference between using \NNLO{} matrix elements and \LO{} matrix elements with six \PDF{} sets at once, for 
example, can be a day of run-time on $10$-$20$ nodes on a cluster or a few minutes to hours on a desktop computer. 

The question then is: when is it possible to use lower order matrix elements for the extraction of \PDF{} uncertainties and 
to examine precise differences between 
\PDF{} sets? While generally a flat $k$-factor in higher perturbative orders might indicate that it is 
acceptable to use the lower order matrix elements, this is neither necessary nor sufficient. A flat $k$-factor is certainty 
beneficial though, since it means that a large portion of the observable dependence essentially factorizes with respect
to the lower order matrix element. In the same way the dependence on the parton momentum 
fractions would have to be very similar at each order of the calculation in order to reliably use the lower order matrix
element to extract \PDF{} uncertainties at a higher order.

In this section we assess the validity of such a procedure by studying the rapidity distribution of the color singlet 
system for all \NNLO{} processes included in this version
of \MCFM{}.  Since the rapidity of the color singlet system
translates directly to the parton momentum fractions, we choose this kinematical distribution as the most 
representative example. Other distributions obtain some form of smearing, but of course can be computed in the same way.

\paragraph{Decay channels and cuts.}

For our demonstrations the set of decay channels and cuts is as follows. In all cases the $W^+$ boson decays
semi-leptonically, the $Z$ boson leptonically and the Higgs boson into $\tau^+ \tau^-$. We do not explicitly present 
results for $W^-$, which are qualitatively the same as the ones shown for $W^+$ production.

Except for the single boson processes $W^+$, $Z$ and $H$, all cuts are standard cuts as in \MCFM{}-8.0 and \MCFM{}-9.0.
For the single boson processes $W^+$ and $Z$ we additionally restrict the rapidity of the leptons to lie
between $2.0$ and $4.25$, motivated by the \LHCb{} experiment. For $H$ production we relax the $\tau$-rapidities $y$ to 
$-5 
\leq y \leq 5$ to probe extreme regions of the phase space. 
We perform all calculations with the default set of \EW{} parameters as in \MCFM{}-9.0 and for the \LHC{} operating at 
$\sqrt s 
= \SI{14}{\TeV}$.  We allow all vector bosons to be off-shell ({\tt zerowidth} is {\tt .false.}) and include their 
decays ({\tt removebr} is {\tt .false.}). For parameters that are set in the input file we use,
\begin{eqnarray}
m_H = 125~\mbox{GeV} \,, \quad
m_t = 173.3~\mbox{GeV} \,, \quad 
m_b = 4.66~\mbox{GeV} \,, 
\end{eqnarray}
and we use $\mu_F = \mu_R = Q^2$ (i.e. we set {\tt dynamicscale} equal to
either {\tt m(34)} or {\tt m(345)} or {\tt m(3456)}, as appropriate).
Our generic set of cuts is,
\begin{eqnarray}
&& p_T(\mbox{lepton}) > 20~\mbox{GeV} \,, \quad
|\eta(\mbox{lepton})| < 2.4 \,, \quad \nonumber \\
&& p_T(\mbox{photon 1}) > 40~\mbox{GeV} \,, \quad
p_T(\mbox{photon 2}) > 25~\mbox{GeV} \,, \quad \nonumber \\
&&|\eta(\mbox{photon})| < 2.5 \,, \quad \Delta R(\mbox{photon 1, photon 2}) > 0.4 \,, \quad \nonumber \\
&& \Delta R(\mbox{lepton}, \mbox{photon}) > 0.7\,, \quad \nonumber \\
&& E_T^{\mbox{miss}} > 30~\mbox{GeV} \,, \quad
\end{eqnarray}
For $Z$ as well as $ZH$ and $Z\gamma$ production we also impose a minimum $Z^*$ virtuality ({\tt m34min}) of 
\SI{40}{\GeV}.

\subsection{Rapidities in single- and diboson production}

To study and exemplify the use and calculation of differential \NNLO{} \PDF{} uncertainties we first show the rapidity
distributions of $W$ and $Z$ bosons in the forward region, where one might
expect to see the most substantial differences between \PDF{}s and the largest uncertainties. This region is of particular
interest for the $W$~\cite{Aaij:2015zlq,Aaij:2016qqz} and $Z$~\cite{Aaij:2015zlq,Aaij:2016mgv,Aaij:2015gna} production
measurements in \LHCb{}.

\paragraph{$W^+$ rapidities.} 

We first compute the $W^+$ rapidity distributions with \PDF{} uncertainties for the six \PDF{} sets
\texttt{ABMP16als118\_5\_nnlo} \cite{Alekhin:2017kpj}, \texttt{CT14NNLO} \cite{Dulat:2015mca}, 
\texttt{MMHT2014nnlo68cl} \cite{Harland-Lang:2014zoa}, 
\texttt{NNPDF30\_nnlo\_as\_0118} \cite{Ball:2014uwa},
\texttt{NNPDF31\_nnlo\_as\_0118} \cite{Ball:2017nwa} and \texttt{PDF4LHC15\_nnlo\_30} \cite{Butterworth:2015oua}. 
We compute the distributions using \NLO{} and \LO{} matrix elements and with full \NNLO{} matrix elements using two 
values of $\taucut$.

\begin{figure}
	\centering
	\includegraphics[]{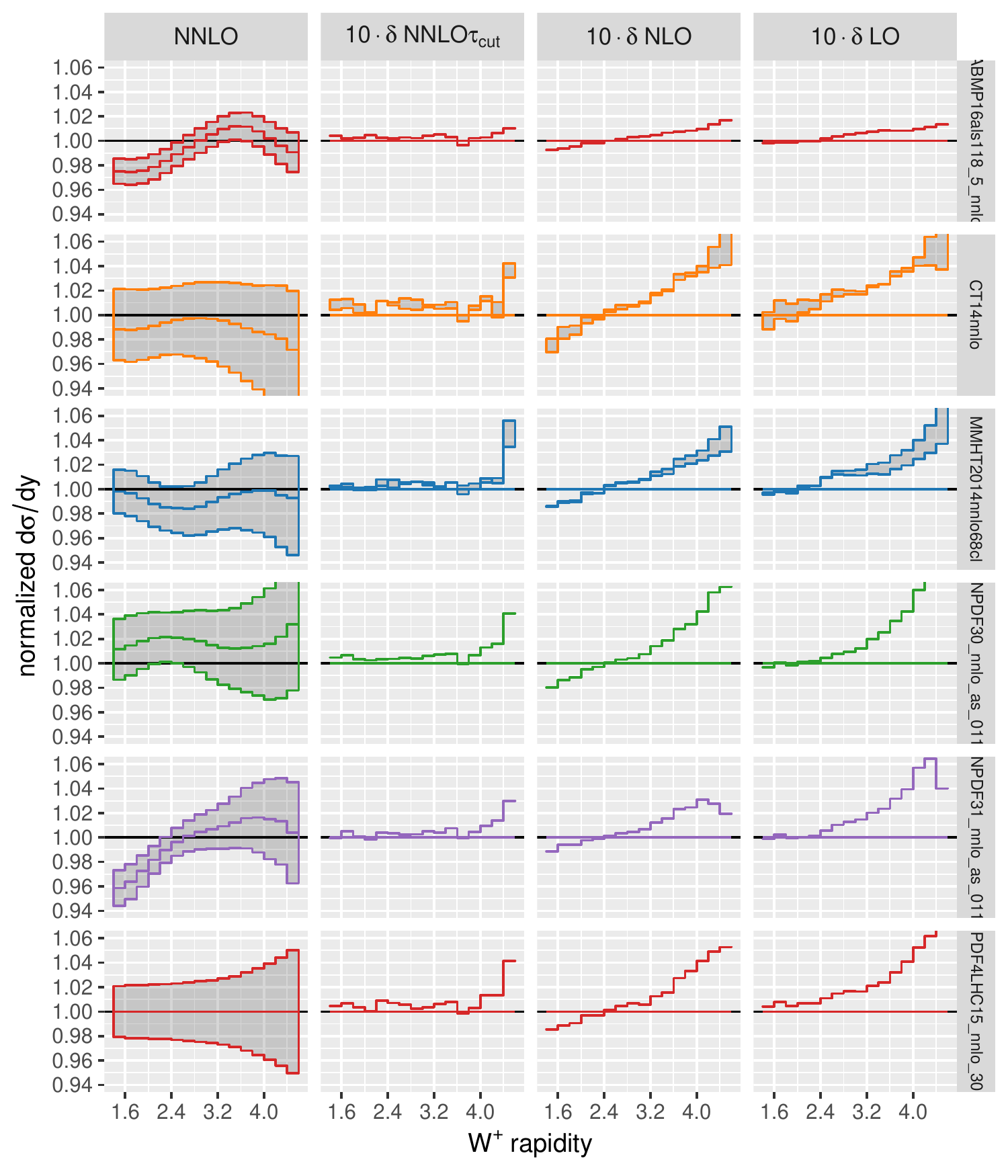}
	\caption{Normalized $W^+$ rapidity distributions in the forward region. The first column shows the \NNLO{} result 
	normalized to the central value of the {\abbrev PDF4LHC} set with a value of $\taucut=2$\,GeV. The other three columns
		show the difference of the \PDF{} uncertainties when using $\NNLO{}~(\taucut=0.1\,\rm{GeV})$, \NLO{} and \LO{} matrix 
		elements, respectively. The differences are  magnified by a factor of 
		$10$. Rows represent different \PDF{} sets.}
	\label{fig:Wrapidity_combination}
\end{figure}

Our results are shown in \cref{fig:Wrapidity_combination} where, as in \cref{fig:wrapidity_nnlo_teaser},
we have normalized all distributions to the {\abbrev PDF4LHC} central value. The
first column shows the normalized \NNLO{} cross section for $\taucut=2$\,GeV, corresponding to slicing cutoff effects of 
about $1\%$ in the total cross section. The second column shows the 
difference between the \NNLO{} $\taucut=2$\,GeV result and the \NNLO{} result with $\taucut=0.1$\,GeV, which corresponds to 
slicing cutoff effects that are smaller than $0.2\%$. \NLO{} and \LO{} matrix 
elements are used for the last two columns, respectively.
All these differences are magnified by a factor of $10$. Difference values above one mean that the nominal \NNLO{} 
uncertainties
used for normalization are larger, whereas values smaller than one mean that the uncertainties are smaller. 
The latter case can be seen, for example, in the comparison with \NLO{} matrix elements at the lowest displayed 
rapidities.

We first observe that the differences between the two \NNLO{} calculations
are at the $1$--$2$ per mille level, entirely compatible with numerical noise. In principle one might only expect half-percent
or so agreement due to the finite cutoff values corresponding to better than $1\%$ and $0.2\%$ effects, but this
does not appear to be the case. The difference between using the \NLO{} and \NNLO{} calculations
reaches about the half a percent level at the largest rapidities, and is only a little larger again when
using just the \LO{} matrix element. Of course the $W^+$ rapidity distribution is not a directly-observable quantity,
but we have checked that the same conclusions apply for the lepton rapidity distribution.

\begin{figure}
	\centering
	\includegraphics[]{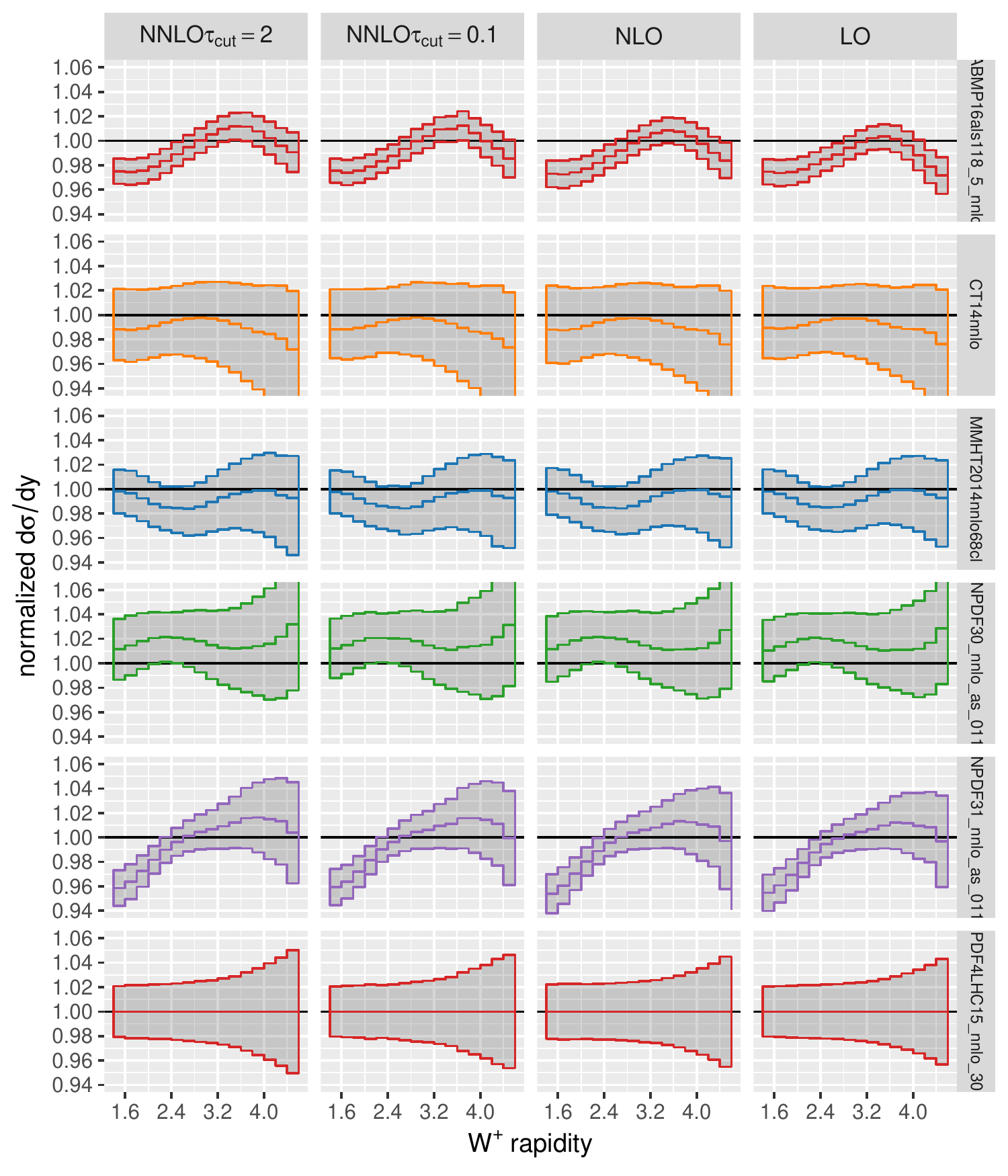}
	\caption{Normalized $W^+$ rapidity distributions in the forward region. The columns represent
	using $\NNLO{}~(\taucut=2\,\rm{GeV})$, $\NNLO{}~(\taucut=0.1\,\rm{GeV})$, \NLO{} and \LO{} matrix elements, respectively, each
one normalized by the {\abbrev PDF4LHC} \PDF{} set.}
	\label{fig:Wrapidity_abs}
\end{figure}

We are now in a position to judge whether, for this process, it is acceptable to use \LO{} or \NLO{} matrix elements
to estimate \NNLO{} \PDF{} uncertainties.  The answer of courses depends on the level of precision that is required.
\LO{} matrix elements do not change the \PDF{} uncertainty predictions in $W^+$ production for 
lower rapidities significantly, but change uncertainties by up to one percent for most \PDF{} sets towards the end of 
the rapidity range shown here. While using \NLO{} matrix elements reduces that difference a little, one ultimately has to 
rely on the full \NNLO{} calculation for uncertainty estimates that are more precise than $0.5\%$.
When using lower order matrix elements one can still compare features of different \PDF{} sets
at the per mille level, since these are relatively unaffected by the shifts induced at higher orders.
This is illustrated in \cref{fig:Wrapidity_abs}, which shows the raw results entering \cref{fig:Wrapidity_combination},
but without taking differences with respect to the \NNLO{} $\taucut=2$\,GeV result.
It is clear that features corresponding to individual \PDF{} sets are retained, both 
qualitatively and quantitatively, when using lower order matrix elements.  

\paragraph{Other processes.}

\begin{figure}
	\includegraphics[]{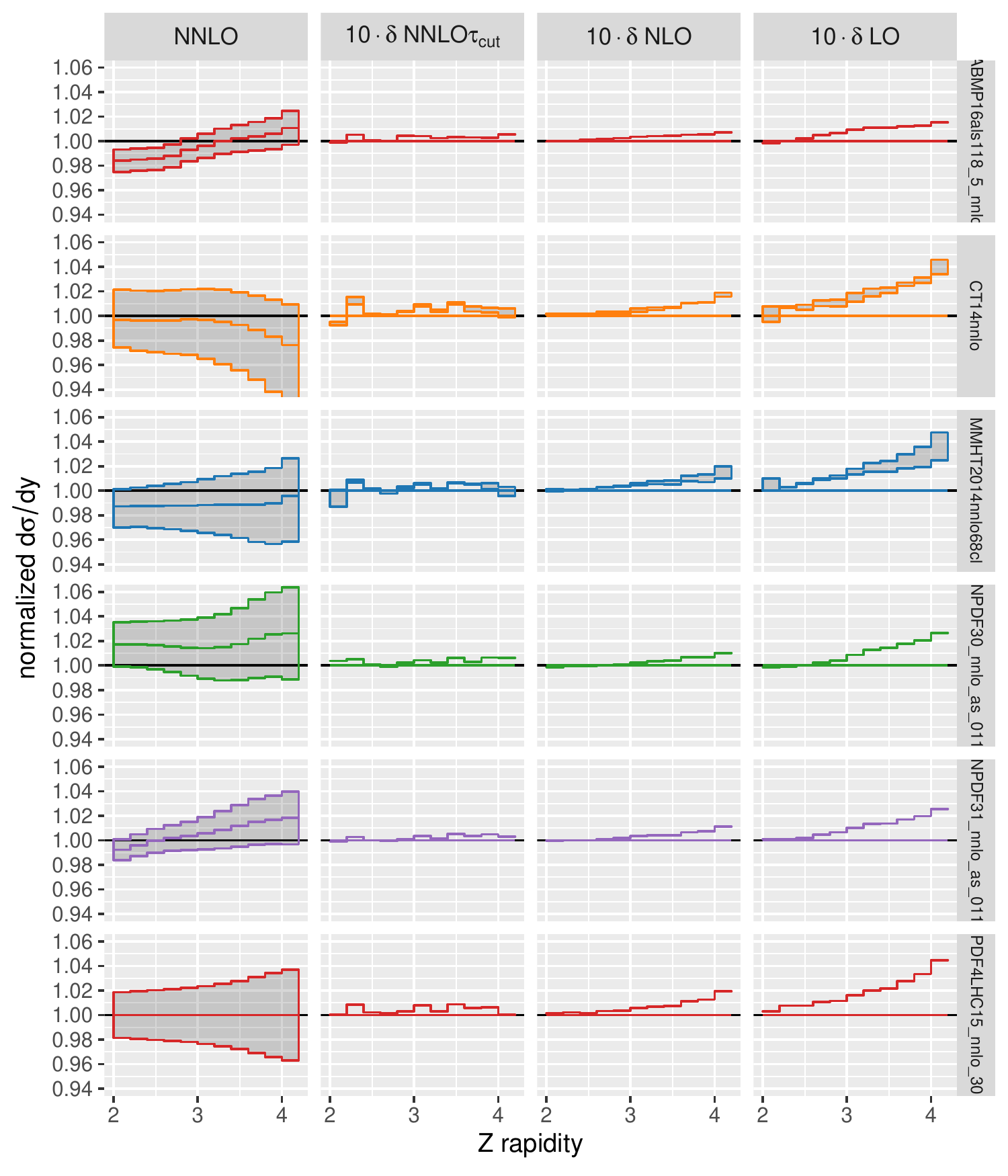}
	\caption{Normalized $Z$ rapidity distributions in the forward region. The first column shows the \NNLO{} result 
		normalized to the central value of the {\abbrev PDF4LHC} set with a value of $\taucut=2$\,GeV. The other three 
		columns
		show the difference of the \PDF{} uncertainties when using $\NNLO{}~(\taucut=0.1\,\rm{GeV})$, \NLO{} and \LO{} matrix 
		elements, respectively. The differences are  magnified by a factor of 
		$10$. Rows represent different \PDF{} sets.}
	\label{fig:Zrapidity_combination}
\end{figure}

\begin{figure}
	\includegraphics[]{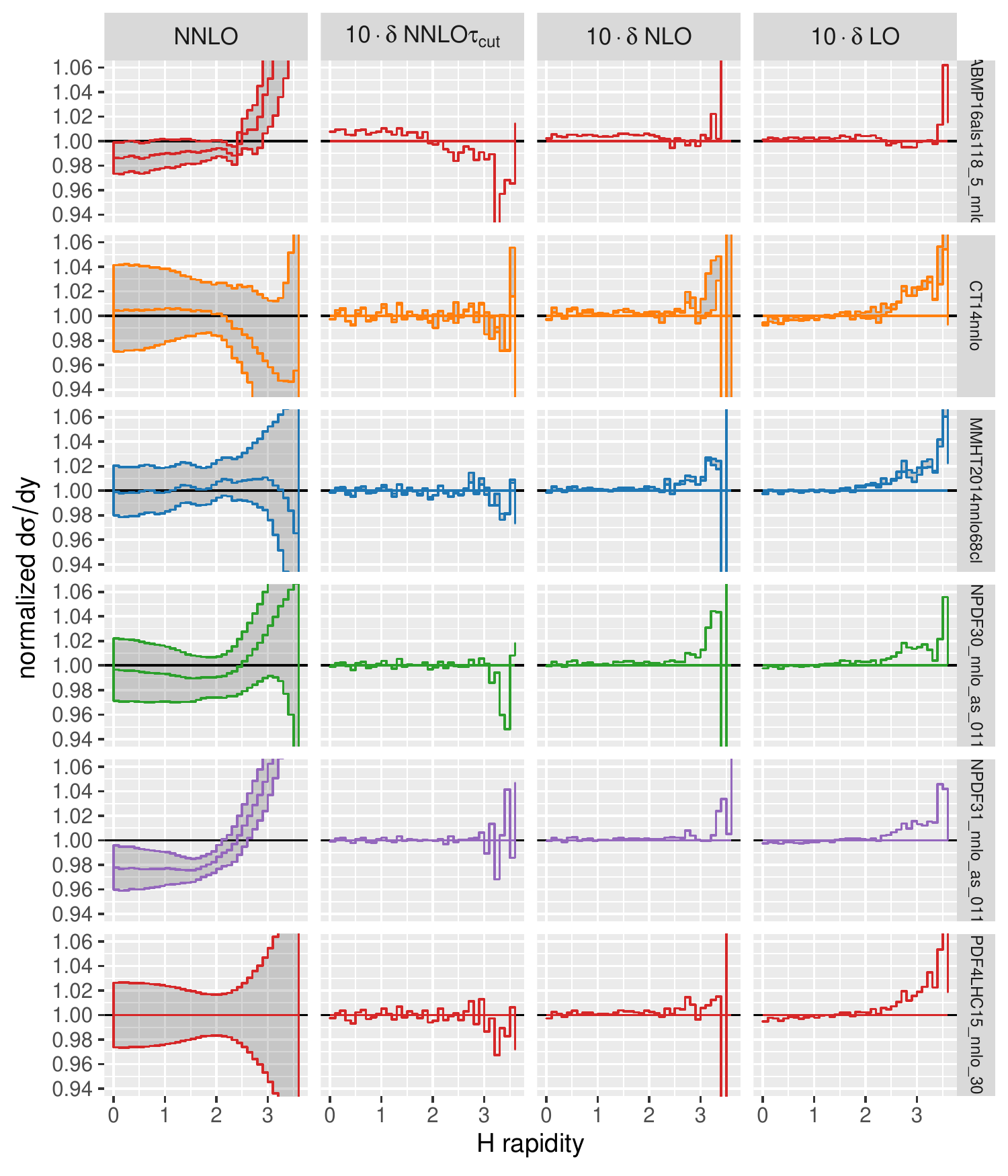}
	\caption{Normalized $H$ rapidity distribution with the lepton rapidities $y$ relaxed to $|y|<5$. The first 
	column 
	shows the \NNLO{} result 
		normalized to the central value of the {\abbrev PDF4LHC} set with a value of $\taucut=0.5$\,GeV (1\%). The other 
		three columns
		show the difference of the \PDF{} uncertainties magnified by a factor of $10$. Rows represent different \PDF{} 
		sets.}
	\label{fig:Higgsrapidity_combination}
\end{figure}

\begin{figure}
	\includegraphics[]{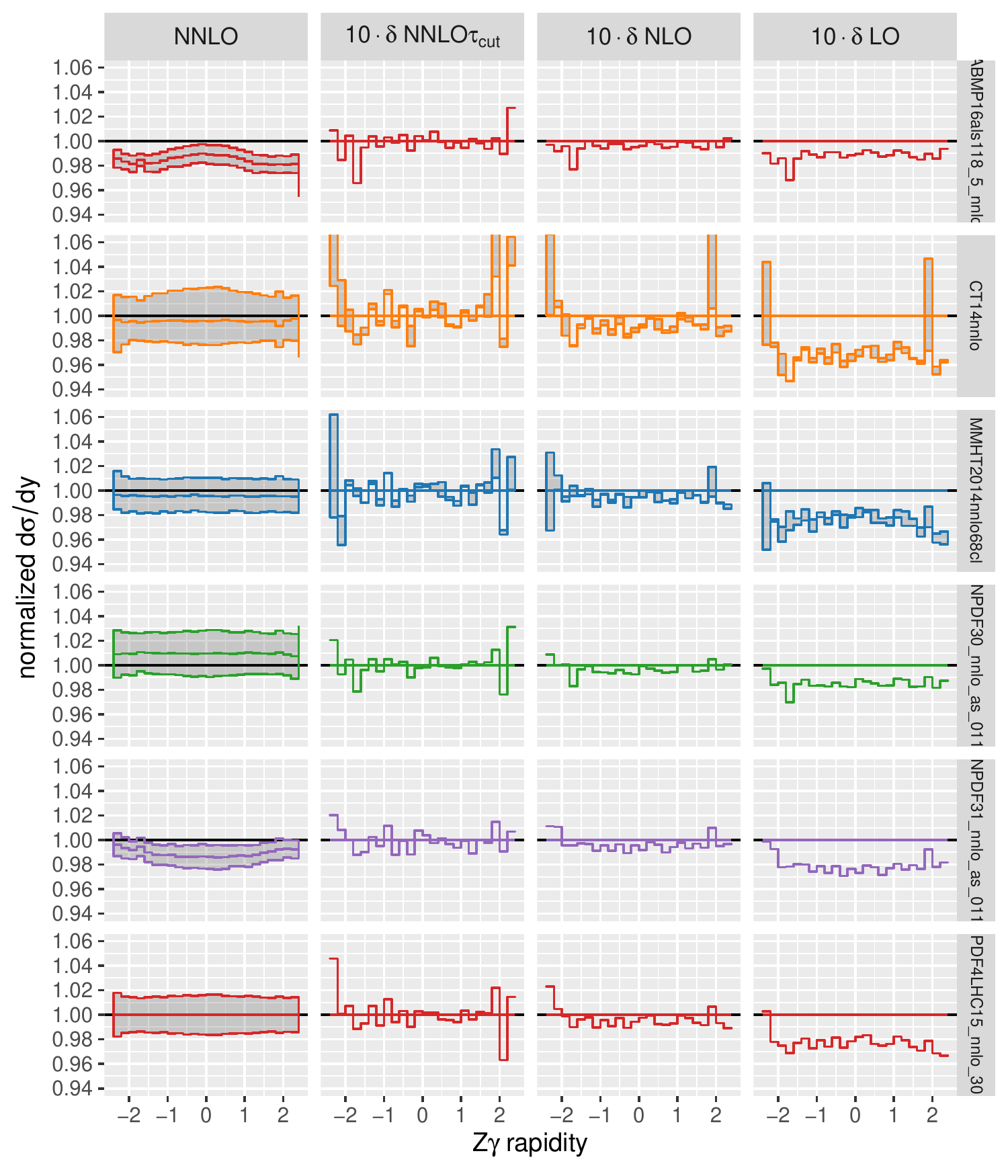}
	\caption{Normalized $Z\gamma$ rapidity distribution. The first 
		column 
		shows the \NNLO{} result 
		normalized to the central value of the {\abbrev PDF4LHC} set with a value of $\taucut/m_{Z\gamma}=3\cdot 10^{-4}$ (1\%). 
		The other 
		three columns
		show the difference of the \PDF{} uncertainties magnified by a factor of $10$. Rows represent different \PDF{} 
		sets.}
	\label{fig:Zgarapidity_combination}
\end{figure}

\begin{figure}
	\includegraphics[]{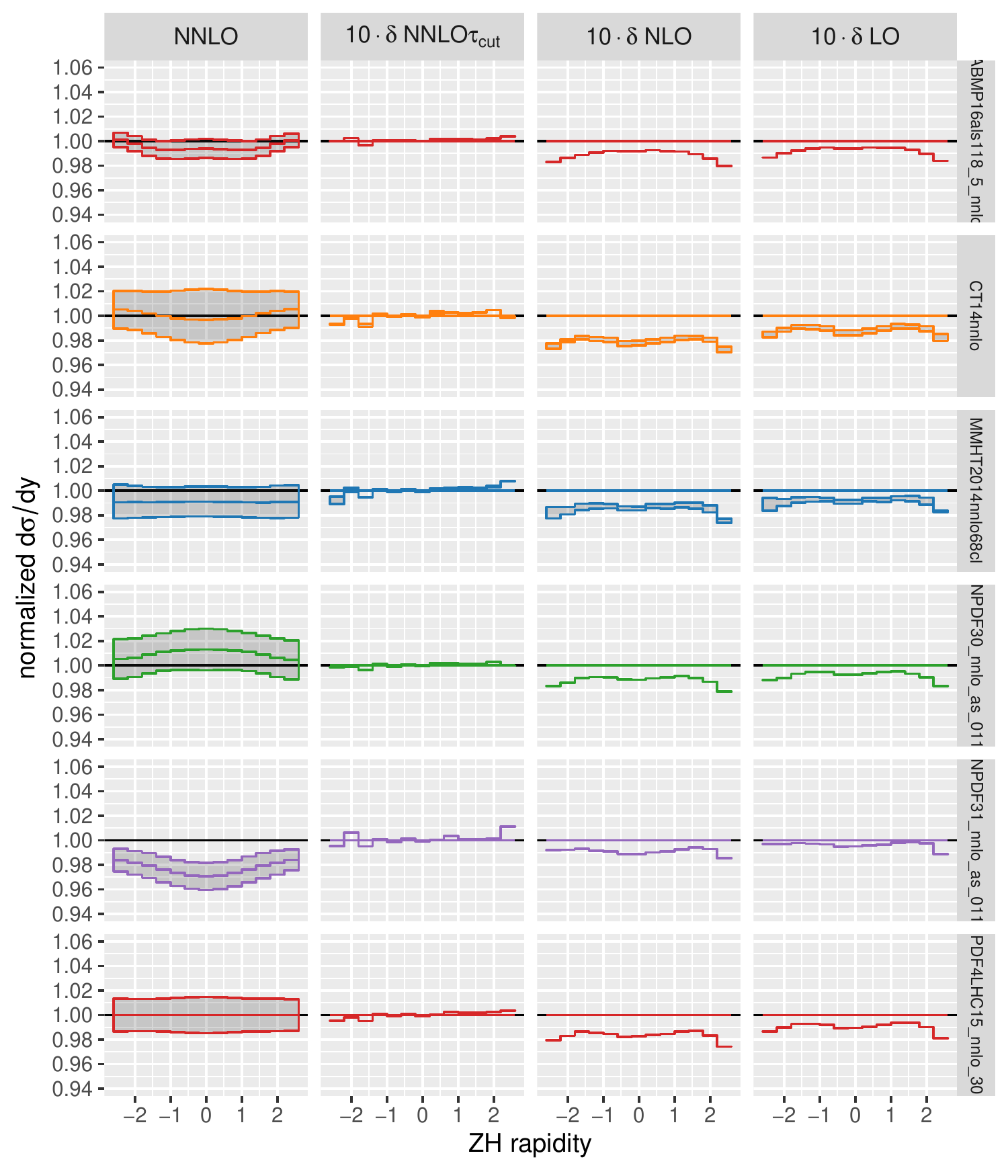}
	\caption{Normalized $ZH$ rapidity distribution. The first 
		column 
		shows the \NNLO{} result 
		normalized to the central value of the {\abbrev PDF4LHC} set with a value of $\taucut=4$\,GeV (1\%). 
		The other 
		three columns
		show the difference of the \PDF{} uncertainties magnified by a factor of $10$. Rows represent different \PDF{} 
		sets.}
	\label{fig:ZHrapidity_combination}
\end{figure}

\begin{figure}
	\includegraphics[]{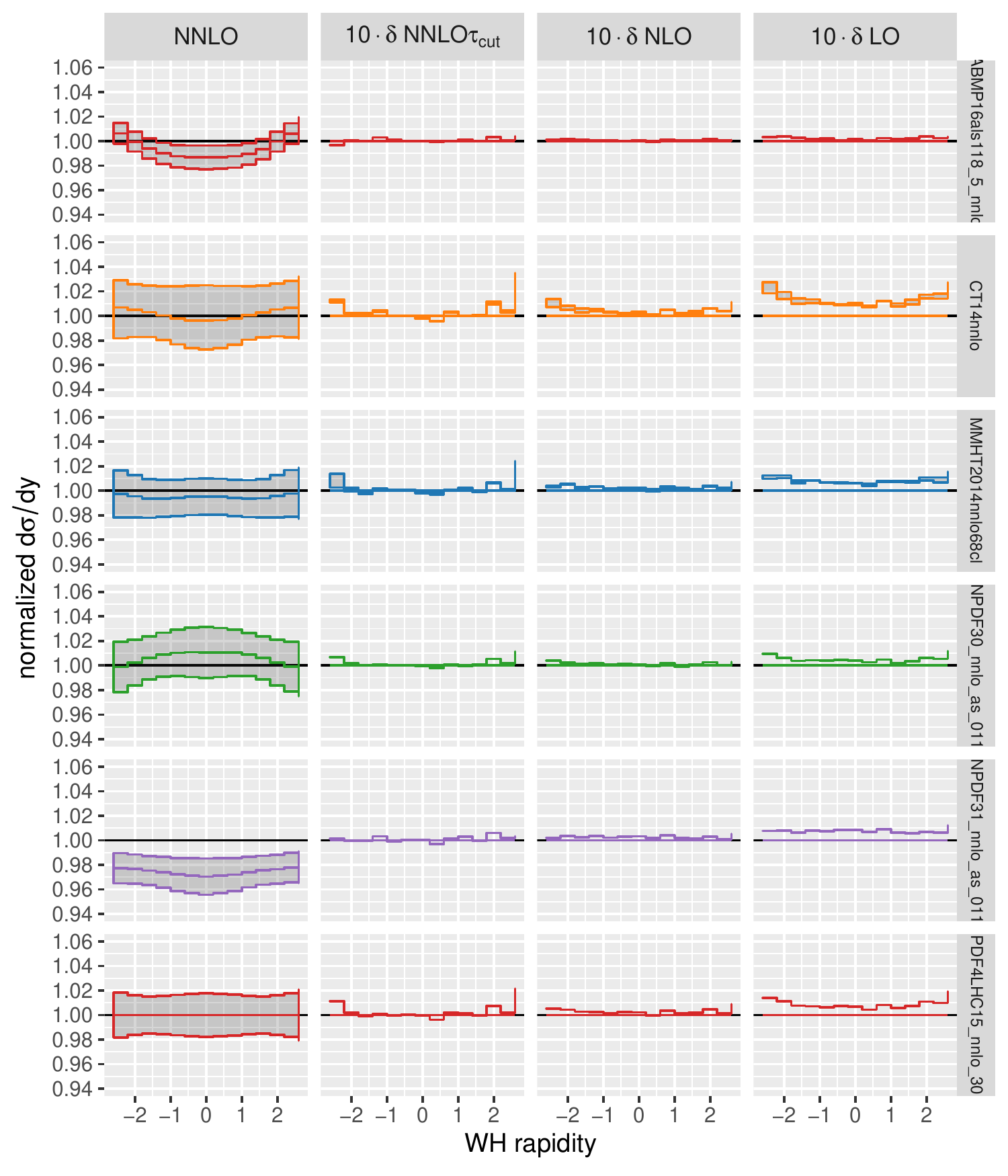}
	\caption{Normalized $WH$ rapidity distribution. The first 
		column 
		shows the \NNLO{} result 
		normalized to the central value of the {\abbrev PDF4LHC} set with a value of $\taucut=4$\,GeV (1\%). 
		The other 
		three columns
		show the difference of the \PDF{} uncertainties magnified by a factor of $10$. Rows represent different \PDF{} 
		sets.}
	\label{fig:WHrapidity_combination}
\end{figure}

\begin{figure}
	\includegraphics[]{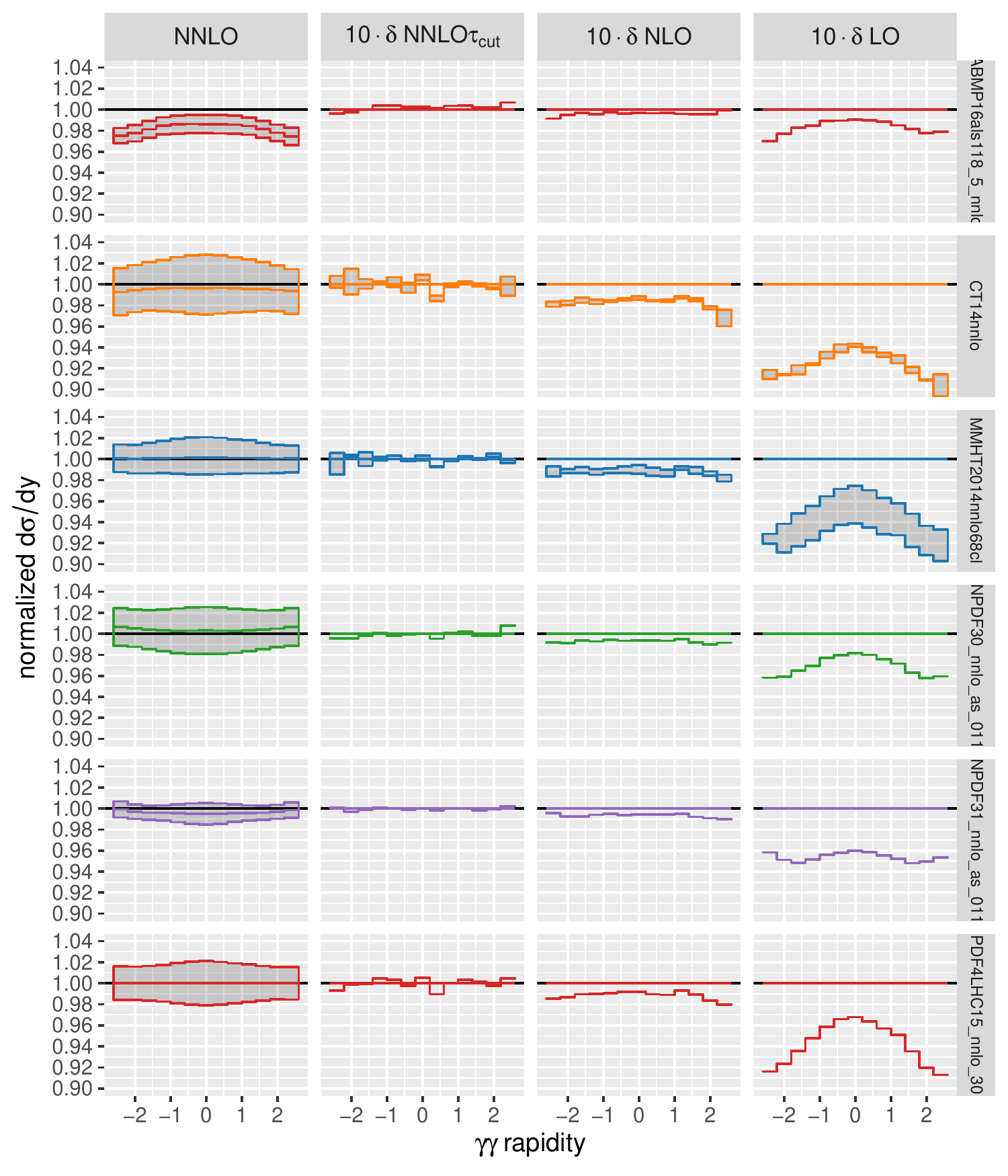}
	\caption{Normalized $\gamma\gamma$ rapidity distribution. The first 
		column 
		shows the \NNLO{} result 
		normalized to the central value of the {\abbrev PDF4LHC} set with a value of $\taucut/m_{\gamma\gamma}=10^{-4}$ 
		(1\%). 
		The other 
		three columns
		show the difference of the \PDF{} uncertainties magnified by a factor of $10$. Rows represent different \PDF{} 
		sets.}
	\label{fig:GamGamrapidity_combination}
\end{figure}

In order to make a more general statement, we repeat this exercise for all the other processes that may be
computed at \NNLO{} in \MCFM{}.  We focus on the equivalent of \cref{fig:Wrapidity_combination} in order to
readily assess the impact of using matrix elements of different orders to estimate \PDF{} 
uncertainties at \NNLO{}.  Our results are shown in
\cref{fig:Zrapidity_combination,fig:Higgsrapidity_combination,fig:Zgarapidity_combination,fig:WHrapidity_combination,fig:ZHrapidity_combination,fig:GamGamrapidity_combination}.
As a reminder, these are all normalized by the \NNLO{} result with a $\taucut$ setting representing $1\%$ 
cutoff effects on the total cross section and include the $0.2\%$ $\taucut$ setting for comparison.

Although there are qualitative differences between the impact of individual \PDF{} sets on $W^+$ and $Z$ production,
the behavior of the uncertainty bands at each order is rather similar, as shown in \cref{fig:Zrapidity_combination}.
We are therefore led to similar conclusions for $Z$ production as already discussed above.
In order to allow for predictions of the Higgs rapidity also in the very forward region, in
\cref{fig:Higgsrapidity_combination} we have relaxed the lepton rapidity to $|y| < 5$.
We then observe that here the differences between using different order matrix elements are also small
and only grow systematically to a few per mille towards large rapidities. In all cases the differences between the
two $\taucut$ values at \NNLO{} are consistent with one per mille numerical noise for moderate rapidities.

The results for the $Z\gamma$~(\cref{fig:Zgarapidity_combination}),
$WH$~(\cref{fig:WHrapidity_combination}) and $ZH$~\cref{fig:ZHrapidity_combination} diboson processes
are more or less similar, with per mille level differences at most.
The situation for $\gamma\gamma$ production~(\cref{fig:GamGamrapidity_combination}) is a
notable exception, where using \LO{} matrix elements leads to noticeable shape and uncertainty differences of up
to one percent compared to the \NNLO{} result.  Of course this is not unexpected since, in this
case, there is a significant contribution from gluon-initiated processes that only enters at \NLO{} and whose
effects are clearly not quite captured with enough precision at \LO{}.  In this case one must use
at least \NLO{} matrix elements in order to properly assess \PDF{} uncertainties, especially since they
themselves are only at the level of $\sim 2\%$.

\paragraph{Summary.}
In general, we find that \NLO{} matrix elements can be used in many cases for precision studies, sometimes even just 
\LO{} matrix elements -- even to determine per mille level differences between different \PDF{} sets.
The difference induced by using a lower order matrix element is far smaller than the differences 
between using different \PDF{} sets. 
Therefore, at least for the processes and observables studied here, it is completely sufficient for per mille level
applications to use \NLO{} matrix elements for the computation of relative \PDF{} uncertainties.
Nevertheless, unless theoretically motivated or plainly demonstrated, we still advocate to use the highest order matrix
element possible given the computational resources available.  Such a calculation can then be
compared with results obtained using lower order matrix elements, which can then be used 
for further runs or more \PDF{} sets, as required. For such a test it might be enough to perform a run using just a
set of central values from different \PDF{} sets, a capability for which \MCFM{}-9.0 allows.

\clearpage
\subsection{Higgs boson transverse momentum}

As a further example calculation we now consider the Higgs boson transverse momentum distribution.
This distribution can be used to constrain the effects of new physics contributions in the Higgs sector, and
has been used previously in the highly boosted regime to measure the $H\to \bar{b} b$ channel at the \LHC{}
\cite{Sirunyan:2017dgc}. 

Perturbative corrections to Higgs+jet production have been computed in an \EFT{} with an integrated-out top-quark up to
\NNLO{}~\cite{Boughezal:2013uia,Chen:2014gva,Boughezal:2015dra,Boughezal:2015aha,Caola:2015wna,Chen:2016zka,Campbell:2019gmd},
and also in the full theory with a finite top-quark mass up to \NLO~\cite{Jones:2018hbb}. Scale uncertainties at 
\NNLO{} are about $10\%$ inclusively and differentially. With theory uncertainties at the $10\%$ level, it is hard to 
imagine that studying \PDF{} uncertainties at the few percent level is a priority. But even for inclusive Higgs 
production, where theory uncertainties are $5-7\%$ and \PDF{} uncertainties have been estimated to be about $3\%$ 
\cite{Anastasiou:2016cez}, recent studies question this precision and suggest \PDF{} uncertainties at the level of 
thirteen percent \cite{Accardi:2016ndt}.

Since predictions for Higgs+jet production at \NNLO{} are computationally highly challenging, one ideally 
wishes to obtain information for different \PDF{}s using lower order matrix elements. Therefore, in this
section we study the Higgs transverse momentum distribution at \NLO{} using \NNLO{} and \NLO{} 
\PDF{}s. The perturbative \NNLO{}/\NLO{} $k$-factors are remarkably flat, being almost constant for the most common 
kinematical distributions. Finite top-quark-mass corrections are also flat with respect to rescaling with the top-quark 
mass dependent Born cross section, even in the threshold region. See ref.~\cite{Jones:2018hbb} for explicit results at 
\NLO{} in the Higgs transverse momentum distribution. The combination of these effects
means that we anticipate the calculation of \PDF{} uncertainties for this process to be excellent in an \EFT{} 
description and even when just using \NLO{} matrix elements. The calculation in \MCFM{} provides \NLO{} predictions 
that go beyond the \EFT{} and take into account mass effects, with residual mass effect uncertainties of $1$-$2\%$ 
\cite{Neumann:2018bsx,Neumann:2016dny}.

\begin{figure}
	\centering
	\includegraphics[]{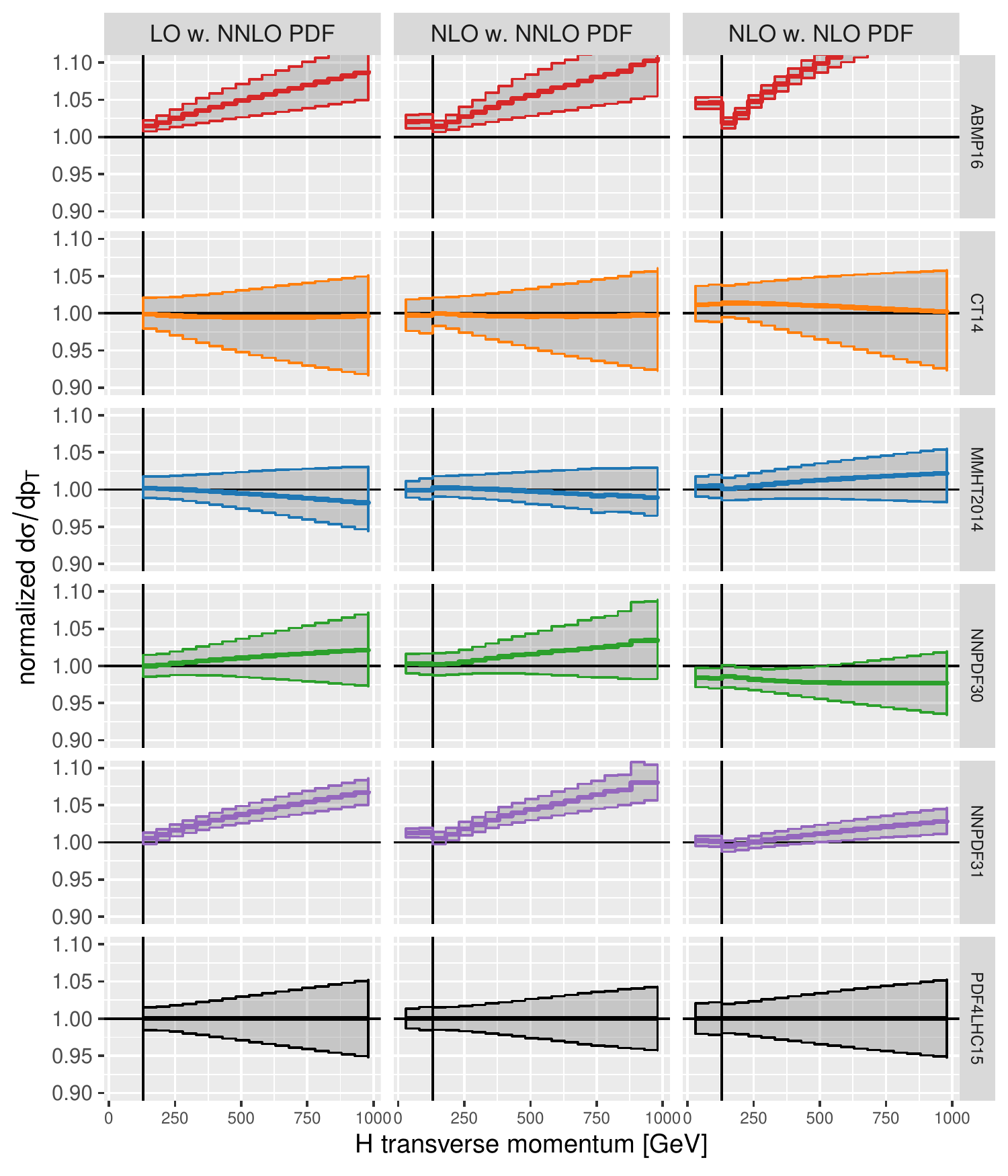}
	\caption{\PDF{} uncertainties for the Higgs transverse momentum distribution normalized to {\abbrev PDF4LHC} with 
	varying matrix element orders and \PDF{} set orders. The left column uses \LO{} matrix 
	elements with \NNLO{} \PDF{} sets, the middle column \NLO{} matrix elements with \NNLO{} \PDF{} sets, and the right
	column uses \NLO{} matrix elements with \NLO{} \PDF{} sets. All \PDF{} sets use $\alpha_s(m_Z)=0.118$.}
	\label{fig:Higgstrans}
\end{figure}

To stress test the calculation of \PDF{} uncertainties across kinematical thresholds, we have applied a jet cut of 
\SI{150}{\GeV}, which means that the region of Higgs boson transverse
momentum below this value is not described at \LO{}.  Our results for the \PDF{} uncertainties in the Higgs boson
transverse momentum distribution up to $p_T = \SI{1}{TeV}$ are shown
in \cref{fig:Higgstrans}.   All \PDF{} sets are chosen to consistently have $\alpha_s(m_Z)=0.118$
at \NLO{} and \NNLO{} to eliminate large differences caused by different central values of $\alpha_s$.
Results are presented in similar fashion to the plots in the last section,
where each column is normalized by the {\abbrev PDF4LHC} central value. 
The left two columns show results with \NNLO{} \PDF{}s but using \LO{}
and \NLO{} matrix elements, respectively, whereas the rightmost column shows consistent \NLO{} results. The left two 
columns indicate small differences of at most two percent between \PDF{} uncertainties estimated using \LO{} and \NLO{} 
matrix elements.

In the right column, for the consistent \NLO{} calculation, the {\abbrev ABMP16} set shows a peculiar behavior:
the predictions are consistently larger than all other sets and reach values that are about $16\%$ larger than the
{\abbrev PDF4LHC} central value. Uncertainties can in no way account for this behavior. This is
despite the fact that using \NNLO{} \PDF{}s (the middle column) the central predictions of {\abbrev ABMP16} are very 
close to those of the {\abbrev NNPDF31} set and the two are mutually consistent within uncertainties.
In that light, only the uncertainty of the {\abbrev ABMP16} \NLO{} 
fit seems to be somewhat underestimated.

We conclude that with current modern \PDF{} sets one observes predictions for central values that differ
by more than ten percent at the largest considered transverse momenta approaching \SI{1}{\TeV}, the region that is the 
most relevant for new physics 
analyses. The differences between the sets that take into account the newest \LHC{} data ({\abbrev ABMP16} and {\abbrev 
NNPDF31}) and the older sets can not be explained by \PDF{} uncertainties at the one sigma level. As long as these
differences are not resolved and are competing with scale uncertainties at the same level, new physics will be
very difficult to constrain. This version of \MCFM{} provides a robust framework for carefully studying such issues
in the light of new data.

\section{Precision studies of $W$ and $Z$ production.}
\label{sec:WZphysics}

Theory predictions for $W$ and $Z$ production are available at \NNLO{} \QCD{} in \MCFM{} \cite{Boughezal:2016wmq} 
and include one-loop electroweak corrections for $Z$ production \cite{Campbell:2016dks}. Remaining perturbative 
truncation uncertainties, estimated by scale variation, are at about the $1\%$ level. In contrast,
experimental uncertainties reach the level of a few per mille.  To achieve a similar accuracy in the
theoretical predictions is highly difficult and even the purely
numerical integration uncertainty can easily surpass the expected systematic theory uncertainties.

These numerical difficulties are illustrated, for example, in ref.~\cite{Aaboud:2016btc} where the 
\emph{numerical} differences between the codes \FEWZ{} and \DYNNLO{}, implementing the same \NNLO{} corrections 
for Drell-Yan production, are found to be $1.2\%$ for fiducial  $W^+$ cross sections.  This is similar to the
estimated truncation uncertainty and double the size of the experimental uncertainty of $0.6\%$.  This has
significant phenomenological impact.
The extracted strange quark density, based on \HERA{} and \ATLAS{}
$W$ and $Z$ data, is affected at the $8\%$ level by this difference, again larger than the $6\%$ experimental
uncertainty \cite{Aaboud:2016btc}.  The measurement of $|V_\text{cs}|$ is similarly impacted by the
\FEWZ{}/\DYNNLO{} difference, requiring an additional theoretical uncertainty of $1\%$, which is again large compared to
the theory truncation uncertainty of less than half a percent. 

In these cases the problem is exacerbated by a choice of experimental cuts that is pathological for at least
one of the calculations, as explained in ref.~\cite{Alioli:2016fum}.  Nevertheless, this highlights the importance of 
precise control over the theoretical calculations, where different techniques for computing higher-order
corrections can result in small but important systematic differences.  In particular, the effect of any type of technical
cutoff or numerical integration artifact must be able to be studied at a level of precision below any other uncertainties,
typically around a few per mille.
These are topics that we have partially addressed in general already, but in this section we illustrate them
in detail by performing a dedicated study of these processes with \MCFMNEW{}.
This follows closely the benchmark comparison study performed in ref.~\cite{Alioli:2016fum}, which collects predictions
from a number of codes including \FEWZ{} and \DYNNLO{}.

\paragraph{\NLO{} results.} As a first point of comparison we have computed the $W^+$ cross section with the settings in 
ref.~\cite{Alioli:2016fum}.\footnote{To do this we had 
	to use the MSTW2008 \NLO{} \PDF{} set even with the \LO{} matrix element, which is not clear from the text.}
First, for the \emph{tuned comparison} cuts and parameters, we find 
a value of \SI{2900.1(1)}{pb} at \NLO{} compatible with the values in table 3 of ref.~\cite{Alioli:2016fum};  for
reference, these range from \SI{2899.0(1)}{pb} to \SI{2899.9(3)}{pb}, with values in-between with larger uncertainties.

The cuts used in ref.~\cite{Alioli:2016fum} have an equal minimum transverse momentum requirement on the lepton and 
neutrino (symmetric cuts). As has been noted in ref.~\cite{Alioli:2016fum}, and much longer ago in ref.~\cite{Frixione:1997ks},
such symmetric cuts are pathological, since one is highly
sensitive to the cancellation of collinear singularities between virtual and real emission corrections.
While being exposed to the edge of the singularity is problematic for an efficient integration in general, calculations
with an intrinsic slicing or cutoff parameter are affected much worse.  Since the jettiness slicing method implemented
in \MCFM{} is just such a calculation it is important for us to study the limitation that this presents in practice.

\begin{figure}
	\centering
	\includegraphics[]{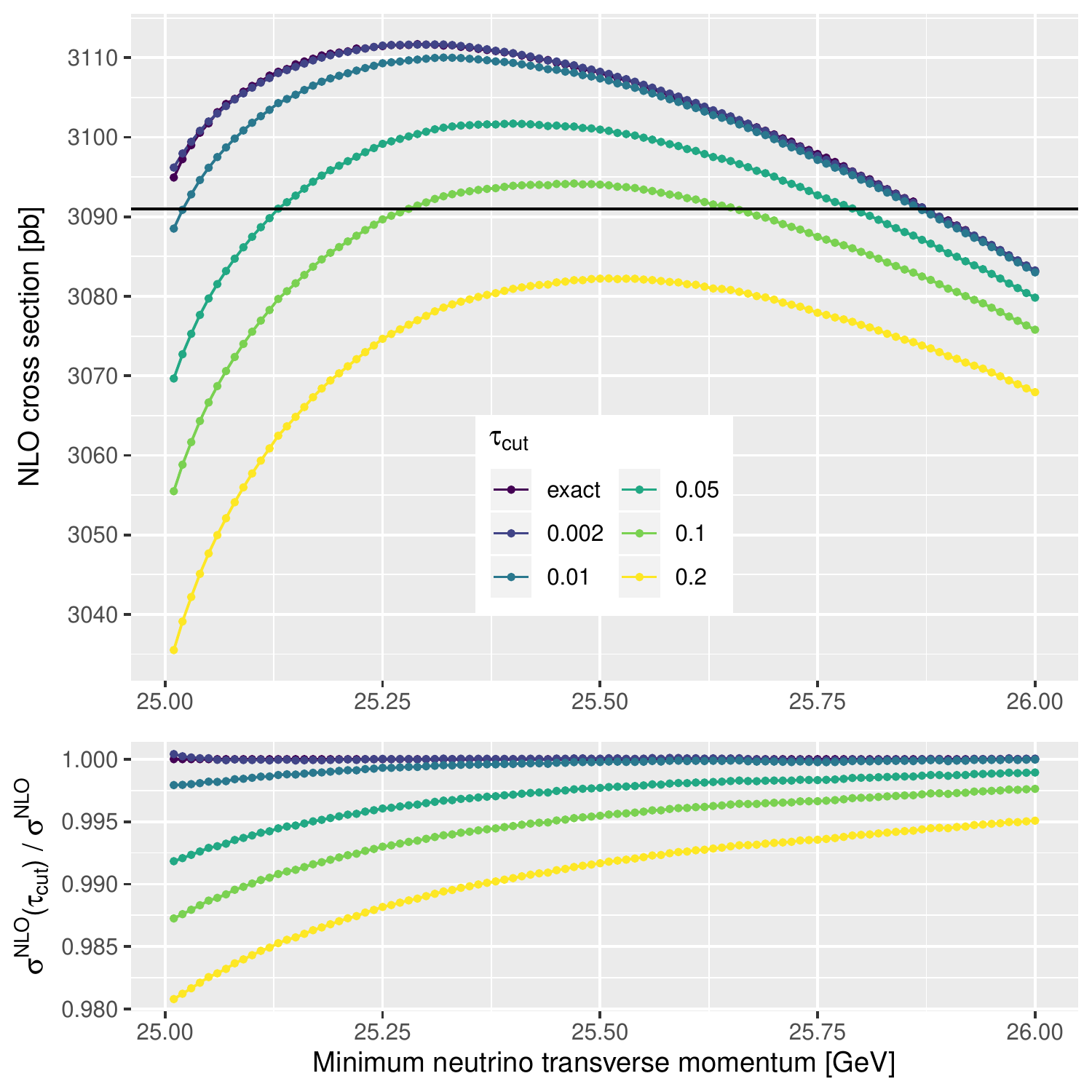}
	\caption{\NLO{} $W^+$ cross section with cuts as in ref.~\cite{Alioli:2016fum}, except that
                the cut on the neutrino transverse momentum is varied (while keeping $p_T^l>\SI{25}{GeV}$).
                The leftmost point corresponds to fully symmetric cuts. The darkest 
		line displays the result obtained using dipole subtraction, the other curves show various 
		choices of $\taucut$ (in GeV).   The upper panel
shows the absolute \NLO{} result and the lower panel the result normalized to the smallest displayed
value of $\taucut=0.002$\,GeV. }
	\label{fig:deltasnlo}
\end{figure}
In the following we use the slightly altered \emph{benchmark} cuts and parameters from 
ref.~\cite{Alioli:2016fum}, used for all their \NNLO{} results. Most importantly, the $p_T$ cuts on lepton and neutrino 
are kept symmetric.
In \cref{fig:deltasnlo} we show the \NLO{} cross section with varying minimum neutrino transverse momentum
between \SIrange[range-phrase=~and~]{25}{26}{\GeV} in steps of \SI{0.01}{\GeV}. To generate this plot we have
computed the neutrino transverse momentum distribution, and subtracted the accumulated $p_T$ bins up to a certain
value from the total result. The darkest line is the exact result 
from a dipole subtraction (exact) calculation. The top panel shows the absolute cross section, while the 
bottom panel shows the ratio to the exact result. The horizontal black line at \SI{3091}{pb} denotes the cross section 
for fully 
symmetric cuts obtained using the exact calculation. The exact line does not meet the black line at \SI{3091}{pb}, 
because the first point with a minimum 
neutrino
transverse momentum of \SI{25.01}{\GeV}, a difference of \SI{0.01}{\GeV} to the symmetric cuts, already leads to a 
shift of about one per mille. As noted in ref.~\cite{Frixione:1997ks}, the approach to the symmetric case has an 
infinite 
slope. The other curves are obtained using jettiness slicing with a finite value of $\taucut$, or correspond to
the result of a fit to these values. The figure clearly
shows that the dependence on $\taucut$ worsens dramatically as the case of symmetric cuts is approached.

It is clear that the case of fully symmetric cuts requires a very small value of $\taucut$ in order to obtain results
that are not affected by the slicing procedure.  Of course, for such a value of  $\taucut$ it is already computationally
very expensive to obtain small numerical uncertainties even at \NLO{}. As we can see, only with a $\taucut$ value of 
$0.002$ is the prediction reliable within one per mille, while for a \SI{1}{\GeV} asymmetry in the cuts one obtains
the same precision for a $\taucut$ value of $0.05$\,GeV. This significantly improves when the asymmetry is increased further.
In practice, the symmetric cuts lead to order of magnitude increases in the computational resources that are required.
Although this alone might not be a reason to abandon symmetric cuts for experimental studies, it emphasizes the importance
of the choice of these cuts for precise numerical comparisons, even at \NLO{}.  Referring to \cref{fig:deltasnlo} again,
we do however notice that -- even for the exact calculation -- the \NLO{} cross section displays an odd behavior below
\SI{25.3}{\GeV}:  the cross section decreases as the cut is reduced.  This is simply the remnant of the problematic
cancellation of singularities, and reflects the fact that the perturbative calculation is not reliable in this region.  This
represents a strong motivation to abandon fully symmetric cuts, as has been done some time ago (for the same reasons)
in the case of diphoton \cite{Catani:2018krb} and dijet \cite{Frixione:1997ks} production.

\begin{figure}
	\centering
	\includegraphics[]{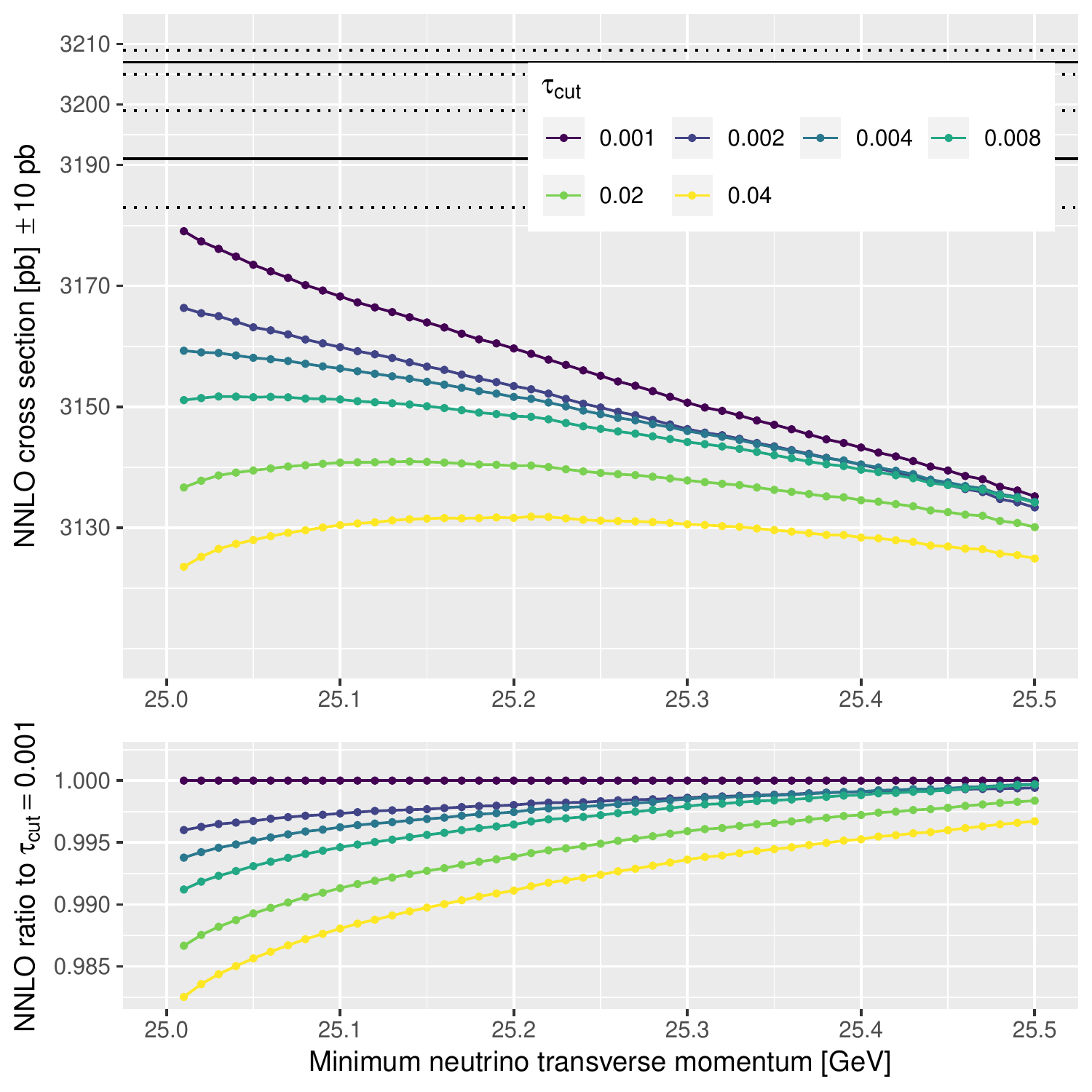}
	\caption{\NNLO{} $W^+$ cross section with cuts as in ref.~\cite{Alioli:2016fum}, $p_T^l>\SI{25}{GeV}$ in dependence 
		of a minimum neutrino transverse momentum cut. The leftmost point corresponds to $p_T^l>\SI{25.01}{GeV}$. Note
		the uncertainty of \SI{10}{pb} on the results.  The upper panel
shows the absolute \NNLO{} result and the lower panel the result normalized to the smallest displayed
value of $\taucut=0.001$\,GeV. The black lines denote the results of \SI{3207(2)}{pb} (FEWZ) and \SI{3191(7)}{pb} (DYNNLO) from 
ref.~\cite{Alioli:2016fum}, solid for the central values and dotted for the uncertainties. 
}
	\label{fig:deltannlo}
\end{figure}
\paragraph{\NNLO{} results.} The equivalent plot for the \NNLO{} calculation is shown in \cref{fig:deltannlo}.
In this case we do not include our $\taucut$ 
fit at \NNLO{}, since for our reasonably reached numerical precision the uncertainties for the 
fitted corrections are 50--100\% due to the presence of the symmetric cuts.
For $\taucut$ values below $0.01$\,GeV it appears that the residual effect of a finite $\taucut$ is at the 
few per mille level.  On the other hand the $\taucut$ dependence for the fully symmetric 
cuts is so large that no apparent convergence towards $\taucut\to0$ is visible. 
This is in contrast to the  $\taucut$ dependence
for $p_T^\nu > \SI{26}{\GeV}$ (not shown in the plot) that is below one percent for $\taucut=0.2$\,GeV.
Since this is still above our expectation from the benchmark cuts (c.f. \cref{fig:taudepnnlo-all}) it
suggests that an asymmetry of at least a few GeV should be used in order to more easily obtain a precision
prediction with this method.

The absolute position of our set of 
curves has an uncertainty of about \SI{10}{pb}, corresponding to $3$ per mille in relative terms, similar
to the result reported from \DYNNLO{} in ref.~\cite{Alioli:2016fum}.  Within this uncertainty, our $\taucut=0.001$\,GeV
result is compatible with the one from \DYNNLO{} but slightly lower than the one from \FEWZ{}.
In order to press the comparison further, say at the 1 per mille level, would require running at even lower
values of $\taucut$ to  better probe the asymptotic logarithmic dependence;  however, we are already limited by the
numerical precision that we are able to reasonably achieve.

Overall we find that it is computationally very expensive to obtain numerically precise values for fully symmetric cuts.
In the presence of these cuts the power corrections are sizeable and force one to use tiny values of $\taucut$, which
amplify cancellations between contributions and challenge all parts of the numerical integration. In addition to this, 
with our binning size of \SI{0.01}{\GeV} and $\taucut=0.001$\,GeV, we already see a difference of $\sim\SI{10}{pb}$ when a 
cut is made at \SI{25.01}{\GeV} instead of \SI{25.00}{\GeV}. This indicates that for the \NNLO{} coefficient
the approach towards fully symmetric cuts is at least as steep as that at \NLO{}.
In light of these findings we believe that not just experimental studies should avoid the symmetric cuts,
but also comparison benchmarks, since no precise comparisons at the per mille level are possible for calculations
with intrinsic cutoff parameters.    We note that all 
observations in this section transfer equivalently to $Z$ production with symmetric cuts on the two leptons.

\begin{figure}
	\centering
	\includegraphics[]{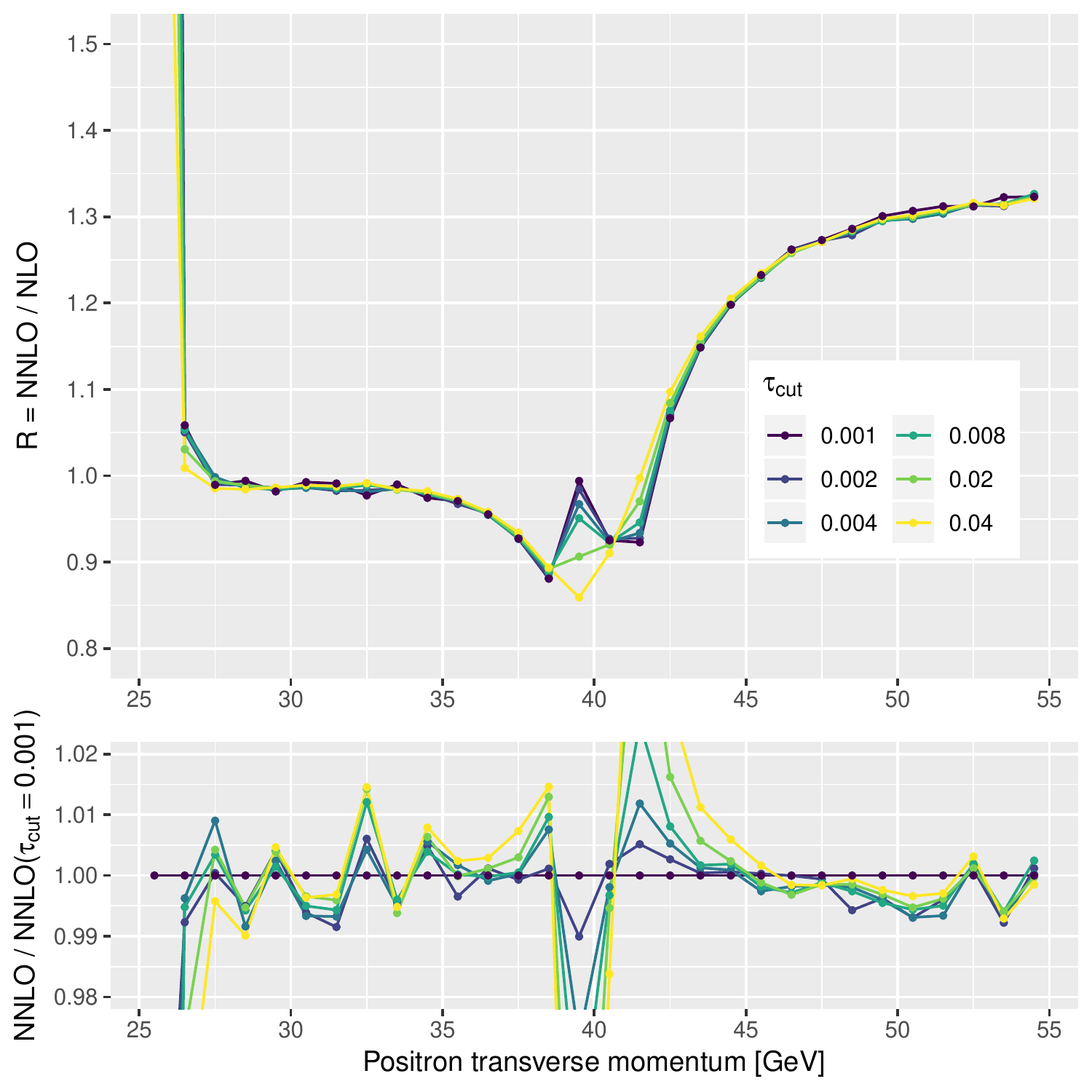}
	\caption{\NNLO{} positron transverse momentum distribution in $W^+$ production. The upper panel shows
		the ratio to the \NLO{} result. The lower panel shows the ratio to the smallest displayed $\taucut$ value of 
		$0.001$.}
	\label{fig:Wplus_lepton_distrib}
\end{figure}

In \cref{fig:Wplus_lepton_distrib} we show the \NNLO{} positron transverse 
momentum 
distribution in $W^+$ production 
normalized to the 
\NLO{} distribution for the \emph{benchmark} cuts and parameters in ref.~\cite{Alioli:2016fum}. This is to be
compared with the corresponding results in fig. 17 of ref.~\cite{Alioli:2016fum}.\footnote{We found that in 
ref.~\cite{Alioli:2016fum} the \NLO{}-normalized \NNLO{} neutrino and lepton distributions use opposite 
\NLO{}-normalization factors. That is, we can fully reproduce their differential distributions for $W^\pm$ when using 
lepton \NLO{} distributions for the normalization of the \NNLO{} neutrino distributions, and vice versa. For example in 
fig.~20 the normalized $l^+$ transverse momentum distribution in the phase 
	space region predominated by $W$+jets shows a $10\%$ difference between the fixed order \NNLO{} result and 
	{\abbrev 
		POWHEG+PYTHIA}, while agreement is found with {\abbrev SHERPA NLO+PS}. With the fixed normalization this 
		situation is exactly reversed. }
The distributions for the neutrino as well as for the positron are virtually indistinguishable, so we only include the 
positron distribution here. Again, the jettiness cutoff effects become sizable towards fully symmetric cuts, but
also in the region around $m_W/2$, which is again sensitive to soft radiation effects and problematic in fixed order
perturbation theory.

\paragraph{Summary.}

In ref.~\cite{Alioli:2016fum} the authors of various codes, including \FEWZ{} and \DYNNLO{}, 
provide benchmark comparisons for $W$ and $Z$ production at \NNLO{} accuracy.
At \NNLO{} the difference between \FEWZ{} and 
\DYNNLO{} inclusively is at the level of 
$0.5\%$, compatible within their respective numerical uncertainties. 
Considering that the experimental uncertainties are at that
level, one now needs to aim for higher precision. Since the benchmark setup uses a choice of cuts
that cause numerical instabilities, this is difficult to achieve in practice.
We therefore advocate the use of a set of cuts that allows for one per mille 
numerical precision, so that numerical errors no longer constitute a substantial uncertainty, and the implementations
and methods can be compared at a more useful level. After having established a common precision baseline, numerically 
difficult cuts could be studied.

To achieve a more useful comparison, we propose the use of cuts that are slightly asymmetric, differing by
at least a few GeV.   Although  highly asymmetric cuts decrease the cross section 
and remove part of the phase space that is calculated at higher orders, the benefits of a small asymmetry
are twofold. First, it removes a region of phase space that is problematic in a fixed order calculation
and results in unphysical predictions (negative cross sections in the transverse momentum distribution). Second, it 
enables precise \NNLO{} predictions with a variety of \NNLO{} codes that in principle should agree.

\clearpage
\section{Conclusions}
\label{sec:conclusions}

With the onset of key hadron-collider measurements at the per mille level, interpretation of the results --
and thus our understanding of nature -- should be limited by uncertainties inherent in the theoretical predictions.
Even with current higher-order predictions, that in some cases have percent-level scale uncertainties,
control of numerical and methodological errors
at the per mille level is required to demonstrate their reliability.  This then allows for the study of input parameters,
and their impact, at the necessary level of precision.
Unfortunately practical resource limitations, set by local workstations 
or even expensive computing clusters, are easily reached by precise calculations at \NNLO{}.
These limitations are even easier to saturate when including scans over additional parameters in the predictions. More 
often than not these limitations result in the introduction of errors, or the use of uncontrolled approximations,
that may lead to a loss of the required precision.  This can have a direct phenomenological 
impact that, for instance, can decide between the advent of a signal for new physics or the continued success
of the \SM{}.

In this paper we have addressed this issue by demonstrating that control at the per-mille level can be achieved
with a new version of the code \MCFM{}.
A key component of the theoretical prediction is the numerical integration over the available phase space,
where any type of technical cutoff or artifact must be able to be controlled below that level of precision. We first ensured 
that the raw numerical predictions can reach these levels of precision and that their errors are reliably estimated.
This has been achieved through our newly implemented fully parallelized \MPI{}+\OMP{} Vegas integration that adaptively 
selects contributions with the largest uncertainties and is fully resumable through automatically written snapshots. Our 
approach allows reliable error estimates for 
precision predictions because it can use a huge number of calls per single integral estimate, i.e. per iteration.
We have compared to a na\"ive parallelization, obtained by combining many independent low-statistics calls,
and find that requires statistical analysis methods to obtain trustworthy error estimates.
For such a situation we recommend the use of the well-known bootstrap/jackknife technique.

A further consideration is that a number of today's \NNLO{} calculations, including those implemented in \MCFM{}, depend on a slicing cutoff to regularize \IR{}
divergences (a jettiness cutoff, $\taucut$, in this paper). Results can only be obtained as an extrapolation $\taucut\to0$, otherwise 
residual finite $\taucut$ effects enter as a systematic error. To estimate the slicing cutoff uncertainties for a 
finite $\taucut$ additional integrations must be performed for a range of $\taucut$ values and the dependence assessed.
This is especially 
importantly differentially, where the residual dependence can be highly non-uniform and large compared to inclusive 
results. This extrapolation is extraordinarily computationally expensive since smaller values of $\taucut$ lead to 
larger numerical cancellation effects. Reaching either the required precision or a small enough $\taucut$ for these 
independent runs is not always possible \cite{}.  To address this we have implemented an automatic correlated
sampling of multiple $\taucut$ values within one single integration run.  This saves orders of magnitude in resources or,
equivalently, improve results with equal resources by orders of magnitude.  Furthermore we have implemented a
boosted jettiness definition for all processes and included leading power corrections, which in combination lead to further order 
of magnitude performance improvements.

Taken together, these improvements ensure that the $\taucut$ dependence can be controlled at the
per mille level on small computing clusters. To assess the reliability and give concrete error 
estimates, we have presented a detailed scheme for estimating the residual $\taucut$ error of \NNLO{} results based on our 
automatic sampling of additional $\taucut$ values and their fitting. Since our fully differential fitting is 
based on the expected behavior in the asymptotic regime, we can reliably exploit it for both improvements and error or reliability 
estimates. We have shown examples where the differential fit improves results by an order of magnitude and furthermore 
makes the differential $\taucut$ dependence uniform. We have also considered the case where the fit is no longer reported to be 
reliable, illustrating the identification of regions that need to be scrutinized further for a valid error estimate.

To avoid introducing the approximation of using \NLO{} matrix elements for the calculation of \PDF{} uncertainties,
our new implementation allows the use of multiple \PDF{} sets and \PDF{} set members simultaneously at \NNLO{}.
They are computed simultaneously in a correlated way, saving many orders of magnitude of computational 
resources compared to uncorrelated integrations.
We have used these improvements to study cases where lower-order matrix elements are used to approximate
full \NNLO{} \PDF{} uncertainties and shown that 
our correlated implementation allows for per mille level comparisons between 
different \PDF{} sets. Studies that discern the impact of different data sets and methods in the fits become directly 
tractable at \NNLO{} at a high precision. We have demonstrated this feature for all \NNLO{} processes in \MCFM{} and 
also for the high-$p_T$ tail of the Higgs boson transverse momentum spectrum at \NLO{}.
In addition we have used our code to attempt to reproduce results contained in the benchmarking exercise of
Ref.~\cite{Alioli:2016fum} but find that the cuts used there prohibit a comparison at the per-mille level.

Modern calculations at the level of \NNLO{} \QCD{} and beyond require the assembly and availability of dozens of 
components that each represents years of work. It is therefore mandatory that such components can be easily and reliably 
reused for further studies. \MCFM{} provides a repository of such work, and has been used several times as a basis for fixed 
order \NNLO{} and \NLO{} \SM{} calculations, resummed calculations, as well as implementations of physics  beyond the \SM{}. 
Its library of amplitudes has found use in dozens of projects and studies. Given this track record, the overhaul of   
all core components of the code described here is an important step to increase its usability and reliability,
and to keep it an important tool for both experimentalists and theorists.
In particular, the features and efficiency gains documented here will enable a public distribution of
\NNLO{} $W^\pm, Z$ and $H$ production processes in association with a jet in the near
future.

\paragraph{Acknowledgments.}

This work was supported by the U.S.\ Department of Energy under award
No.\ DE-SC0008347.  This document was prepared using the resources of
the Fermi National Accelerator Laboratory (Fermilab), a
U.S. Department of Energy, Office of Science, HEP User
Facility. Fermilab is managed by Fermi Research Alliance, LLC (FRA),
acting under Contract No.\ DE-AC02-07CH11359.

\appendix

\section{Detailed description of new features}
\label{sec:newfeatures-app}

In this section we present the new and modified features in \MCFM{} and describe how to use them on a technical level, 
complementing the new \MCFM{} manual. With all re-implemented 
and newly implemented components we strive for Fortran 2008 compliance, making explicit use of its features. Following 
the Fortran standard furthermore allows us to achieve compatibility with not just the GNU compiler.
In previous versions of \MCFM{} the licensing was unclear, since none was specified. We now license all code 
under the GNU GPL 3 license\footnote{See \url{https://www.gnu.org/licenses/gpl-3.0.en.html}.}.

\paragraph{Improved input file mechanism.}

We have implemented a new input file mechanism based on the configuration file parser \texttt{config\_fortran} 
\cite{JTeunis}.
This INI-like file format no longer depends on a strict ordering of configuration elements, allows easy access to
configuration elements through a single global configuration object, and makes it easy to add new configuration
options of scalar and array numerical and string types. Using the parser package also allows one
to override or specify all configuration options as command line arguments to \MCFM{}, for example running
\MCFM{} like \texttt{./mcfm\_omp input.ini -general\%nproc=200 -general\%part=nlo}. This is useful for batch
parameter run scripts. Settings can also be overridden with additional input files that specify just a subset of options.

\paragraph{New histogramming.}

We replaced the previous Fortran77 implementation of histograms, that used routines from 1988 by M. Mangano,
with a new suite of routines.
The new histogram implementation allows for any number of histograms with any number of bins,
each of which is dynamically allocated. Furthermore, everything is also handled in a fully multi-threaded approach with 
the integration. For each \OMP{} thread temporary 
histograms are allocated which are then reduced to a single one after each integration iteration, so that
no \OMP{} locks (critical regions) are required. 

\paragraph{New Vegas integration, part-adaptive and resumable.}

The previous implementation of the Vegas routine was based on Numerical Recipes code. We have re-implemented
Vegas and the surrounding integration routines. All parts of a \NLO{} or \NNLO{} calculation are now
chosen adaptively based on the largest absolute numerical uncertainty. A precision goal can be set
in the input file as well as a $\chi^2/\text{it}$ goal and a precision goal for the warmup run. If
the goals for the warmup are not reached, the warmup repeats with twice the number of calls. With the
setting \texttt{writeintermediate} one can control whether histograms are written in intermediate
stages during the integration. Enabling the setting \texttt{readin} allows one to resume the integration
from any point from a previous run. Snapshots saving the whole integration state are saved automatically.
When resuming, the only parameter that the user can safely officially change is the \texttt{precisiongoal}. Further
tweak configuration options to control the stages of the integration have been introduced, which can provide
benefits over the default settings in certain situations.

The section \texttt{integration} in the configuration file allows for tweaks in the following way. The precision
goal can be adjusted by setting \texttt{precisiongoal} to a relative precision that should be reached. Similarly,
the settings \texttt{warmupprecisiongoal} and \texttt{warmupchisqgoal} control the minimum relative precision and
$\chi^2/\text{it}$ for the warmup phase of \texttt{iterbatchwarmup} (default 5) iterations. If the warmup criterion
fails, the number of calls is increased by a factor of two. The calls per iteration get increased by a factor of 
\texttt{callboost} (default 4) after the warmup. From then on the number of calls per iteration is 
increased by a factor of \texttt{itercallmult} (default 1.4) for a total of \texttt{iterbatch1} iterations. After these 
first \texttt{iterbatch1} iterations, the increase happens for every \texttt{iterbatch2} iterations. The setting 
\texttt{maxcallsperiter} controls the cap for the number of calls per iteration. The 
number of Vegas grid subdivisions can be controlled with \texttt{ndmx} (default 100).

The purpose of these settings is a fine control in certain situations. For example to compute expensive \PDF{} 
uncertainties, one wants a relatively precise warmup run (where additional \PDF{} sets are not sampled) and as few 
calls as necessary afterwards: For the plots in this paper we thus chose a relative warmup precision goal of $10\%$, 
and set \texttt{callboost} to $0.25$. This means that the first \texttt{iterbatch1} iterations after the warmup run 
only 
with a quarter of the calls than during the warmup. This precision is sufficient to compute precise \PDF{} 
uncertainties, when making use of the strong correlations as in \MCFMNEW{}. Any further iterations come in batches of 
\texttt{iterbatch2}, which we set to $1$. It allows for a quick switching to parts of the \NNLO{} cross section that 
have the largest uncertainty. For normal applications one wants to boost the number of calls after the warmup 
significantly, so a default value of \texttt{callboost=4} is chosen.

We provide default settings for the initial number of calls per iteration for all components of a \NNLO{} calculation. 
They can be overridden with the following settings in the \texttt{integration} section: \texttt{initcallslord}, 
\texttt{initcallsnlovirt}, \texttt{initcallsnloreal}, \texttt{initcallsnlofrag} for parts of a \NLO{} calculations,
\texttt{initcallssnlobelow}, \texttt{initcallssnloabove} for parts of a \SCET{} based \NLO{} calculation, and 
\texttt{initcallsnnlobelow}, \texttt{initcallsnnlovirtabove}, as well as \texttt{initcallsnnlorealabove} for the parts 
of the \NNLO{} coefficient.

\paragraph{Low discrepancy sequence.}
\MCFM{}-8.0 and prior relied on a linear congruential generator implementation from Numerical Recipes for the 
generation of a pseudo-random sequence. With newer versions the \MT{} implementation of the C++ standard library is 
used, and with this version of \MCFM{} we include an implementation of the Sobol low discrepancy sequence based on the 
code sobseq \cite{Vugt2016} with initialization numbers from ref.~\cite{Joe2010}. The Sobol sequence is 
used by default and can be toggled using the flag \texttt{usesobol = .true.} in the \texttt{integration} 
section of 
the input file. \cref{sec:integrationuncertainties}. When running in \MPI{} mode, the number of nodes has to be a power 
of two for the Sobol sequence, because we use it in a strided manner. Otherwise the code will automatically fall back 
to 
using the \MT{} sequence with seed value \texttt{seed} in the integration section of the input file. A \texttt{seed} 
value of $0$ denotes a randomly initialized seed.

\paragraph{Fully parallelized OMP+MPI use of LHAPDF.}

In previous versions of \MCFM{} calls to \LHAPDF{} were forced to access from only a single OMP thread
through a lock. This is because the interface was based on the old LHAglue interface, part
of \LHAPDF{}. We have written an interface to \LHAPDF{} from scratch based on the new object oriented treatment
of \PDF{}s in \LHAPDF{} 6. For each OMP thread we initialize a copy of the used \PDF{} members which
can be called fully concurrently. The amount of \PDF{} sets with or without \PDF{} uncertainties is only limited
by the available system memory. The memory usage of \MCFM{} can then range from roughly 20MB when only one central 
\PDF{} grid is being used, to $\sim 7.4$ GB when 32 OMP threads fully load
all members of the \PDF{} sets \texttt{CT14nnlo}, \texttt{MMHT2014nnlo68cl}, \texttt{ABMP16als118\_5\_nnlo},
 \texttt{NNPDF30\_nnlo\_as\_0118}, \texttt{NNPDF31\_nnlo\_as\_0118} and \texttt{PDF4LHC15\_nnlo\_30} for
 \PDF{} uncertainties. The total number of members for these grids is 371, each loaded for every of the
 32 OMP threads.
 
Since each OMP thread allocates its own copy of \PDF{} members and histograms we have no need to introduce
any OMP locks. On the other hand the memory usage increases and one runs into being CPU cache or DRAM
bandwidth bound earlier. In practice, we find that this is still faster than having OMP locks, which directly
decrease the speedup in the spirit of Amdahl's law. Ideally the \LHAPDF{} library should be improved to allow for 
thread-safe calls with just one memory allocation.

\paragraph{Histograms for additional values of $\taucut$, $\mu_R,\mu_F$ and multiple \PDF{}s.}
When using the automatic scale variation, in addition to the normal histograms, additional
histograms with filenames \texttt{\_scale\_XY\_} are generated, where \texttt{X} is a placeholder for the 
renormalization scale variation and \texttt{Y} for the factorization scale variation. \texttt{X} and \texttt{Y} can 
either be \texttt{u} for an upwards variation by a factor of two, \texttt{d} for a downwards variation by a factor of 
two, or just \texttt{-} if no change of that scale was made. The envelope of maximum and minimum can then easily be 
obtained.

For the sampling of additional values of $\taucut$ for \NLO{} and \NNLO{} calculations using jettiness subtractions, 
additional histograms with filenames \texttt{\_taucut\_XXX\_} are written. Here \texttt{XXX} is a placeholder for the 
chosen $\taucut$ values in the optional array \texttt{taucutarray}, if specified, or one of the five automatically 
chosen values. These additional files 
only contain the \emph{differences} to the nominal choice of 
$\taucut$, so that $\Delta\sigma(\tau_\text{cut,nominal}) - \Delta\sigma(\tau_\text{cut,i})$ is stored. If 
\texttt{taucutarray} has not 
been specified, the automatic choice of additional
$\taucut$ values is enabled based on the default nominal $\taucut$ for the process or the users choice of the nominal 
$\taucut$ value as specified in \texttt{taucut}.
In addition a file with \texttt{\_taucutfit\_} is generated, which in addition to the fitted corrections and its 
uncertainty includes columns for the maximum relative integration uncertainty for the additionally sampled $\taucut$ 
values and the 
reduced $\chi^2$ of the fit. With the procedure in \cref{sec:benchmark}, the fit together with the individual $\taucut$ 
histograms allows the user to assess the systematic $\taucut$ error and possibly improve results.

When multiple \PDF{} sets are chosen, additional files with the names of the \PDF{} sets are generated. In case
\PDF{} uncertainties are enabled, the histograms also include the upper and lower bounds of the \PDF{} uncertainties.

\paragraph{User cuts, histograms and re-weighting.}

Modifying the plotting routines in the files \texttt{src/User/nplotter*.f} allows for modification of the pre-defined 
histograms and addition of any number of arbitrary observables. The routine \texttt{gencuts\_user} can be adjusted  in 
the file
\texttt{src/User/gencuts\_user.f90} for additional cuts after the jet algorithm has performed the 
clustering. In the same file the routine \texttt{reweight\_user} can be modified to include a manual re-weighting
for all integral contributions. This can be used to obtain improved uncertainties in, for example, tails of distributions.
One example is included in the subdirectory \texttt{examples}, where the \texttt{reweight\_user} function approximately
flattens the Higgs transverse momentum distribution, leading to equal relative uncertainties even in the tail at 
\SI{1}{TeV}.

\subsection{Compatibility with the Intel compiler and benchmarks}

Previous versions of \MCFM{} were developed using \texttt{gfortran} as a compiler. \MCFM{} contained code that did not 
follow 
a specific Fortran standard, and was only compatible with using \texttt{gfortran}. We fixed code that did not compile 
or work with the recent Intel Fortran compiler \texttt{ifort} 19.0.1. This does not mean that we claim to be strictly 
standards 
compliant with a specific Fortran version, but we aim to be compliant with Fortran 2008. We now fully support GCC 
versions newer than $7$ and Intel compilers newer than $19$. There might still be compatibility issues with other 
Fortran compilers, but we are happy to receive bug reports for any issues regarding compilation, that are not due to a 
lack of modern Fortran 2008 features. To use the Intel compiler one has to change the USEINTEL flag in the files 
\texttt{Install} and \texttt{makefile}
to \texttt{YES}.

To see whether \MCFM{} can make use of potential Intel compiler improvements over the GNU compiler 
collection (GCC) we benchmarked 
the double
real emission component of Higgs production at \NNLO{}. We perform tests on our cluster with
Intel Xeon 64-bit X5650 2.67 GHz Westmere CPUs, where two six-core CPUs are run in a dual-socket mode with a total
of twelve cores. Similarly, we have an AMD 6128 HE Opteron 2GHz quad-socket eight-core setup, thus each having
32 cores per node.

We benchmark both the Intel and GCC compilers on both the Intel and AMD systems. On the Intel system we use 16 MPI 
processes each with 12 OMP threads, 
and on the AMD system we have 8 MPI processes using 32 OMP threads. With this we have the same total
clockrate of \SI{512}{GHz} for each setup. For all benchmarks we find that the scaling is perfect up to this size, that 
is if we use half the number of MPI or OMP threads we double our run-time.

We first try both the Intel fortran compiler 19.0.1 and GCC 9.1.0 on the Intel system with the highest generic
optimization flags \texttt{-O3 -xsse4.2} and \texttt{-O3 -march=westmere}, respectively. Furthermore,
we lower the optimizations to \texttt{-O2} each and remove the processor specific optimization flags
\texttt{-xsse4.2} and \texttt{-march=westmere}, respectively. All our benchmark run-times in the following are consistent
within $\pm \SI{0.5}{\s}$.

We do not support enabling unsafe math operations with \texttt{-ffast-math}, since the code 
relies on the knowledge of NaN values and checks on those. Such checks would be skipped with the meta 
flag\texttt{-ffast-math} which sets \texttt{-ffinite-math-only}.

\begin{table}[]
	\centering
	\caption{Benchmark results on the Intel system with $10\cdot25$M calls distributed over 16 MPI processes, each 
	using 12 
	OMP 
	threads. The GCC version is 9.1.0 and the Intel Fortran compiler 19.0.1}
	\vspace{1em}
	\begin{tabular}{@{}ll@{}}
		\toprule
		\multicolumn{1}{c}{\textbf{Compiler/flags}} & \multicolumn{1}{c}{\textbf{wall time $\pm$ 0.5s}} \\ \midrule
		ifort -O3 -xsse4.2                          & 90s                                           \\
		ifort -O2 -xsse4.2                          & 86s                                           \\
		ifort -O2									& 90s											\\
		ifort -O1									& 103s 											\\
		gfortran -O3 -march=westmere                & 101s                                          \\ 
		gfortran -O2 -march=westmere		        & 105s											\\
		gfortran -O2								& 105s											\\
		gfortran -O1								& 110s											\\
		\bottomrule
	\end{tabular}
	\label{tab:benchintel}
\end{table}

The benchmark results in \cref{tab:benchintel} show that using the Intel compiler, performance benefits of $\simeq 
10-20\%$ can be achieved. Our goal here is not to go beyond this and check
whether exactly equivalent optimization flags have been used in both cases. Enabling optimizations beyond \texttt{-O2} 
have little impact, but come with a penalty for the Intel compiler and with a slight benefit for gfortran. We also
notice that processor specific optimizations play no significant role. This might also be in part due to the fact
that \MCFM{} does not offer much space for (automatic) vectorization optimizations. To summarize, the default 
optimization flags of 
\texttt{-O2} should be sufficient in most cases. We do not expect that the conclusions from these benchmarks
change for different processes. On the other hand if computing \PDF{} uncertainties, the majority of time
is used by \LHAPDF{} and different optimization flags for \LHAPDF{} might play a role then.
We performed the same benchmark with an older version of GCC, version 7.1.0 using \texttt{-O2} optimizations, and found 
that the run-times are the same as for the newer version.

Finally, we performed some benchmarks
on our AMD setup and found that it is $\simeq 2.5$ times slower for the same total clockrate. Using the Intel compiler
for the AMD setup decreased the performance by another $\simeq 30\%$. This is likely due to the fact that the Intel
compiler already optimizes for the general Intel architecture. 

These benchmarks try to give a general impression and might depend in detail on the process, the
number of histograms and whether to compute \PDF{} uncertainties, for example. Especially when computing
\PDF{} uncertainties the perfect scaling we tested here might break down since the computation can become
memory bound. We discuss this caveat in more detail in \cref{subsec:performance}.

\subsection{Remarks on memory bound performance issues}
\label{subsec:performance}
To get numerically precise predictions at the per mille level for \NNLO{} cross sections,
already hundreds of million of calls are necessary. Obtaining \PDF{} uncertainties using
those \NNLO{} matrix elements significantly increases the computational time. In a simplified view
the total computational time composes as $N_\text{calls}*(T + N_\PDF{}\cdot T_\PDF{})$, where $T$ is the
computational effort for the matrix element piece, and the \PDF{} part is proportional
to the time calling the \PDF{} evolution $N_\PDF{}$ times and code related to performing the convolutions.
For tree level matrix element evaluations, usually also $T \lll T_\PDF{}$ holds, so the computational cost
grows linearly with the number of \PDF{}s.

This naive picture breaks down in practice when a lot of \PDF{}s
are sampled together with a lot of histograms or histogram bins. The total memory necessary
to store all the histogram information grows like $N_\PDF{} \cdot N_\text{bins} \cdot N_\text{thr.}$,
where $N_\PDF{}$ is the number of \PDF{} members, $N_\text{bins}$ the number of histogram bins
summed over all histograms and $N_\text{thr.}$ is the number of \OMP{} threads. The factor $N_\text{thr.}$
enters since we have thread-local storage to avoid \OMP{} locks.
The values are stored in double precision, so the total memory used is
$N_\PDF{} \cdot N_\text{bins} \cdot N_\text{thr.} \cdot 8 \text{ bytes}$.

 Assuming for example, 300 \PDF{} members,
10 histograms with each 20 bins and 12 threads, this sums up to \SI{720}{kb} of memory. For the
virtual corrections and \LO{} pieces, one has to update this amount of memory once for each call. For the real
emission matrix elements one has to accumulate all dipole contributions, so this number additionally scales
with the number of dipole contributions. All the histogram updates are usually fully vectorized for modern 
superscalar processors with SSE and/or AVX extensions. But if this used memory is too large and does not easily fit 
into the 
CPU core caches anymore, a transfer to and from DRAM happens, which now is the limiting factor and significantly slows
down the computation. Because for that reason, one should work with a minimal number of necessary histograms when 
working
with a lot of \PDF{} members. This is especially important for cluster setups that are not optimized towards
memory bound applications, non-NUMA systems. For example in our cluster we have relatively old AMD Opteron quad-socket 
eight-core nodes with little CPU cache, and with above numbers we are already limited in wall-time improvements with 
using $\sim16$ cores. Then reducing 
the number of histograms will \emph{significantly} improve the performance. In principle one can reduce the histogram 
precision to single precision and cut memory transfer and storage in half, while doubling the computational speed. This 
might lead to problems with accumulated rounding errors though, and we have not investigated this further, since in
practice one can sufficiently limit the number of histograms or \PDF{} sets.

\section{Supporting plots for the jackknife-after-bootstrap procedure}
\label{app:bootstrap}

\begin{figure}
	\centering
 	\includegraphics[width=\columnwidth]{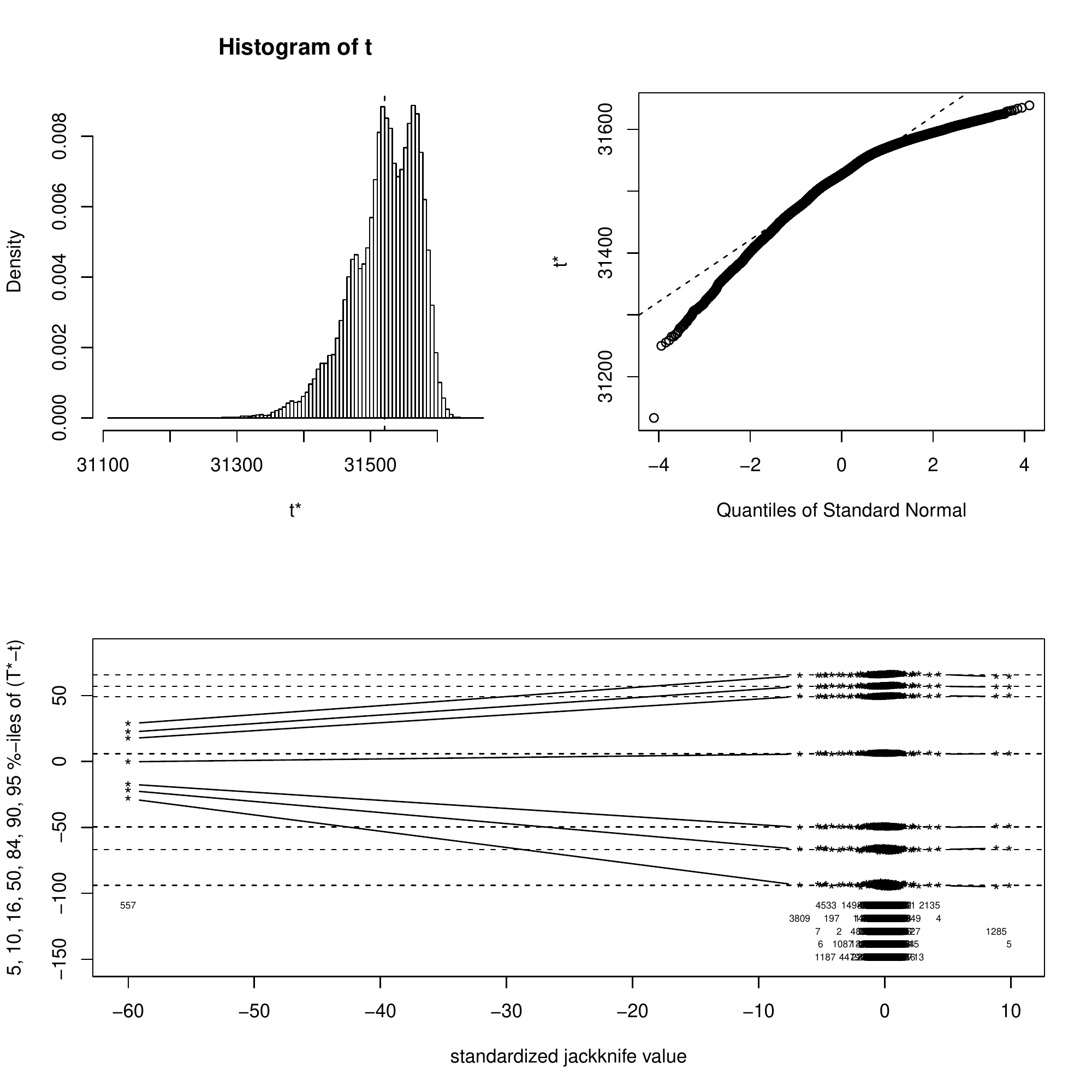}
 	\caption{Result of applying the bootstrap technique to our \MT{} data set of about 4500 data points with 10 million 
 		calls each. The sample size is not Gaussian due to one significant outlier. }
 	\label{fig:bootstrap_mt_first}
 \end{figure}

\begin{figure}
 	\centering
 	\includegraphics[width=\columnwidth]{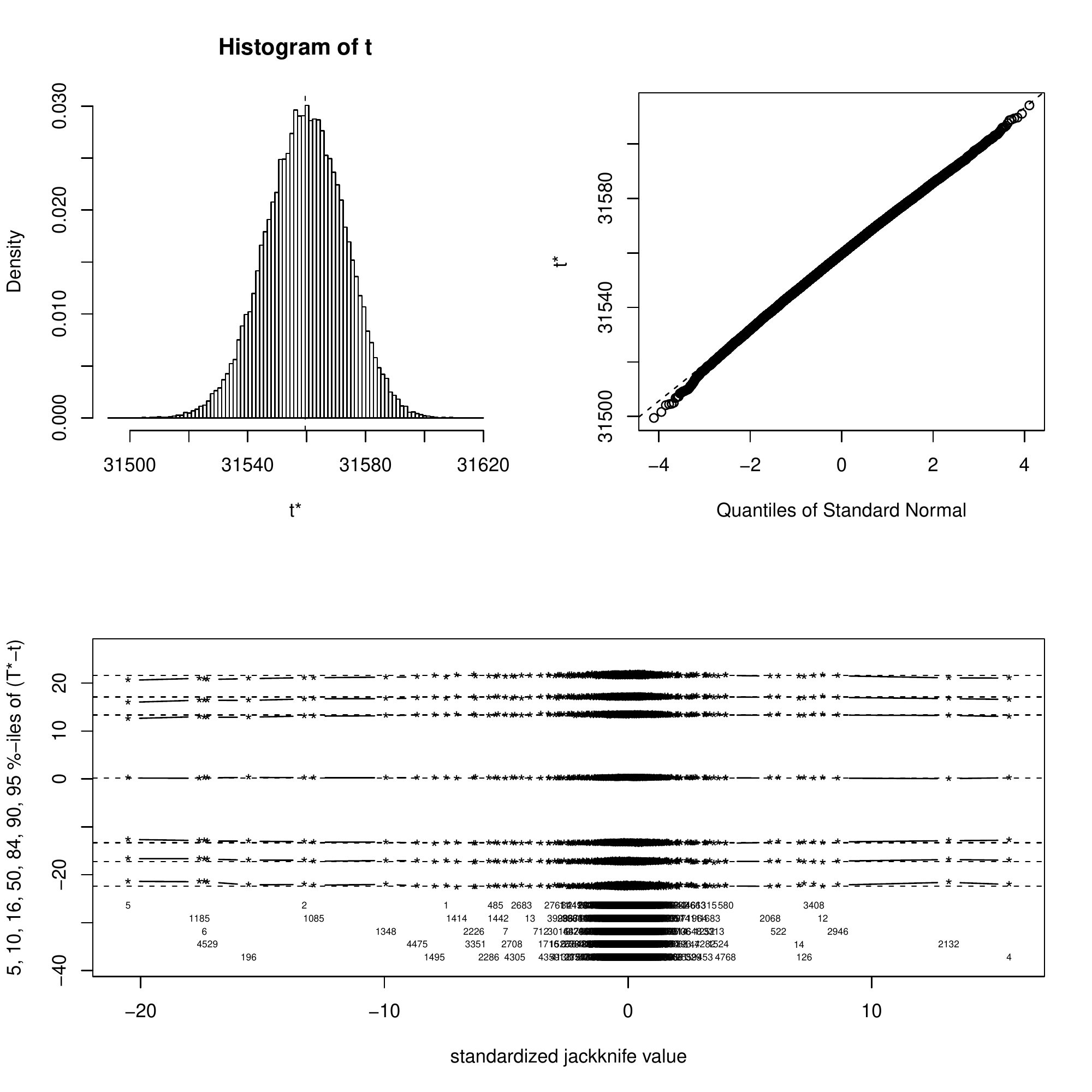}
 	\caption{Result of the applying to bootstrap technique to our \MT{} data set of about 4500 data points with 10 
 	million 
 		calls each.  The worst four outliers as shown in the jackknife plot in \cref{fig:bootstrap_mt_first} have been 
 		removed. The result is $31559 \pm 13$. }
 	\label{fig:bootstrap_mt_second}
 \end{figure}

In \cref{fig:bootstrap_mt_first} and \cref{fig:bootstrap_mt_second} the lower panel displays a visualization of the 
jackknife 
after bootstrap technique. Each point represents one of the $N$ data points that is being left out. The dashed lines 
represent quantiles of the bootstrap distribution. \Cref{fig:bootstrap_mt_first} shows that leaving out the single 
point 
number $557$ would significantly shrink the percentiles and make the Gaussian distribution symmetric. After removing
the outlier \cref{fig:bootstrap_mt_second} is obtained, where now no single point would significantly modify the 
bootstrap 
distribution. For more details we refer to \cite{boot3}.

\bibliographystyle{JHEP}
\bibliography{refs}

\end{document}